\documentclass[multphys,vecphys]{promult}
\pdfoutput=1

\usepackage{makeidx}     
\usepackage{graphicx}    
\usepackage{multicol}    

\usepackage{amsmath}
\usepackage{amssymb}
\usepackage{epsfig}
\usepackage{graphicx}
\usepackage{cite}
\usepackage{multirow}
\usepackage{longtable}
\usepackage{lscape}
\usepackage{bbm}
\usepackage{dcolumn}
\usepackage{rotating}
\usepackage[squaren]{SIunits}

\makeindex             

\newcommand{\bi}{\begin{itemize}}
\newcommand{\ei}{\end{itemize}}

\newcommand{\be}{\begin{equation}}
\newcommand{\ee}{\end{equation}}
\newcommand{\bea}{\begin{eqnarray}}
\newcommand{\eea}{\end{eqnarray}}

\newcommand{\ie}{{\it i.e.}}

\newcommand{\eg}{{\it e.g.}}

\newcommand{\cf}{{\it cf.}}

\newcommand{\etc}{{\it etc.}}
\newcommand{\eq}{Eq.}

\newcommand{\fig}{Fig.}

\newcommand{\Ref}{Ref.}
\newcommand{\Refs}{Refs.}
\newcommand{\Sec}{Sec.}
\newcommand{\Secs}{Secs.}

\newcommand{\equ}[1]{\eq~(\ref{equ:#1})}
\newcommand{\figu}[1]{\fig~\ref{fig:#1}}

\newcommand{\rhat}{{\widehat R}}

\newcommand{\xhat}{{\widehat X}}

\begin{document}

\title*{Neutrinos from Cosmic Accelerators \newline Including Magnetic Field and Flavor Effects}
\titlerunning{Neutrinos from Cosmic Accelerators ...} 


\author{Walter Winter\thanks{E-mail: {\tt winter@physik.uni-wuerzburg.de}}
}
\institute{ Institut f{\"u}r Theoretische Physik und Astrophysik,  Universit{\"a}t W{\"u}rzburg,  \\
       97074 W{\"u}rzburg, Germany} 

\maketitle

We review the particle physics ingredients affecting the normalization, shape, and flavor composition of astrophysical neutrinos fluxes, such as different production modes, magnetic field effects on the secondaries (muons, pions, kaons), and flavor mixing, where we focus on $p\gamma$ interactions. We also discuss the interplay with neutrino propagation and detection, including the possibility to detect flavor and its application in particle physics, and the use of the Glashow resonance to discriminate $p\gamma$ from $pp$ interactions in the source.
We illustrate the implications on fluxes and flavor composition with two different models: 1) the target photon spectrum is dominated by synchrotron emission of co-accelerated electrons and 2) the target photon spectrum follows the observed photon spectrum of gamma-ray bursts.  In the latter case, the multi-messenger extrapolation from the gamma-ray fluence to the expected neutrino flux is highlighted.

%
%
%

\section{Introduction}

In addition to gamma-ray and cosmic ray instruments, neutrino telescopes, such as IceCube~\cite{Ahrens:2003ix} or ANTARES~\cite{Aslanides:1999vq}, provide interesting data on the sources of the highest-energetic particles found in the universe, so-called ``cosmic accelerators''; see \Refs~\cite{Learned:2000sw,Halzen:2002pg,Chiarusi:2009ng,Katz:2011ke} for reviews. In particular, neutrinos are a prominent way to search for the origin of the cosmic rays, or to discriminate between leptonic and hadronic models describing the observed spectral energy distribution of photons.
There are numerous possible sources, see \Ref~\cite{Becker:2007sv} for an overview and \Ref~\cite{Rachen:1998fd} for the general theory. Very interesting extragalactic candidates for neutrino and cosmic ray production may be gamma-ray bursts (GRBs)~\cite{Waxman:1997ti} and active galactic nuclei (AGNs)~\cite{Stecker:1991vm,Mannheim:1993,Stecker:2005hn}.
The most stringent bounds for these sources, which are expected to be roughly uniformly distributed over the sky, so far come from IceCube, which  has recently released data on time-integrated~\cite{Abbasi:2010rd} and time-dependent~\cite{Abbasi:2011ara} point source searches, GRB neutrino searches~\cite{Abbasi:2011qc}, and diffuse flux searches~\cite{Abbasi:2011jx}. So far, no astrophysical neutrinos have been detected, which has been for a long time consistent with generic Waxman-Bahcall~\cite{Waxman:1998yy} and Mannheim-Protheroe-Rachen~\cite{Mannheim:1998wp} bounds. However, data from IC40 and IC59, referring to the 40 and 59 string configuration of IceCube, respectively, start to significantly exceed these bounds, see \Refs~\cite{Abbasi:2011qc,IC59ProcICRC,Abbasi:2011jx}, which is in tension with the corresponding  neutrino production models, such as \Refs~\cite{Waxman:1997ti,Guetta:2003wi,Abbasi:2009ig} for GRBs. For example, neutrino data may soon challenge the paradigm that GRB fireballs are the sources of the ultra-high energy cosmic rays (UHECR)~\cite{Ahlers:2011jj}. For constraints to AGN models, see, \eg,  \Ref~\cite{Arguelles:2010yj}.
As a consequence, the age of truth has come for neutrino astrophysics, which is also the age of precision: Especially since data are available now, it is necessary to critically review the underlying assumptions from both the astrophysics and particle physics perspectives, and to develop the models from rough analytical estimates into more accurate numerical predictions. 

In this review, we focus on the minimal set of particle physics ingredients for the neutrino production, which must be present in virtually  all sources, using several specific examples. We focus on photohadronic ($p\gamma$) interactions for the meson production, with the exception of \Sec~\ref{sec:glashow}. We do not only discuss the predicted neutrino flux, but also the flavor and neutrino-antineutrino composition at source and detector. The discussed effects include: 
\begin{itemize}
\item
 Additional pion production modes, such as t-channel (direct) and multi-pion production; see, \eg, \Refs~\cite{Mucke:1998mk,Mucke:1999yb,Mucke:2000rn,Murase:2005hy,Hummer:2010vx,Baerwald:2010fk,Baerwald:2011ee}.
\item
 Neutron and kaon production; see, \eg, \Refs~\cite{Asano:2006zzb,Kachelriess:2006fi,Kachelriess:2007tr,Hummer:2010ai,Moharana:2010su,Baerwald:2010fk,Moharana:2011hh,Baerwald:2011ee}.\footnote{In general, there is also an additional contribution from charmed meson production, see \Refs~\cite{Kachelriess:2006fi,Kachelriess:2007tr} for a detailed comparison.}
\item
 The cooling and decay of secondaries (pions, muons, and kaons); see, \eg, \Refs~\cite{Kashti:2005qa,Lipari:2007su,Kachelriess:2007tr,Reynoso:2008gs,Hummer:2010ai,Baerwald:2010fk,Baerwald:2011ee}
\item
 Flavor mixing and possible new physics effects; see \Ref~\cite{Pakvasa:2008nx} for a review.
\item
 The helicity-dependence of the muon decays; see, \eg, \Refs~\cite{Lipari:2007su,Hummer:2010vx}.\footnote{This effect has been discussed earlier in the context of atmospheric neutrinos, see, for instance, \Refs~\cite{Barr:1988rb,Barr:1989ru,Lipari:1993hd}.}
\item
 Spectral effects, such as the energy dependence of the mean free path of the protons, and their impact on the prediction; see, \eg, \Refs~\cite{Hummer:GRBFireball,Li:new}.
\item
 The impact of the maximal proton energy on the neutrino spectrum; see, \eg, \Ref~\cite{Moharana:2011hh}.
\item
 Deviations from the frequently used $E_\nu^{-2}$ neutrino flux assumption; see, \eg, \Ref~\cite{Winter:2011jr}.
\end{itemize}
While many of these effects have been studied elsewhere in the literature, we mainly show examples generated with the NeuCosmA (``Neutrinos from Cosmic Accelerators'') software in this review to present them in a self-consistent way.

The structure of this review is as follows: In \Sec~\ref{sec:overview}, we give a simplified picture for the connection among neutrinos, cosmic rays, and gamma-rays. Then  in \Sec~\ref{sec:production}, we review the minimal set of ingredients for neutrino production from the particle physics perspective. In \Sec~\ref{sec:propdet}, we discuss neutrino propagation and detection, including the possibility to detect flavor and the use of the Glashow resonance, where we illustrate how new physics can be tested in the neutrino propagation in \Sec~\ref{sec:newphys}. We furthermore present two specific applications: a generic AGN-like model in \Sec~\ref{sec:gensource} and a model for GRBs in \Sec~\ref{sec:grb}, where the main difference is the model for the target photons. Then we finally summarize in \Sec~\ref{sec:summary}.

\section{Neutrinos and the multi-messenger connection}
\label{sec:overview}

Here we outline a simplified picture of the neutrino or cosmic ray source, as often used in the literature, whereas we add extra ingredients in the next section. In this approach, charged mesons originate from $pp$ or $p \gamma$ interactions, where we focus on $p \gamma$ (photohadronic) interactions in this work; see, \eg,  \Refs~\cite{Kelner:2006tc,Vissani:2011ea} for $pp$ interactions, which may be dominant for particular source classes, such as supernova remnants.  In the simplest possible picture, charged pions are produced by the $\Delta(1232)$-resonance 
\begin{equation}
	p + \gamma \rightarrow \Delta^+ \rightarrow \left\{\begin{array}{lc} n + \pi^+ & \frac{1}{3} \text{ of all cases} \\[0.2cm]  p + \pi^0 & \frac{2}{3} \text{ of all cases} \end{array} \right.  . \label{equ:Delta}
\end{equation}
While this process is not sufficient for state-of-the-art models for neutrino production, it is very useful to illustrate a few qualitative points common to many cosmic ray and neutrino production models.
The protons on the l.h.s. of \equ{Delta} are typically assumed to be injected into the interaction volume with an $(E_p')^{-\alpha}$ spectrum\footnote{Here primed parameters refer to the shock rest frame (SRF), where as unprimed parameters to the observer's frame.} coming from Fermi shock acceleration, where $\alpha \sim 2$. They interact with the photons on the l.h.s. of \equ{Delta} with energy  $\varepsilon' \sim (0.2 - 0.3) \, \mathrm{GeV}/E_p'$. While the assumptions for the injected protons are similar for most models (except from the minimal and maximal energies), the target photons are typically described in a model- and source-dependent way, such as by:
\begin{enumerate}
\item
 synchrotron emission from co-accelerated electrons or positrons,
\item
 thermal emission, such as from an accretion disk,
\item
 a more complicated combination of radiation processes, 
\item
 an estimate inferred from the gamma-ray observation,
\end{enumerate}
just to name a few examples.
While any realistic simulation of a particular source will imply option 3),  the other options typically rely on fewer parameters and may be good approximations in many cases. In particular, a reliable prediction for the photon density in the source may be obtained from the gamma-ray observation, option 4), if the photons can  escape. In fact, we will use option 1) in \Sec~\ref{sec:gensource} and option 4) in \Sec~\ref{sec:grb}.

After an interaction between proton and photon, the particles on the r.h.s. of \equ{Delta} are produced with the given branching ratios. The neutrinos then originate from $\pi^+$ decays via the decay chain
\begin{eqnarray}
\pi^+ & \rightarrow & \mu^+ + \nu_\mu \, ,\nonumber \\
& & \mu^+ \rightarrow e^+ + \nu_e + \bar{\nu}_\mu \, , \label{equ:piplusdec}  
\end{eqnarray}
where in this standard picture $\nu_e:\nu_\mu:\nu_\tau$ are produced in the ratio $1:2:0$ if the polarities (neutrinos and antineutrinos) are added. In addition, high-energy gamma-rays are produced by 
\begin{equation}
 \pi^0 \overset{98.8\%}{\longrightarrow} \gamma + \gamma \, . \label{equ:pizerodec}
\end{equation}
These are typically emitted from the source at lower energies due to electromagnetic cascades,  in addition to   gamma-rays escaping from the interaction volume (the ones contributing on the l.h.s. of  \equ{Delta}).

From \equ{Delta}, we can also illustrate the production of cosmic ray protons, ignoring for the moment that the composition of cosmic rays may be heavier at high energies~\cite{Abraham:2010yv}. First of all, some of the protons injected into the interaction volume on the l.h.s. of \equ{Delta} may escape, leading to cosmic ray production. However, even if the protons are magnetically confined, the neutrons on the r.h.s. of \equ{Delta}, which are electrically neutral, can easily escape if the source is optically thin to neutron escape. After decay (typically outside the source)
\begin{equation}
n \rightarrow  p + e^- + \bar{\nu}_e  \label{equ:ndec}  \, ,
\end{equation}
they lead to cosmic ray flux and an additional $\bar \nu_e$ neutrino flux which is an unavoidable consequence of the interactions in \equ{Delta}. The cosmic ray protons with energies above  $6 \cdot 10^{19} \, \mathrm{eV}$  interact with the cosmic microwave background (CMB) photons by \equ{Delta}, leading to the so-called Greisen-Zatsepin-Kuzmin (GZK) cutoff~\cite{Greisen:1966jv,Zatsepin:1966jv}. However, according to \equ{Delta}, charged pions are produced in these interactions as well, which means that an additional neutrino flux should come with that, which is often called ``cosmogenic neutrino flux''.

In summary, the photohadronic interaction in \equ{Delta} offers a self-consistent picture for a cosmic ray source, with a possible connection among cosmic ray, neutrino, and gamma-ray escape. In specific models, however, that does not mean that a large neutrino flux is guaranteed for every cosmic ray source. For instance, the interaction rate for the process in \equ{Delta}, which depends on the photon density, may be low. 

\section{Simulation of neutrino sources}
\label{sec:production}

Here we give a more detailed generic picture of the simulation of neutrino sources from the particle physics perspective with the minimal set of ingredients, without using a specific model. A flowchart summarizing the contents of this section, which can be followed during the reading, is given in \figu{flowchart}.

\subsection{Photohadronic interactions}

In order to describe the processes within an interaction volume (one zone in the simplest case), two kinds of spectra are needed: 
 $Q'(E')$ (in units of $\mathrm{GeV^{-1} \, cm^{-3} \, s^{-1}}$) describes the number of particles injected or ejected per volume element and energy interval, and $N'(E')$ (in units of $\mathrm{GeV^{-1} \, cm^{-3}}$) describes the particle density per energy interval.  The secondary meson injection rate $Q_b'(E_b')$  for a pion or kaon species $b$ produced in photohadronic interactions is given by (following \Ref~\cite{Hummer:2010vx})
\begin{equation}
Q_b'(E_b') = \int\limits_{E_b'}^{\infty} \frac{dE_p'}{E_p'} \, N_p'(E_p') \, \int\limits_{0}^{\infty} c \, d\varepsilon' \, N_\gamma'(\varepsilon') \,  R_b( x,y )  \, .
\label{equ:prodmaster}
\end{equation}
Here $x=E_b'/E_p'$ is the fraction of energy going into the secondary, $y \equiv (E_p'\varepsilon')/m_p$,\footnote{ Here $y$ can be related to the center-of-mass energy by $s=m_p^2+2 m_p (1-\cos \theta_{p\gamma}) \, y$, where $\theta_{p\gamma}$ is the angle between proton and photon momentum; $\theta_{p\gamma}=\pi$ corresponds to heads-on collisions.} and $R_b( x,y )$ is the ``response function''. If many interaction types are considered, the response function can be quite complicated. However, if it is known from particle physics,  \equ{prodmaster} can be used to compute the secondary injection for arbitrary proton and photon spectra.
The important point here is that the secondary production depends on the product normalization of the proton density $N_p'(E_p')$ and the target photon density $N_\gamma'(\varepsilon')$ within the interaction volume. Thus a higher proton density can be compensated by a lower photon density, and vice versa.\footnote{Strictly speaking, this degeneracy only holds in the absence of any other radiation process. For example, inverse Compton scattering depends on the photon density individually, which means that the input spectral shapes (especially $N'_\gamma$ in \equ{prodmaster}) will be modified if that process contributes significantly.}  Another implication of \equ{prodmaster} is that the secondary production depends on the densities within the source $N'$, not the injection rates $Q'$. Of course, one cannot look into the source, but can only observe cosmic messengers escaping from the source. As we will demonstrate later, the observed/ejected photon or cosmic ray spectrum $Q'$ is only  directly representative for the corresponding density spectrum within the source $N'$ if ``trivial'' escape is the leading process, \ie, $Q'=N'/t'_{\mathrm{esc}}$ with $t'_{\mathrm{esc}} \sim R'/c$ and $R'$ the size of the interaction region. For this section,  \equ{prodmaster} is used as a starting point for the computation of the neutrino fluxes, where we do not discuss the origin of spectral shape and normalization of $N_p'$ and $N_\gamma'$. In practice, typically an $(E_p')^{-2}$ injection spectrum is assumed for the protons, as mentioned above, where the maximal energy is limited by synchrotron and adiabatic losses. The photon density may be a consequence of a complicated interplay of radiation processes. In either case, the derivation of these densities depends on the model, and we will show several examples in \Secs~\ref{sec:gensource} and~\ref{sec:grb}.

\begin{figure}[t]
\begin{center}
\includegraphics[width=\textwidth]{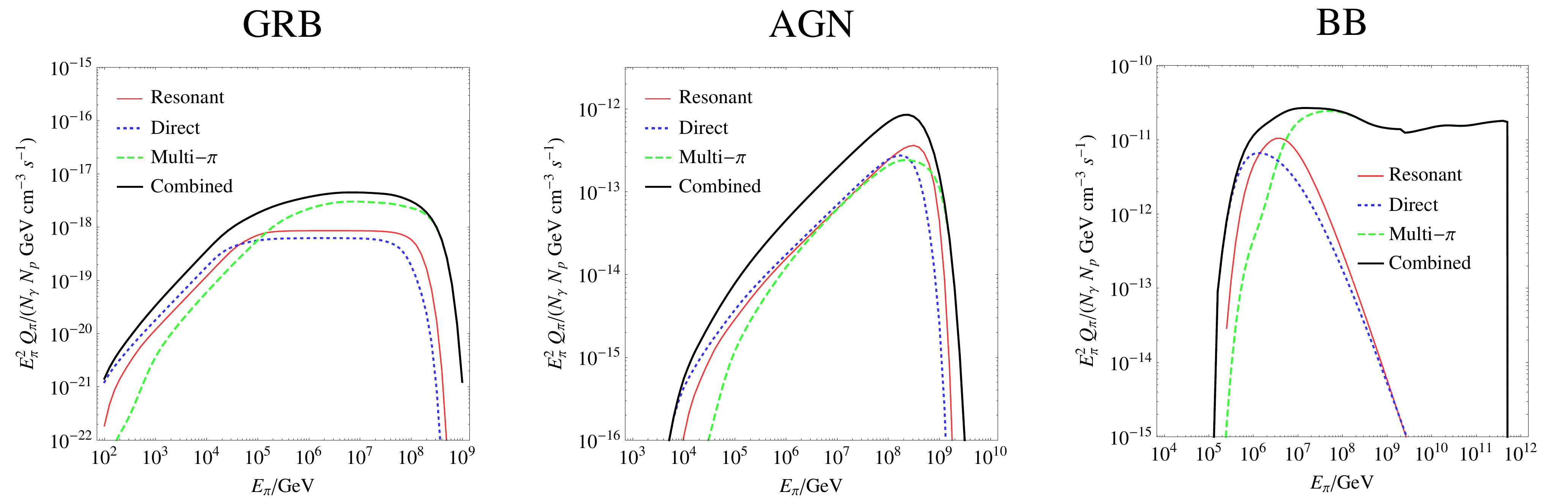}
\end{center}
\caption{\label{fig:photo} Contributions of different $\pi^+$ production modes to the total spectrum for a typical GRB, an AGN, and a ($10 \, \mathrm{eV}$) black body (BB)  target photon density. Figure taken from \Ref~\cite{Hummer:2010vx}.}
\end{figure}

Compared to a numerical approach, \equ{Delta} has limitations to describe the meson production. First of all, additional pion production modes contribute, such as higher resonances, direct (t-channel), and multi-pion production, which will also lead to $\pi^-$ production (\cf, \figu{flowchart}). These are not as easy to describe as the $\Delta$-resonance because of different shapes of the cross sections and more complicated kinematics. The Monte Carlo package SOPHIA~\cite{Mucke:1999yb} can deal with these interactions. In order to increase the efficiency, often parameterizations of SOPHIA are used, such as \Refs~\cite{Kelner:2008ke,Hummer:2010vx}. In the following, we use \Ref~\cite{Hummer:2010vx} (model Sim-B), because the secondary muons and pions are needed explicitly. We show the impact of the resonant production (including higher resonances) on $\pi^+$ production in \figu{photo} for a typical GRB, an AGN, and a ($10 \, \mathrm{eV}$) black body (BB)  target photon field. As one can read off from this figure, the resonances always give a reasonable first estimate for the actual pion production, but quantitatively they only dominate at the breaks. In addition, multi-pion processes can change the spectral shape significantly, such as for the GRB example, which is a consequence of the cross section dependence on the center-of-mass energy. As a further limitation, note that \equ{Delta} does not describe kaon production and subsequent decay into neutrinos, where the leading modes are given by
\begin{eqnarray}
 p + \gamma & \rightarrow & K^+ + \Lambda/\Sigma \, ,  \label{equ:kprod} \\
& & K^+ \rightarrow  \mu^+ + \nu_\mu  \label{equ:kplusdec} \, . 
\end{eqnarray}
The branching ratio for the leading channel in \equ{kplusdec} is about 64\%.
The second-most-important decay mode is $K^\pm \rightarrow \pi^\pm +  \pi^0$ (20.7\%).
The other decay modes account for 16\%, no more than about 5\% each.
Because interesting effects can only be expected in the energy range with the most energetic neutrinos, we only use the direct decays from the leading mode.

In the literature, the $\Delta$-resonance approximation in \equ{Delta} is, even in analytical approaches, typically not taken literally. For example, a simple case is the approximation by Waxman and Bahcall~\cite{Waxman:1997ti} (``WB $\Delta$-approx.''), for which one can write the response function as
\begin{equation}
 R_{\pi^\pm}(x,y) = 0.5 \times \delta(x-0.2) \times 500 \,  \mu \mathrm{barn} \times  \left\{ \begin{array}{ll} 
0 & 2 y < 0.2 \, \mathrm{GeV} \\[0.2cm]
 \, 1- \left( \frac{0.2 \, \mathrm{GeV}}{2 y} \right)^2  & 0.2 \, \mathrm{GeV} \le 2 y \\ & < 0.4 \, \mathrm{GeV} \\[0.2cm]
\frac{0.12 \, \mathrm{GeV}^2 }{(2 y)^2} & 2 y \ge 0.4 \, \mathrm{GeV}
\end{array} \right.
\label{equ:photosimp}
\end{equation}
which implies that charged pions are produced in 50\% of all cases, and these take 20\% of the proton energy. In addition, the width of the $\Delta$-resonance is taken into account.\footnote{Note that this description is slightly more accurate than \Ref~\cite{Waxman:1997ti}, which uses an additional integral approximation.} In fact, this function peaks at $2y \simeq 0.4 \, \mathrm{GeV}$, which is higher than the threshold for photohadronic interactions -- and it is even a little bit higher in the numerical calculation. The reason is that for the threshold often head-on collisions are assumed ($\theta_{p \gamma} = \pi$), whereas these only contribute a small part to the total number of interactions. Using \equ{photosimp} in \equ{prodmaster} and re-writing the integral over $\varepsilon'$ in one over $y$, it is easy to show that for power law spectra $N_p' \propto (E_p')^{-\alpha}$ and $N_\gamma' \propto (\varepsilon')^{-\beta}$ 
\begin{equation}
 Q'_{\pi}(E'_ {\pi}) \propto (E'_\pi)^{-\alpha + \beta -1} \, . \label{equ:qsimple}
\end{equation}
This means that the pion spectral index depends on both the proton and photon spectra, where it is inversely proportional to the photon spectral index.
As we will see below, the neutrino spectrum follows the pion spectrum, which means that the assumption of an $E_\nu^{-2}$ spectrum for the neutrinos, as it is often used in data analysis and many models in the literature, is only a valid assumption for $\beta \simeq 1$ -- which is roughly observed for GRBs below the break. On the other hand, if the target photons come from synchrotron emission, such a hard photon spectrum is not possible, and $\beta \gtrsim 3/2$ may be more plausible if the electrons are injected with a spectral index similar to the protons. As a consequence, the neutrino spectrum becomes harder. In addition, multi-pion processes in the photohadronic interactions will act in the same direction and make the neutrino spectrum even harder, \cf, \figu{photo}; see also \Ref~\cite{Baerwald:2010fk} for a detailed comparison between the approximation in \equ{photosimp} and the numerics. Note that for $pp$ interactions with ``cold'' (non-relativistic) protons, the $E_\nu^{-2}$ assumption may be plausible~\cite{Kelner:2006tc}. We discuss the implications of the $E_\nu^{-2}$ assumption for the detector response in \Sec~\ref{sec:propdet}.

\subsection{Decays of secondaries}

The weak decays of pions and muons are described in detail in \Ref~\cite{Lipari:2007su}.
 In general, in case of ultra-relativistic parents of type $a$, the distribution of the daughter particle of type $b$ takes a scaling form in order to obtain for the energy spectra
\begin{equation}
 Q_b'(E_b')=\sum_a\int_{E_b'}^\infty\,dE_a'\,N_a'(E_a')\,(t_\mathrm{dec}')^{-1}\,\frac{1}{E_a'}\,F_{a\rightarrow b}\left(\frac{E_b'}{E_a'}\right) 
\label{equ:weakdecay}
\end{equation}
summed over all parent species.
 The functions $F_{a\rightarrow b}$ for pion, kaon and helicity dependent muon decays can be read off from \Ref~\cite{Lipari:2007su} (Sec. IV). Consider the simplified case of a $\delta$-function for $F_{a\rightarrow b}$. For instance, for decays of neutrons in \equ{ndec}, one may approximate
\begin{equation}
 F_{n\rightarrow \bar{\nu}_e}=\delta\left(\frac{E_\nu'}{E_n'}-\chi_n \right)
\end{equation}
with $\chi_n=5.1\times10^{-4}$.  If the neutrons do not interact, $Q_n'=N_n' (t_\mathrm{dec}')^{-1}$ (see below), and we find for the neutrino injection
\equ{weakdecay}
\begin{equation}
 Q_{\bar{\nu}_e}'(E_\nu')=\frac{1}{\chi_n} Q_n'\left(\frac{E_\nu'}{\chi_n}\right) \, .
\label{equ:ninj}
\end{equation}
In this case, the neutrino spectrum follows the neutron spectrum. If the neutrinos originate from pion decays, the neutrino injection follows the pion injection spectrum by similar arguments.

\begin{figure}[t]
\begin{center}
\includegraphics[width=\textwidth]{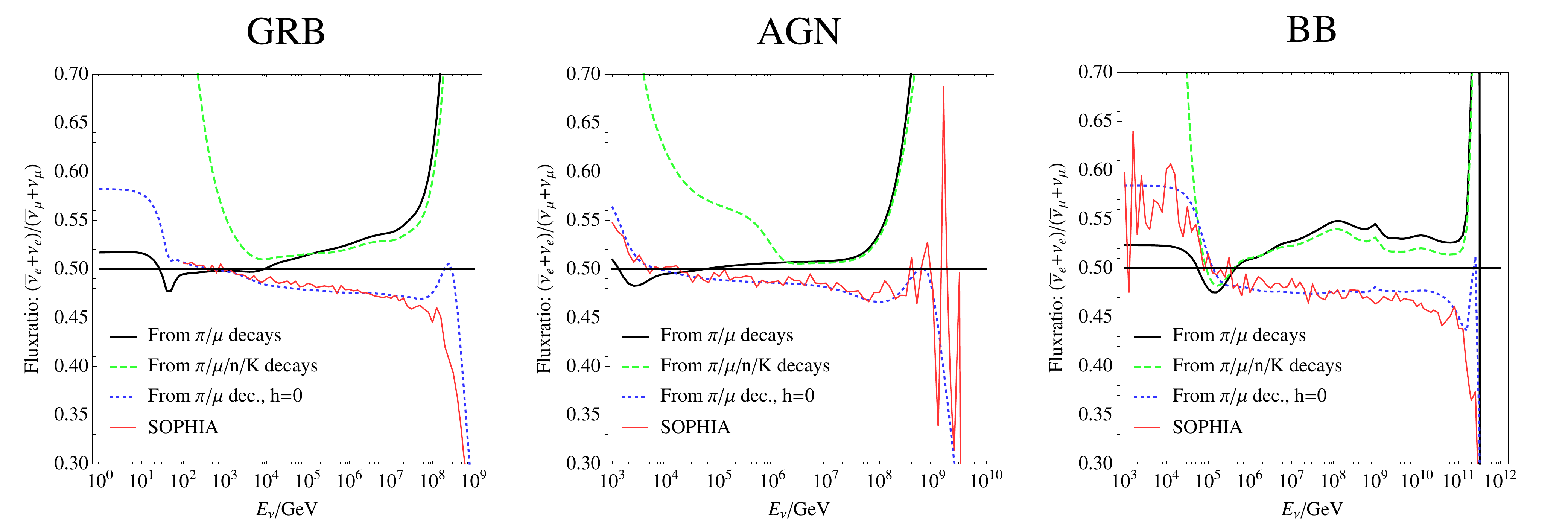}
\end{center}
\caption{\label{fig:helicity} Flux ratio at the source between electron and muon neutrinos as a function of the neutrino energy, for the same objects as in \figu{photo}.  No secondary cooling is included in this figure. Figure taken from \Ref~\cite{Hummer:2010vx}.}
\end{figure}

Since the decays of muons are helicity dependent, \ie, $F_{a\rightarrow b}$ is different for left- and right-handed muons, it is necessary to keep track of these two species separately (\cf, \figu{flowchart}). Although the effects of the helicity-dependent muon decays on the fluxes are probably small, the flavor composition is slightly affected, depending on the parameters of the source.  We illustrate this effect  in \figu{helicity} for the GRB, AGN, and black-body examples in \figu{photo}. In this figure, the horizontal line corresponds to the standard assumption, \ie, neutrinos being produced in the flavor composition $\nu_e:\nu_\mu:\nu_\tau$ of $1:2:0$. Including the scaling of the secondary decays, the dotted (blue) curves are obtained if the helicity of the muons is averaged over. From the comparison with the light (red) solid curves, it is clear that this assumption is implemented in SOPHIA. On the other hand, as pointed out in \Ref~\cite{Lipari:2007su}, keeping track of the muon helicity slightly changes the flavor composition, see dark (black) solid curves, which are significantly different from the standard assumption and the helicity-averaged version. However, it is also clear from \figu{helicity} that the deviation from the standard prediction depends on the input spectra, as it is smaller for the AGN than the GRB example.
The dashed (green) curves show the contributions of the neutron and kaon decays, which affect the flavor composition at very low and high energies, respectively.  

\subsection{Cooling of secondaries}
\label{sec:cooling}

In order to describe the cooling of the secondary pions, muons, and kaons, we use the steady state approach, \ie, we do not allow for an explicit time-dependence since the statistics of neutrino observations is typically expected to be low. The steady state equation for the particle spectrum, assuming continuous energy losses, is given by
\begin{equation}
\label{equ:steadstate}
Q'(E')=\frac{\partial}{\partial E'}\left(b'(E') \, N'(E')\right)+\frac{N'(E')}{t'_\mathrm{esc}} \, ,
\end{equation}
with $t'_\mathrm{esc}(E')$ the characteristic escape time, $b'(E')=-E' \, t'^{-1}_{\mathrm{loss}}$ with
$t'^{-1}_{\mathrm{loss}}(E')=-1/E' \, dE'/dt'$ the rate characterizing energy losses. This differential equation balances the particle injection on the l.h.s. with energy losses and escape on the r.h.s. of the equation. Note again that the steady density $N'(E')$ is needed for the photohadronic interactions in \equ{prodmaster}, not the actual injection spectrum. In addition, note that if there are no energy losses ($b'=0$), one has immediately $Q'(E')=N'(E')/t'_\mathrm{esc}$ from \equ{steadstate}, which we have already used above. If decay is the dominant escape mechanism, one finds $Q'(E') \propto N'(E')/E'$.

\begin{figure}[t]
\begin{center}
 \includegraphics[width=0.5\textwidth]{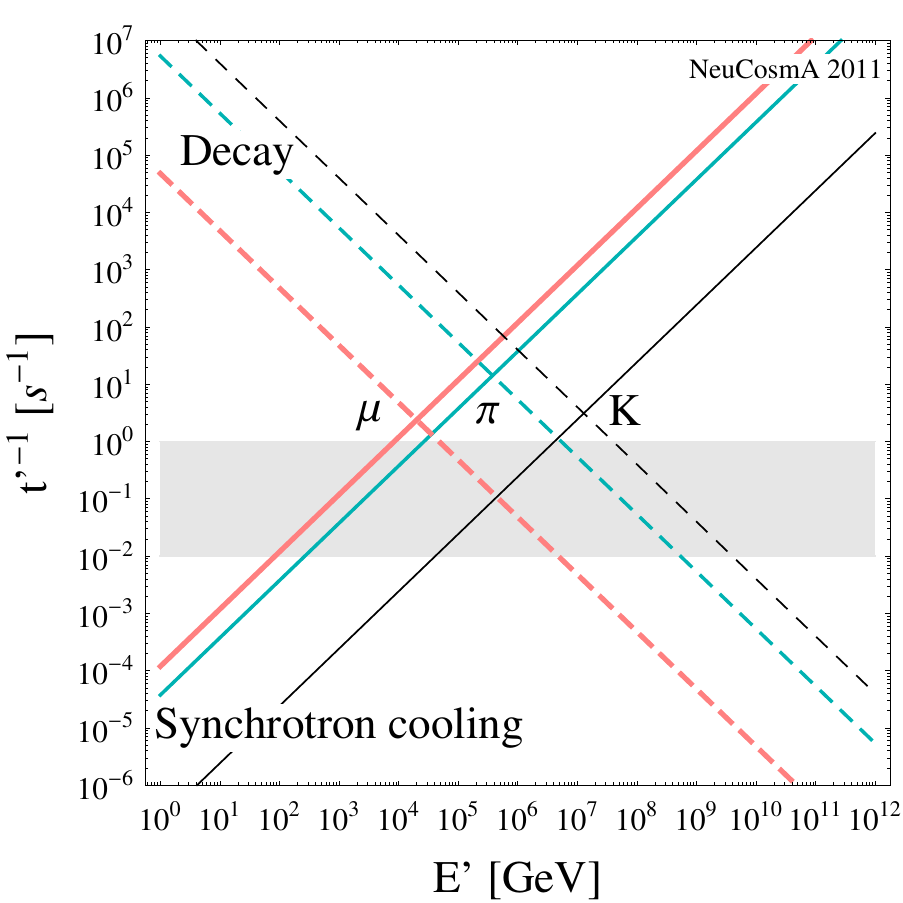}
\end{center}
 \caption{\label{fig:times} Synchrotron cooling and decay rates for pions, muons, and kaons as a function of energy for a magnetic field of $B' \simeq 300\,\mathrm{kG}$. The gray-shaded region shows the estimated range for the escape (or adiabatic cooling) rate for GRBs for a variability timescale ranging from $0.01 \, \mathrm{s}$ to $1 \, \mathrm{s}$. Figure taken from \Ref~\cite{Baerwald:2011ee}.}
\end{figure}

While the primary proton and photon spectra in \equ{prodmaster} could be affected by a number of radiation processes, for the neutrino fluxes and flavor compositions, at least the processes of the secondaries (pions, muons, kaons) are important. We illustrate the synchrotron cooling and decay rates for pions, muons, and kaons as a function of energy in \figu{times}. As one can easily see in the figure, for any species, decay dominates at low energies, while synchrotron cooling dominates at high energies. Other cooling or escape processes are often sub-dominant, as illustrated by the gray-shaded region for an adiabatic cooling components in GRBs. 
The two curves meet at a critical energy $E'_c$ for each species, which is different depending on the particle physics parameters.
As a consequence of \equ{steadstate}, the corresponding steady spectra $N'$ are loss-steepened by two powers above
\begin{equation}
E'_c = \sqrt{ \frac{9 \pi \epsilon_0 m^5 c^7}{\tau_0 e^4 B'^2}} \, ,
\label{equ:ec}
\end{equation}
where synchrotron and decay rates are equal.
These critical energies depend on the particle physics properties of the parent, \ie,  the mass $m$ and the rest frame lifetime $\tau_0$, and the magnetic field $B'$ as the only astrophysical parameter. It is therefore a very robust prediction, and might allow for the only direct measurement of $B'$.  Re-scaling the magnetic field shifts the critical energies by a constant amount on the horizontal axis, but does not change the spacing in the logarithmic picture. In \figu{times}, we also show the estimated range for adiabatic cooling as additional cooling component (shaded region), which may have some impact, especially on the muons, in extreme cases. In these cases, the height of the spectral peaks will be somewhat reduced, but the qualitative picture does not change.

\begin{figure}[t]
 \includegraphics[width=0.5\textwidth]{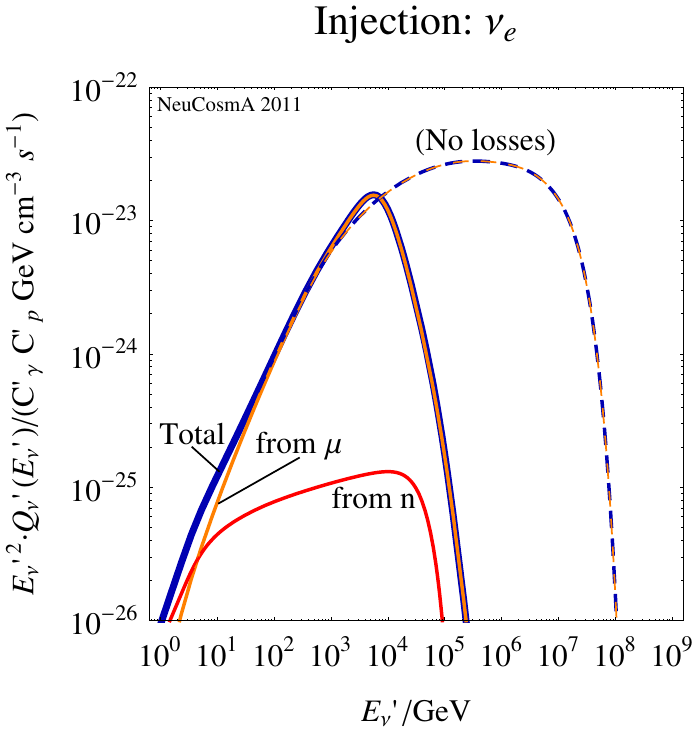} \includegraphics[width=0.5\textwidth]{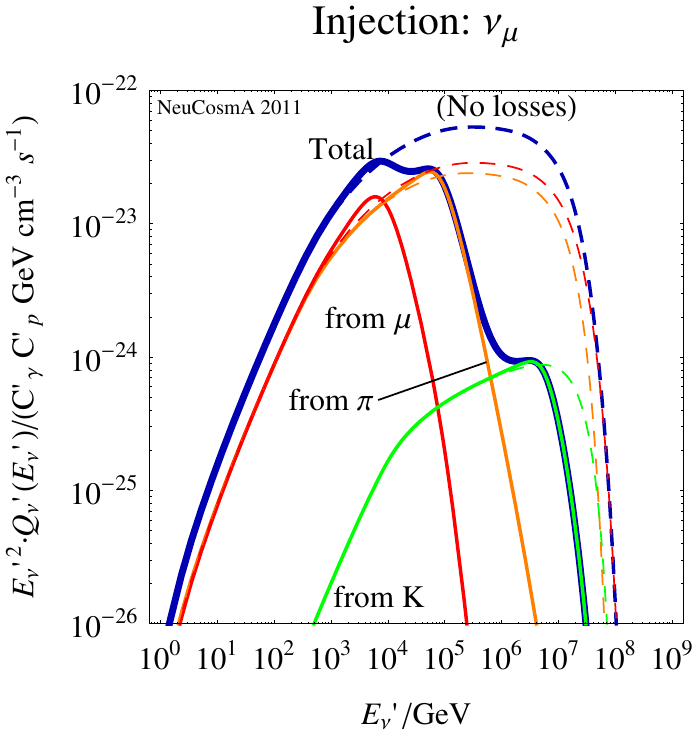}
 \caption{\label{fig:nuinj} Injection spectra for electron neutrinos (left panel) and muon neutrinos (right panel) at the source for a GRB example, where the individual contributions from the parents are shown (neutrinos and antineutrinos added). Figure taken from \Ref~\cite{Baerwald:2011ee}.}
\end{figure}

We show in \figu{nuinj} the consequences for the spectral shape at one GRB example. Here the injection spectrum of electron neutrinos (left panel) and muon neutrinos (right panel) is shown, including the individual contributions from the parents. First of all, one can read off from this figure that in the case of no losses (dashed curves) the spectral shapes of all contributions are very similar, and the neutrino fluxes add in a trivial manner. A change of the primary spectra $N_p'$ and $N_\gamma'$ in \equ{prodmaster} may change the shape of the dashed curves, but almost in the same way for all curves. If the synchrotron losses are switched on (solid curves), the spectral split predicted by \equ{ec} (see also \figu{times}) among the neutrino spectra coming from different parent species can be clearly seen in the right panel. One can also see a small pile-up effect coming from the muon decays, \ie, a small region where the cooled muons coming from  higher energies pile up and dominate, and lead to a higher flux than in the ``no losses'' case. In the left panel, only two spectra are shown, since only muon or neutron decays may produce the electron flavor. Because of the very small $\chi_n$ in \equ{ninj}, the neutron decays only show up at low energies. Since the decay of pions, muons, or kaons, which dominate in different energy ranges, lead to different neutrino flavor compositions, the flavor composition at the source will be changed as a function of energy. In the literature, often the following source classes in terms of $\nu_e:\nu_\mu:\nu_\tau$ are distinguished:\footnote{See \Ref~\cite{Pakvasa:2010jj} for a summary, including other options leading to a neutron beam or muon beam-like source.}
\begin{description}
\item[{\bf Pion beam source}] Pion decays and muon decays equally contribute, flavor composition $1:2:0$.
\item[{\bf Muon damped source}] Strong muon cooling, which means that pion decays dominate; flavor composition $0:1:0$.
\item[{\bf Muon beam source}] Pile-up muons dominate; flavor composition $1:1:0$. 
\item[{\bf Neutron beam source}] Neutron decays dominate; flavor comp. $1:0:0$. 
\item[{\bf Undefined source}] Several of these processes compete; flavor composition $X:1-X:0$ ($X \notin \{ 0, \frac{1}{3}, \frac{1}{2}, 1 \}$). 
\end{description}
Note that in none of these cases $\nu_\tau$ are produced in significant quantities, which means that it is sufficient to use the ratio between the $\nu_e$ and $\nu_\mu$ production at the source, sometimes called ``flavor ratio''.
From the preceding discussion, it is clear that the above classifications can only hold in specific energy ranges.
\begin{figure}[t]
\begin{center}
 \includegraphics[width=0.8\textwidth]{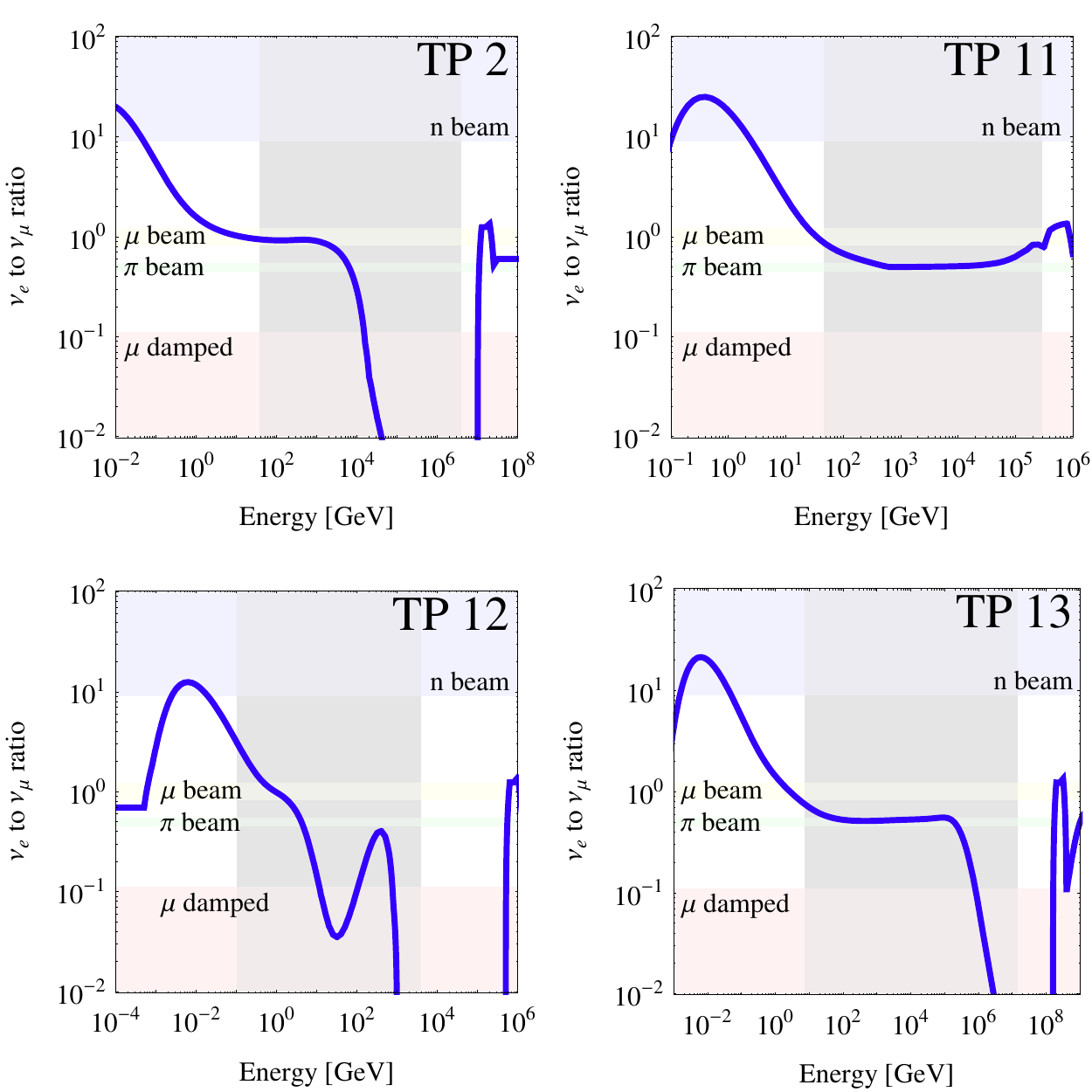} 
\end{center}
 \caption{\label{fig:flratio} Energy-dependent flavor ratio at the source for  several selected parameter sets (test points ``TP''; \cf, \figu{hillas})  for a model where $N_\gamma'$ is generated by synchrotron losses of co-accelerated electrons.  The (dark) gray-shaded areas mark the regions where the fluxes peak. Figure taken from \Ref~\cite{Hummer:2010ai}.}
\end{figure}
However, all of these sources can be recovered in a numerical simulation. In \figu{flratio}, four examples for energy dependent flavor ratios at the source are shown for different parameter sets, where  in this model $N_\gamma'$ is generated by synchrotron losses of co-accelerated electrons. In this figure, also the relevant flavor ratio ranges for the different sources introduced above are shown. The upper right panel shows the classical pion beam source, which is typically found for low magnetic fields. Nevertheless, the contribution of neutron decays at low energies can be clearly seen. The lower right panel shows a pion beam evolving in a muon damped source at high energies, as, \eg, in \Ref~\cite{Kashti:2005qa}. The upper left panel depicts a muon beam to muon damped source. In this case, the cooled muons pile up at lower energies, where the muon decays dominate. And the lower left panel shows an undefined source, where several processes compete.

Of course, not only the secondaries are affected by synchrotron (or adiabatic) losses, but also the primary protons. Depending on the model, one can use these losses to determine the maximal proton energy, or one can put in the maximal proton energy by hand. From \figu{nuinj}, it is interesting to discuss the impact of the maximal proton energy on the neutrino fluxes. In the ``no losses'' case (dashed curves), the maximal neutrino energy is directly determined by the maximal proton energy $E'_{\nu, \mathrm{max}} \simeq 0.05 \, E'_{p, \mathrm{max}}$. There is, however, one exception: the neutrinos from neutron decays are limited by $E'_{\nu, \mathrm{max}} \simeq 10^{-4} \, E'_{p, \mathrm{max}}$, \cf, \equ{ninj}. If the synchrotron losses are switched on (solid curves), the neutrino spectrum from neutron decays still follows the proton spectrum, since the neutrons are electrically neutral, whereas the maximal neutrino energies for the other production modes are determined by \equ{ec}. As a consequence, the neutron decay spectrum strongly depends on the assumptions for $E'_{p, \mathrm{max}}$, whereas the other spectral shapes are entirely unaffected by $E'_{p, \mathrm{max}}$ as long as $E'_{p, \mathrm{max}} \gtrsim 6 \, E'_c$ for kaons (see \equ{ec}, factor six from kaon production and decay kinematics). It is therefore not surprising that for strong enough magnetic fields one can find parameter sets for which the neutron decays dominate (\cf, \Refs~\cite{Moharana:2010su,Moharana:2011hh}), but one should keep in mind that this depends on the assumptions for the maximal proton energy (and the inclusion of multi-pion processes \etc, which may mask this effect).

\subsection{Transformation into observer's frame}

The transformation of the injection spectrum of the neutrinos $Q'_{\nu_\alpha}$  from the source to the observable flux $\phi_\beta$  of $\nu_\beta$ (in units of $\mathrm{GeV^{-1} \, cm^{-2} \, s^{-1}}$) at the Earth
is given by
\begin{equation}
 \phi_\beta = \sum\limits_{\alpha=e,\mu,\tau} \hat N \, P_{\alpha \beta} \,  \frac{(1+z)^2}{4 \pi d_L^2} \, Q'_{\nu_\alpha} \, , \qquad E_\nu=\frac{\Gamma}{1+z} \, E_\nu' \, , \label{equ:boost}
\end{equation}
where a simple Lorentz boost $\Gamma$ is used (instead of a viewing angle-dependent Doppler factor). Here $P_{\alpha \beta}$ is the transition probability $\nu_\alpha \rightarrow \nu_\beta$, discussed in \Sec~\ref{sec:propdet}, and  $\hat N$ is a (model-dependent) normalization factor. For example, if an isotropically emitting spherical zone is boosted with $\Gamma$ towards the observer, then $\hat N = (4/3) \, R'^3 \pi \Gamma^2$ since the emission is boosted into a cone with opening angle $1/\Gamma$. For a relativistically expanding fireball, it is simpler to perform the transformation in a different way, see \Sec~\ref{sec:grb}.
Furthermore, $d_L(z)= (1+z) \, d_{\mathrm{com}}(z)$ is the luminosity distance, and $d_{\mathrm{com}}(z)$ is the comoving distance.
  From \equ{boost}, one can read off that the redshift dependence of the neutrino luminosity scales as $E_\nu^2 \phi \propto 1/d_L^2$ independent of the model, which is expected. Note that in \equ{boost} the neutrino and antineutrino fluxes are often added if the detector cannot distinguish these.

\subsection{Summary of ingredients, and limitations of the approach}

\begin{figure}[t]
 \includegraphics[width=1\textwidth,viewport=20 192 696 548]{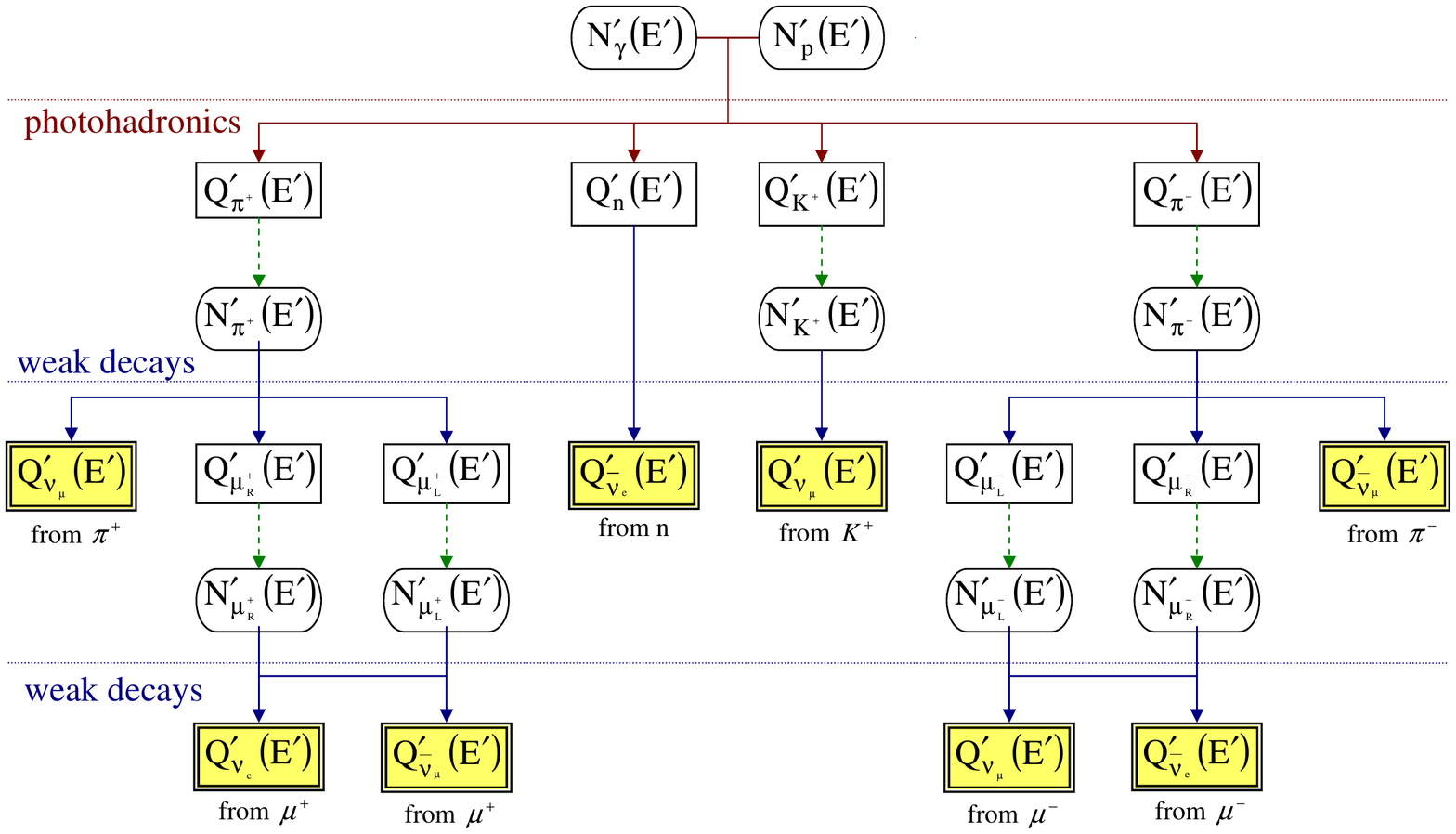}
 \caption{\label{fig:flowchart}Flowchart describing the model (in SRF). The functions $Q'(E)$ denote (injection) spectra per time frame $[\mathrm{\left( GeV\,cm^3\,s\right)^{-1}}]$ and $N'(E)$ steady spectra $[\mathrm{\left( GeV\,cm^3\right)^{-1}}]$ derived from the balance between injection and losses or escape. Dashed arrows stand for solving the steady state differential equation \equ{steadstate}. Figure taken from \Ref~\cite{Baerwald:2011ee}.}
\end{figure}

We summarize the generic neutrino production in \figu{flowchart}. As one can see in this flowchart, the starting point are the proton and photon densities within the source. Once these (and $B'$) are fixed, the rest is just a particle physics consequence. Therefore, for the computation of specific neutrino fluxes, the main effort is actually to determine $N_p'(E_p')$, $N_\gamma'(\varepsilon')$, and $B'$. 

Of course, there are some processes not taken into account in this picture, which may add to the ingredients discussed above for specific source classes. For instance, secondary neutrons, produced in the photohadronic interactions, may interact again if the source is optically thick to neutrons, the secondary pions, muons, and kaons may be re-accelerated~\cite{Koers:2007je}, synchrotron photons of the secondaries may add to $N_\gamma'$, \etc\ 
In addition, the neutrino spectrum may be more complicated in multi-zone models, since these naturally allow for more freedom.

From a particle physics perspective, additional kaon and charmed meson production modes may be added, and also the secondaries may interact again (see \Refs~\cite{Kachelriess:2006fi,Kachelriess:2007tr}). However, compared to analytical computations, the numerical approach described in this section already takes into account the secondary cooling in a self-consistent way, additional neutrino production modes can be easily included, and the full energy dependencies can be accounted for. For example, it has been demonstrated in \Ref~\cite{Hummer:GRBFireball}, that all necessary ingredients to reproduce the analytical GRB fireball (neutrino) calculations in \Refs~\cite{Waxman:1997ti,Guetta:2003wi,Abbasi:2009ig,Abbasi:2011qc} are contained.
It should represent the minimal set of of ingredients for neutrino production which are present in every source in the spirit of constructing the simplest possible model first. Of course, if $B'$ is small, the secondary cooling effects will be small as well, which is automatically included.

\section{Neutrino propagation and detection}
\label{sec:propdet}

In this section, we discuss several aspects of neutrino propagation and detection from the theoretical perspective. 

\subsection{Neutrino propagation and observables}

It is well known that neutrinos may change flavor from the production to the detection point. While this phenomenon is in general described by neutrino oscillations, astrophysical neutrinos are typically assumed to suffer from decoherence over very long distances. This means that effectively (in most practical cases) only flavor mixing enters the astrophysical neutrino propagation (see \Ref~\cite{Farzan:2008eg} for a more detailed discussion). In that case, $P_{\alpha \beta}$ in \equ{boost} becomes
\begin{equation}
 P_{\alpha \beta} = \sum\limits_{i=1}^3 |U_{\alpha i}|^2 | U_{\beta i}|^2  \, 
\label{equ:flmix}
\end{equation}
for three active neutrinos, where $U_{\alpha i}$ are the usual PMNS mixing matrix elements in the standard parameterization; see \eg\ \Ref~\cite{Schwetz:2011zk} for recent values of the mixing angles.
This implies that neutrino oscillations, \ie, the $\Delta m^2 L/E$-dependence, are averaged out. An initial flavor composition $\nu_e:\nu_\mu:\nu_\tau$ of $1:2:0$ will therefore evolve (approximately) into $1:1:1$ at the detector, see, \eg, \Refs~\cite{Learned:1994wg,Rodejohann:2006qq}. In \Sec~\ref{sec:newphys}, we will see that \equ{flmix} can significantly change in the presence of new physics effects, which opens new possibilities to test such effects. However, \equ{flmix} also implies that there could be some sensitivity to standard flavor mixing, which may be complementary to Earth-based experiments, see discussions in \Refs~\cite{Serpico:2005sz,Serpico:2005bs,Winter:2006ce,Xing:2006xd,Majumdar:2006px,Blum:2007ie,Awasthi:2007az,Hwang:2007na,Pakvasa:2007dc,Donini:2008xn,Quigg:2008ab,Choubey:2008di,Xing:2008fg}. In the light of the current bounds for astrophysical neutrino fluxes from IceCube, however, such applications might be unlikely. 

The main observable in neutrino telescopes are muon tracks from charged current interactions of muon neutrinos, producing Cherenkov light, which can be detected in so-called digital optical modules (DOMs). Because of the long muon range which is increasing with energy, the muon track does not have to be fully contained in the detector volume, which leads to excellent statistics increasing with energy. In addition, 
muon tracks have a very good directional resolution (order one degree). Additional event topologies include electromagnetic (mostly from electron neutrinos) and hadronic (from tau neutrinos) cascades, as well as neutral current cascades for all flavors. For even higher energies, the tau track may be separated, leading to so-called double bang or lollipop events; see \Ref~\cite{Beacom:2003nh} for an overview. In practice, the main ``flavor'' analysis so far performed by the IceCube collaboration has been a cascade analysis~\cite{IcCascade:2011ui}. To see that, consider that electromagnetic (from $\nu_e$) and hadronic (from $\nu_\tau$) cascades cannot  be distinguished. A useful observable is therefore the ratio of muon tracks to cascades~\cite{Serpico:2005sz}
\begin{eqnarray}
\rhat &\equiv& \frac{\phi_{\mu}}{\phi_{e}+\phi_{\tau}} \, .
 \label{equ:R}
\end{eqnarray}
 Note that neutral current events will also produce cascades, and $\nu_\tau$ will also produce muon tracks in 17\% of all cases, which, in practice, have to be included as backgrounds. In \Ref~\cite{IcCascade:2011ui}, the contribution of the different flavors to the cascade rate for a $E_\nu^{-2}$ extragalactic test flux with equal contributions of all flavors at the Earth was given as: electron neutrinos 40\%, tau neutrinos 45\%, and muon neutrinos 15\% (after all cuts). This implies that charged current showers dominate and that electron and tau neutrinos are detected with comparable efficiencies, \ie,  that \equ{R} is a good first approximation to discuss flavor at a neutrino telescope.
The benefit of this flavor ratio is that the
  normalization of the source drops out. In addition, it represents the experimental flavor measurement with the simplest possible assumptions. For a pion beam source, one finds  $\hat R \simeq 0.5$ at the detector, for a muon damped source $\hat R \simeq 0.6$, and for a neutron beam source $\hat R \simeq 0.3$, with some dependence on the mixing angles. In principle, this and other observables, such as the ratio between electromagnetic and hadronic showers, can therefore be used to determine the flavor composition of the sources, see, \eg, \Refs~\cite{Xing:2006uk,Lai:2009ke,Choubey:2009jq,Esmaili:2009dz,Lai:2010tj} (unless there are new physics effects present, see \eg\ \Ref~\cite{Bustamante:2010nq}, which may yield similar ratios at the detector for different flavor compositions at the source).

\begin{figure}[t!]
\begin{center}
\includegraphics[width=0.5\textwidth]{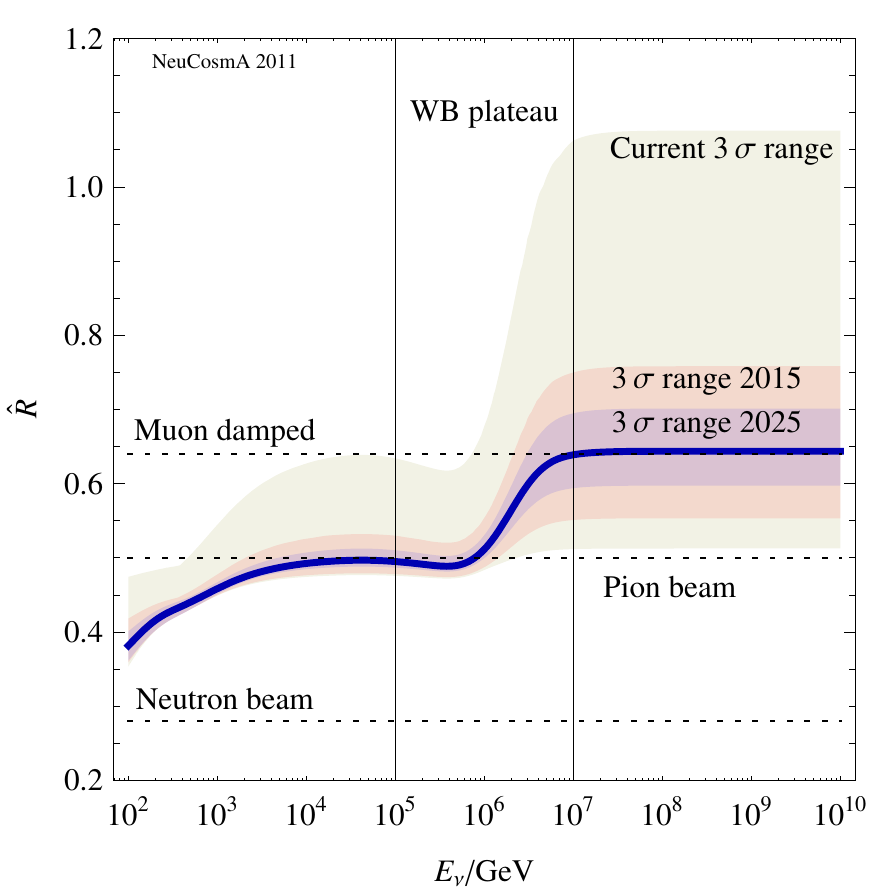} 
\end{center}
\caption{\label{fig:flrobs} Flavor ratio $\rhat$ as a function of $E_\nu$ at the detector for a GRB neutrino flux with the Waxman-Bahcall (WB) ``plateau'' (in $E^2 \phi_\mu$) between $10^5$ and $10^7$~GeV (\cf, \Ref~\cite{Waxman:1998yy}).  
The shaded regions show the impact of the $3 \sigma$ mixing angle uncertainties now (``current'')~\cite{Schwetz:2008er}, in about 2015 (next generation, dominated by Daya Bay and T2K~\cite{Huber:2009cw}), and in about 2025 (for a neutrino factory~\cite{Huber:2003ak,Tang:2009na}). Here $\theta_{13}=0$ is assumed for the sake of simplicity. Figure taken from \Ref~\cite{Baerwald:2011ee}.}
\end{figure}

We show the ratio $\hat R$  as a function of energy for a GRB neutrino flux in \figu{flrobs}. At the ``plateau'' of the flux, where the largest number of events is expected from, one can clearly see the flavor transition between a pion beam and a muon damped source at higher energies. In addition, the neutron decays have an impact at very low energies. In this figure, the uncertainty on $\hat R$ ($3\sigma$) coming from the current mixing angle uncertainties is shown, as well as the expectation for the next generation of reactor and long-baseline experiments (``2015''), and for a high-precision neutrino oscillation facility (``2025''). Obviously, the current uncertainties on the mixing angles are still too large to allow for a clear identification of the flavor composition at the source, whereas already the knowledge from the next generation will allow for a flavor ratio discrimination -- at least in principle.

\subsection{Detector response and impact of spectral shape}
\label{sec:detector}

For time-integrated point source searches in IceCube~\cite{Abbasi:2010rd}, the simplest possible approach to describe the event rate of muon tracks is to use the exposure $\mathrm{Exp}(E_\nu,\delta) \equiv A_\nu^{\mathrm{eff}}(E_\nu,\delta) \, t_{\mathrm{exp}}$, where $A_\nu^{\mathrm{eff}}$ is the neutrino effective area and $t_{\mathrm{exp}}$ is the observation (exposure) time. Here  $A_\nu^{\mathrm{eff}}(E_\nu,\delta)$ is a function of the flavor or interaction type (which we do not show explicitely), the incident neutrino energy $E$, and the declination of the source $\delta$. The neutrino effective area already includes Earth attenuation effects (above PeV energies) and event selection cuts to reduce the backgrounds, which depend on the type of source considered, the declination, and the assumptions for the input neutrino flux, such as the spectral shape. Normally, the cuts are optimized for an $E_\nu^{-2}$ flux.
The total event rate of a neutrino telescope can be obtained by folding the input neutrino flux with the exposure as
\begin{equation}
 N = \int dE_\nu  \, \mathrm{Exp}(E_\nu,\delta) \, \phi_\mu(E_\nu) = \int dE_\nu A_\nu^{\mathrm{eff}}(E_\nu, \delta) \, t_{\mathrm{exp}} \, \phi_\mu(E_\nu)  \, .
\label{equ:N}
\end{equation} 
Here $\phi_\mu(E_\nu)$ is, for point sources, given in units of $\mathrm{GeV^{-1} \,  cm^{-2} \, s^{-1}}$ for neutrinos and antineutrinos added. If backgrounds are negligible, the 90\% (Feldman-Cousins) sensitivity limit $K_{\mathrm{90}}$ for an arbitrarily normalized input flux used in \equ{N} can be estimated as $K_{\mathrm{90}} \sim 2.44/N$~\cite{Feldman:1997qc}. This imples that a predicted flux at the level of the sensitivity limit, irrespective of the spectral shape, would lead to the same number (2.44) of events. The 90\% confidence level differential limit in terms of  $E_\nu^2 \phi_\mu$ can be defined as $2.3 \, E_\nu/\mathrm{Exp}(E_\nu,\delta)$, see, \eg, \Ref~\cite{Abraham:2009uy}.

\begin{figure}[tp]
\begin{center}
\includegraphics[width=0.9\textwidth]{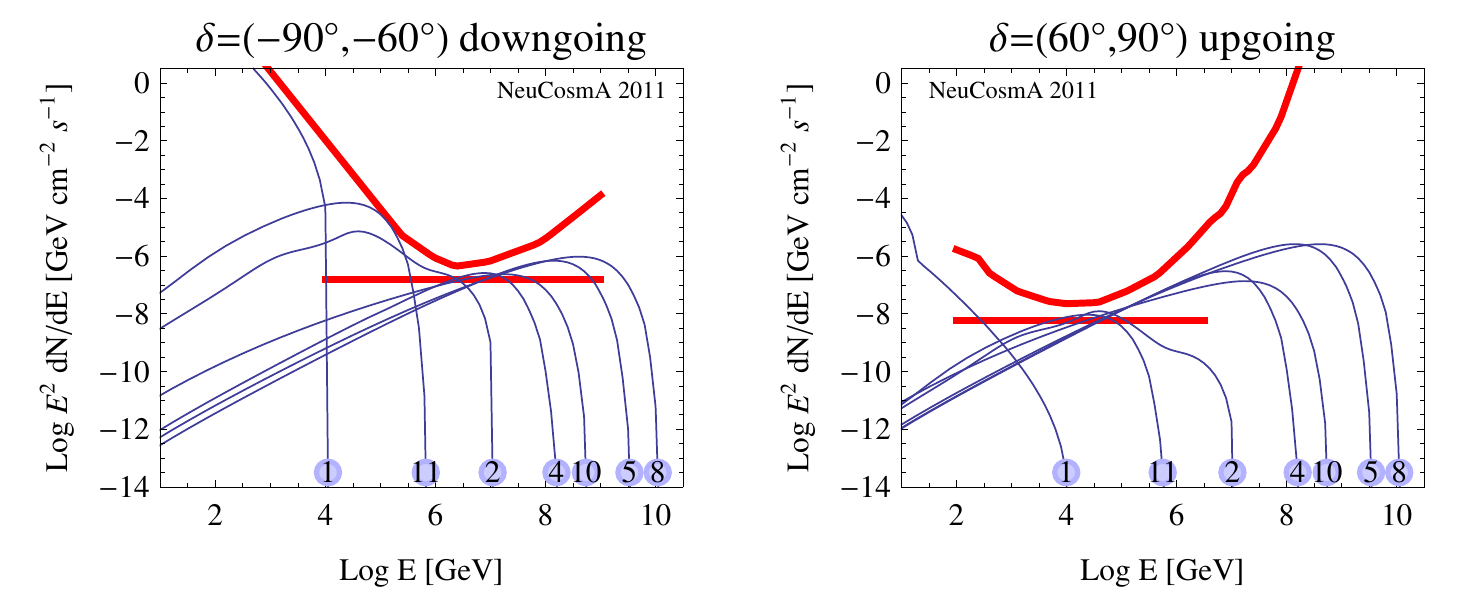}
\end{center}
\caption{\label{fig:spectra} Limits for selected muon neutrino spectra (including flavor mixing) from \Ref~\cite{Hummer:2010ai} for different declination bands for the time-integrated point source search in IC40 (90\% CL)~\cite{Abbasi:2010rd}.  The thick lines show the limits for an $E_\nu^{-2}$ flux (in the dominant energy range), and the thick curves the differential limits. Figure taken from \Ref~\cite{Winter:2011jr}. }
\end{figure}

The interplay between spectral shape and detector response has, for instance, been studied in \Ref~\cite{Winter:2011jr}. In order to illustrate that, we show in \figu{spectra} the limits for selected muon neutrino spectra and two different source declinations, corresponding to downgoing and upgoing events, for the time-integrated point source search in IC40. In this figure, the thick horizontal lines show the limits for an $E_\nu^{-2}$ flux (in the dominant energy range), and the thick curves the differential limits. One can easily see that the differential limits are useful, since any ``well behaved'' neutrino flux will stay below these limits. In addition, the differential limit shows the energy range where the instrument is most sensitive to a flux, whereas for the horizontal lines, representing an $E_\nu^{-2}$ flux limit, only the contribution close to the differential limit minimum contributes to \equ{N}. In \figu{spectra}, the minimal and maximal energies for the horizontal lines are indeed arbitrarily chosen, since the limit hardly depends on these.\footnote{For the chosen example, this holds as long as the energy range around the differential limit minimum within about 1.5 orders of magnitude in energy is included. Sometimes the range where 90\% of the events come from is shown.} From the different fluxes in \figu{spectra} it is clear that the interplay between spectral shape and detector response is important. For example, spectrum \#4 will be better constrained in the left panel (downgoing events) than in the right panel (upgoing events) because the differential limit peaks at higher energies -- in spite of the lower absolute performance at the differential limit minimum. The reason is the coincidence between spectral peak and differential limit minimum. This picture changes completely in the presence of strong magnetic field effects on the secondaries, see spectrum \#2, where even the imprint of these effects becomes important. The comparison to the $E_\nu^{-2}$ test flux (horizontal line) clearly demonstrates that the  $E_\nu^{-2}$ assumption is insufficient for all practical purposes, since it does not take into account the energy range of the flux. In addition, as we discussed around \equ{qsimple} earlier, the $E_\nu^{-2}$ assumption only holds for a very special case for the target photon density. Therefore, detector response and source model are intimately connected, and it is not quite clear if some sources may be missed just because the detector response does not match the source prediction. We will come back to this issue in \Sec~\ref{sec:gensource}.

\figu{spectra} is also useful to illustrate the impact of the source declination, which is, for IceCube, a measure for the direction of the muon track in terms of the nadir angle because of the location at the South Pole. Note, however, that for other neutrino telescopes, such as ANTARES, this relationship is non-trivial. Obviously, the differential limit is different in the left and right panels of this figure, corresponding to downgoing and upgoing events, respectively. First of all, note that the downgoing events have to fight the atmospheric muon  background, which leads to a worse performance at low energies because of appropriate cuts. However, the upgoing events suffer from the Earth attenuation for energies $\gg$~PeV, which leads to the better sensitivity of the downgoing events for high energies, and the shift of the differential limit minimum to higher energies. From the discussion above it is clear that both event types (upgoing and downgoing) are complementary not only because they test a different part of the sky, but also because they test different energy ranges.

Finally, let us briefly comment on the statistics expected for neutrinos. First of all, it is clear from the discussion above that per definition any current IC40 limit, obtained over about one year of data taking, is compatible with about 2.4 events at the current 90\% confidence limit. Assuming that the effective area increases by about a factor of three to four from IC40 to IC86~\cite{Karle:2010xx}, one can extrapolate that the current limits are compatible with $2.4 \cdot 4 \cdot 10 \simeq 100$ events over ten years of full IceCube operation if the current bounds are saturated. Any further non-observation of events will reduce this maximal expectation. Therefore, one can already say right now that any conclusion about the astrophysical neutrino sources, the sources of the cosmic rays, or leptonic versus hadronic models for $\gamma$-ray observations will most likely be based on the information from many sources of one class. A typical example is the stacking of GRBs using their gamma-ray counterparts, such as in \Ref~\cite{Abbasi:2011qc}. The aggregation of fluxes, no matter if diffuse or stacked, will however imply new systematics and model-dependent ingredients, see discussion in \Ref~\cite{Baerwald:2011ee}. 

\subsection{Glashow resonance to discriminate $pp$ from $p\gamma$?}
\label{sec:glashow}

 A useful observable may be the Glashow resonance $\bar{\nu}_e + e^- \to W^- \to
anything$ at around $6.3 \, \text{PeV}$~\cite{Learned:1994wg,Anchordoqui:2004eb, Bhattacharjee:2005nh,Maltoni:2008jr,Hummer:2010ai,Xing:2011zm,Bhattacharya:2011qu}
to distinguish between neutrinos and antineutrinos in the detector, since this process is
only sensitive to $\bar \nu_e$. For photohadronic ($p\gamma$) interactions, however, mostly $\pi^+$ and therefore $\nu_e$ are produced at the source, see \equ{Delta} and \equ{piplusdec}, which means that no excess of events should be seen at this resonance energy -- at least in the absence of flavor mixing. On the other hand, for $pp$ interactions in the source, $\pi^+$ and $\pi^-$ are produced in about equal ratios, which increases the $\bar{\nu}_e$ production rate.  Therefore, the Glashow resonance is frequently proposed as a discriminator between $pp$ and $p\gamma$ sources. Note that this argument is especially interesting for $E_\nu^{-2}$ neutrino fluxes, whereas any other, (significantly different) spectral index may be a clear indicator for $p \gamma$ interactions.

In \Ref~\cite{Hummer:2010ai}, the electron neutrino-antineutrino ratio at source (Sec.~3.3) and detector (Sec.~4.3) has been computed explicitely. At the source, the following observations have been made:
\begin{itemize}
\item
 Additional production modes (such as direct and multi-pion production) in photohadronic interactions produce $\pi^-$ in addition to $\pi^+$. The pure $\pi^+$ source therefore does not exist, and an up to 20\% contamination from $\bar \nu_e$ at the source has to be accepted even in this case. As a consequence, the Glashow resonance must be seen to some degree, even for the $p \gamma$ source.
\item
 Since the photohadronic interaction in \equ{Delta} also exists for neutrons producing $\pi^-$ in this case, any optical thickness to neutron escape will lead to a $\pi^+$-$\pi^-$ equilibration. This means that the optically thick $p\gamma$ source cannot be distinguished from the $pp$ source by the Glashow resonance.
\item
 Neutron decays, which are inherently present in any photohadronic source, lead to $\bar \nu_e$ flux, faking a $\pi^-$ contribution and therefore a $pp$ source in particular energy intervals (determined by the maximal proton energy).
\end{itemize}
As a consequence, only the ``$p\gamma$ optically thin'' (to neutron escape) source might be uniquely identified if less than 20\% of $\bar \nu_e$ contamination are found. On the other hand, one cannot uniquely identify a $pp$ source. Note that modern approaches take the composition between $pp$ and $p\gamma$ interactions as a variable, see \Refs~\cite{Maltoni:2008jr,Xing:2011zm,Bhattacharya:2011qu}. However, these approaches can typically not describe the contamination from neutron decays, because this depends very much on the model.

At the detector, the use of the Glashow resonance becomes even more complicated for the following reasons:
\begin{itemize}
\item
 Flavor mixing re-introduces a $\bar \nu_e$ component from $\bar \nu_\mu$ produced in $\mu^+$ decays even for  pure $\pi^+$ production, \cf, \equ{piplusdec}. 
\item
 As a consequence, the flavor composition at the source is important, and has to be determined at the same time. For example, a muon damped $pp$ source may be easily mixed up with a pion beam $p \gamma$ source. 
\item
 The Glashow resonance occurs at a specific energy (6.3 PeV), which means that for this process the energy dependence is not important and only one particular energy matters. However, the transformation of the energy from source to detector depends on redshift and a possible Lorentz boost,  \cf, \equ{boost}, which have to be known to draw any conclusions.
\end{itemize}
In summary, the discovery of a neutrino signal with significantly suppressed Glashow resonance events will be interesting, and may allow for possible conclusions about the source if flavor composition, $z$, and $\Gamma$ are known. On the other hand, the detection of a pronounced  Glashow resonance events is probably non-conclusive for the physics of the source.

\subsection{Testing new physics in the neutrino propagation}
\label{sec:newphys}

In the Standard Model, the transition probability $P_{\alpha \beta}$ in \equ{boost} is described by the usual flavor mixing in \equ{flmix}, which is independent of energy. New physics effects may lead to deviations from this picture, where the effects discussed in the literature include sterile neutrinos, neutrino decay, quantum decoherence, Lorentz invariance violation, among others, see \Ref~\cite{Pakvasa:2010jj} for a review. In some of these cases, $P_{\alpha \beta}$ may  even be a function of energy $P_{\alpha \beta}(E)$. We choose neutrino decay as an example in this section, see \Refs~\cite{Farzan:2002ct,Beacom:2002vi,Beacom:2003zg,Meloni:2006gv,Lipari:2007su,Majumdar:2007mp,Maltoni:2008jr,Bhattacharya:2009tx,Bhattacharya:2010xj,Bustamante:2010nq,Mehta:2011qb}, where in \Refs~\cite{Bhattacharya:2009tx,Bhattacharya:2010xj,Mehta:2011qb} energy dependent effects have been considered. 

Following \Ref~\cite{Mehta:2011qb}, for decay into {\em invisible} particles\footnote{See \eg\ \Refs~\cite{Lindner:2001fx,Maltoni:2008jr} for the systematical discussion/classification of possible scenarios. In general, neutrinos may decay into particles {\em invisible} for the detector, or other {\em visible} active flavors. Furthermore, the decay may be {\em complete}, \ie, all particles have decayed, or {\em incomplete}, where the energy dependence of the decay can be seen. For the case of complete decays, there are only eight different effective decay scenarios (each mass eigenstate can be stable or unstable), no matter if the decays are visible or invisible or if intermediate (unstable) invisible states are involved~\cite{Maltoni:2008jr}. Likewise, there are only eight scenarios for invisible incomplete decays, whereas the treatment of incomplete visible decays is more complicated, see, \eg\ \Refs~\cite{Lindner:2001fx,Lindner:2001th}.}, the transition probability can be described by a modified version of \equ{flmix}:
\begin{equation}
 P_{\alpha \beta} = \sum\limits_{i=1}^{3}  |U_{\beta i}|^2 \, |U_{\alpha i}|^2 \,
 \, D_i(E) \,  \quad \text{with} \quad D_i(E) =
\exp \left( - \hat\alpha_i \frac{L}{E} \right)\,, \label{equ:pdecay}
\end{equation} as
the {\em damping coefficient}~\cite{Blennow:2005yk}.
Here $\hat\alpha_i=m_i/\tau^0_i$ with $\tau^0_i$ the rest frame lifetime for mass eigenstate $\nu_i$. Typically
the neutrino lifetime is quoted as $\tau_i^0/m_i$   since $m_i$ is unknown, see \Ref~\cite{Pakvasa:2010jj} for a review. Note that this transition probability is energy dependent, compared to \equ{flmix}.
In this case, the flavor ratio $\hat R$ in \equ{R} can be rewritten as
\begin{equation}
\hat R = \frac{P_{e\mu}(E)  \, \xhat (E) + P_{\mu\mu}(E)  }{[P_{ee}(E)+P_{ e \tau}(E) ]\, \xhat (E) + [P_{\mu
e}(E)+P_{\mu \tau}(E)]} \, 
 \label{equ:Rspecial}
\end{equation}
if $\xhat (E) \equiv Q_e(E)/Q_\mu(E)$ is the ratio between electron neutrinos and muon neutrinos ejected at the source (assuming that hardly any tau neutrinos are produced). \equ{Rspecial} carries now two energy dependencies: the energy dependence of the flavor composition at the source $\xhat (E)$, and the energy dependence of the new physics effect $P_{\alpha \beta}(E)$, which have to be intrinsically disentangled. On the other hand, the energy dependence of the new physics effect may provide a unique signature~\cite{Bhattacharya:2009tx,Bhattacharya:2010xj}, and the energy dependence of the flavor composition at the source may help to disentangle different scenarios~\cite{Mehta:2011qb}.

\begin{figure}[tp]
\begin{center}
\includegraphics[width=0.45\textwidth]{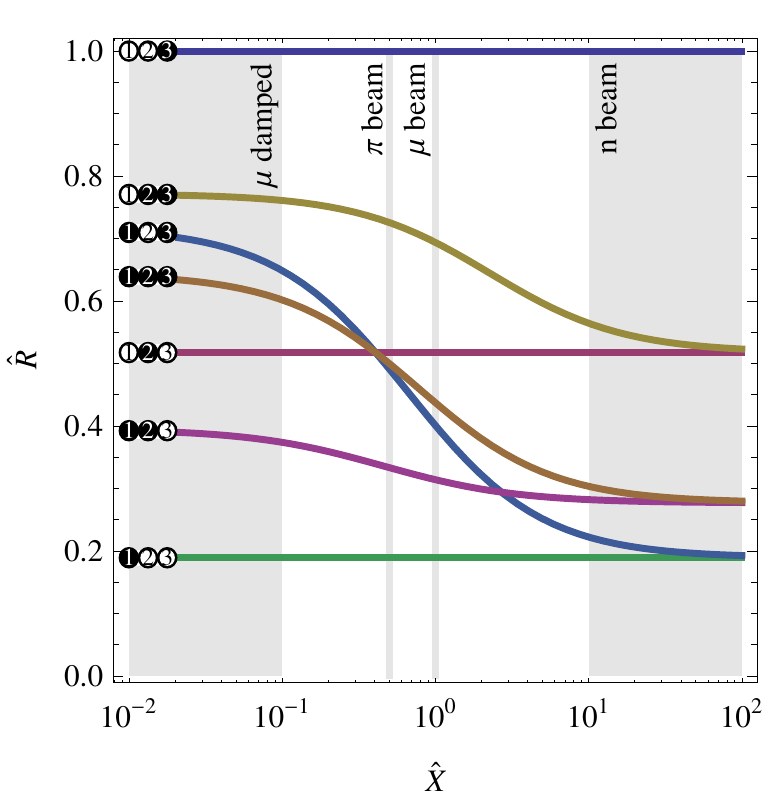} \hspace*{0.5cm}
\includegraphics[width=0.45\textwidth]{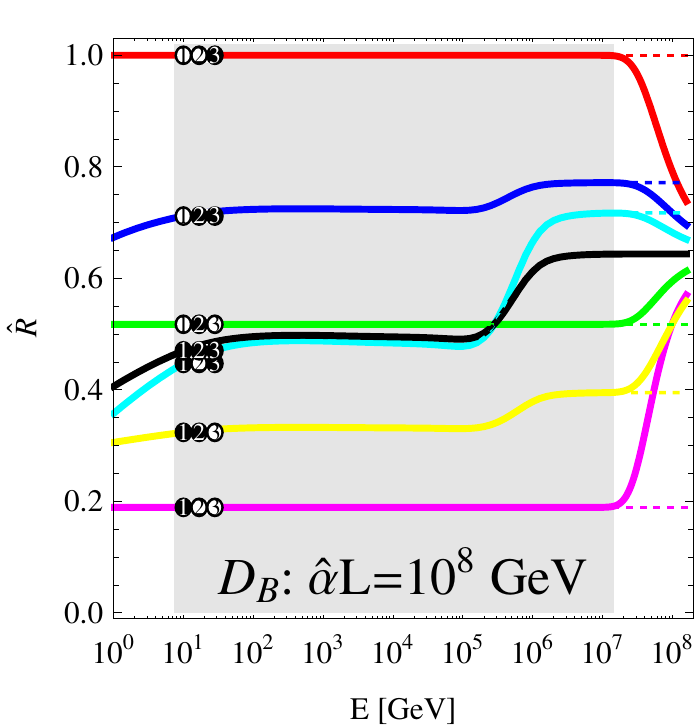}
\end{center}
\caption{\label{fig:xdecay} Left panel: $\hat R$ as a function of the initial flavor composition $\xhat$ for different decay scenarios (complete decays), where filled disks correspond to stable mass eigenstates and unfilled disks to unstable mass eigenstates. Right panel:  $\hat R$ as a function of energy for a pion beam to muon damped source 
(\cf, \figu{flratio}, TP~13) for a specific  decay parameter (same for all unstable states), as given in the panel. The dashed curves stand for complete decays. Figures adopted from \Ref~\cite{Mehta:2011qb}. }
\end{figure}

In general, there are $2^3=8$ decay scenarios for invisible incomplete decays, since every (active) mass eigenstate may be either stable or unstable. In \figu{xdecay}, left panel, $\hat R$ is shown as a function of the initial flavor composition $\xhat$ for these different (complete) decay scenarios, where filled disks correspond to stable mass eigenstates and unfilled disks to unstable mass eigenstates (the scenario with only unstable states is of course not shown, since no neutrinos can be detected in this case). In this panel, the different types of sources are also marked. One can clearly see that especially for a pion beam source, three of the scenarios cannot be distinguished, whereas these can be disentangled in principle for any other type of source. In addition, for three scenarios (the scenarios where only one mass eigenstate is stable), $\hat R$ is independent of the initial flavor composition. Note that the scenario with the largest $\hat R$ ($m_1$ and $m_2$ unstable) faces the strongest constraints because of the observation of neutrinos from supernova 1987A.

In \figu{xdecay}, right panel, $\hat R$ is shown as a function of energy for a particular choice of the decay parameter times distance and a specific source with a flavor transition from pion beam to muon damped source at about $10^6 \, \mathrm{GeV}$. First of all note that because of the exponential damping in \equ{pdecay}, the decays are practically complete for energies $E \ll 10^8 \, \mathrm{GeV}$, and the neutrinos are stable for energies $E \gg 10^8 \, \mathrm{GeV}$ for the chosen decay parameters. This is why the scenarios deviate from the complete decay curves (dashed) at very high energies and converge into the standard scenario there (all neutrinos stable). The pion beam (low energies) cannot distinguish three of the scenarios at low energies, as expected. However, above $10^6 \, \mathrm{GeV}$, where the flavor transition occurs into a muon damped source, the corresponding curves split up (and the neutrino decays are still complete), before they converge into the standard case. This example illustrates that the energy dependent flavor transition of a specific source may be a useful for new physics tests, provided that enough statistics can be collected.

\section{Application to generic (AGN-like) sources}
\label{sec:gensource}

As it is illustrated in \figu{flowchart} (see also \equ{prodmaster}), the proton $N'_p$ and photon densities $N'_\gamma$ within the source control the secondary meson, and hence the neutrino production. In this section, 
we follow the ansatz in \Ref~\cite{Hummer:2010ai}: we assume that the target photons are produced in a self-consistent way, by the synchrotron emission of co-accelerated electrons (positrons); see, \eg, \Ref~\cite{Muecke:2002bi} for a corresponding specific (BL Lac) AGN blazar model. The main purpose of this model is the prediction of spectral shape and flavor composition of the neutrino source, while it cannot predict the normalization of the neutrino flux in the form presented here.  In addition, no neutrinos have been observed yet, which means that an ansatz tailor-made for neutrinos may be useful to test the interplay between detector response and source model -- after all, there may be sources for which the optical counterpart is absorbed, so-called ``hidden sources'', see, \eg, \Refs~\cite{Razzaque:2004yv,Ando:2005xi,Razzaque:2005bh,Razzaque:2009kq}.
The ingredients to this model are comparable to the conventional GRB neutrino models, discussed in \Sec~\ref{sec:grb}, where however the origin of the target photons is different.

\subsection{Additional model ingredients}

The primaries, in this case protons and electrons, are assumed to be injected with an $(E')^{-\alpha}$ injection spectrum, where the (universal) injection index $\alpha$ is one of the important model parameters. They are assumed to lose energy dominantly by synchrotron radiation, controlled by $B'$, and adiabatic cooling, controlled by $R'$, which also determine their maximal energies. This means that \equ{steadstate} is applied to the primaries, which implies that the acceleration and radiation zones are different in this model. As a consequence, the electron spectrum becomes loss-steepened by one power.

The synchrotron photons (the target photons in \equ{prodmaster}) are computed in the  Melrose-approximation~\cite{Melrose:1980gb} averaged over the pitch angle. The power radiated per photon energy $\varepsilon'$ by one particle with energy $E'$, mass $m$ and charge $q$ in a magnetic field $B'$ is given by:
\begin{equation}
\label{equ:singlesyn}
 P'_s(\varepsilon',E')=1.8\cdot\frac{\sqrt{3}\,q^3\,B'}{16\,\varepsilon_0\,m\,c\,h}\cdot\left(\frac{\varepsilon'}{\varepsilon'_c}\right)^{1/3}e^{-\varepsilon'/\varepsilon'_c}\quad\text{with}\quad \varepsilon'_c=\frac{3\,q\,B'}{16\,m}\left(\frac{E'}{m\,c^2}\right)^2.
\end{equation}
We have to convolute this with the spectrum of radiating electrons as
\begin{equation}
 P'(\varepsilon')=\int_0^\infty dE' \, N'_e(E')\, P'_s(\varepsilon',E') \, .
\end{equation}
The number of produced photons per time can be computed with:
\begin{equation}
 Q'_\gamma(\varepsilon')=\frac{P'(\varepsilon')}{\varepsilon'}.
\end{equation}
Assuming that the photons escape (and hardly interact again), the steady photon spectrum, which is needed for the computation of photohadronic interactions, can be estimated by multiplying $Q'_\gamma$  with escape time $t'_{\mathrm{esc}} \simeq R'/c$ of the photons.  The synchrotron spectral index $(\varepsilon')^{-\alpha_\gamma}$ obtained from this approach is $\alpha_\gamma=(\alpha_e-1)/2+1=\alpha/2+1$, which means that the dependence on the primary injection index $\alpha$ is small. The rest of the computation follows  \Sec~\ref{sec:production}.

\begin{figure}[tp]
\centering
\includegraphics[width=\textwidth]{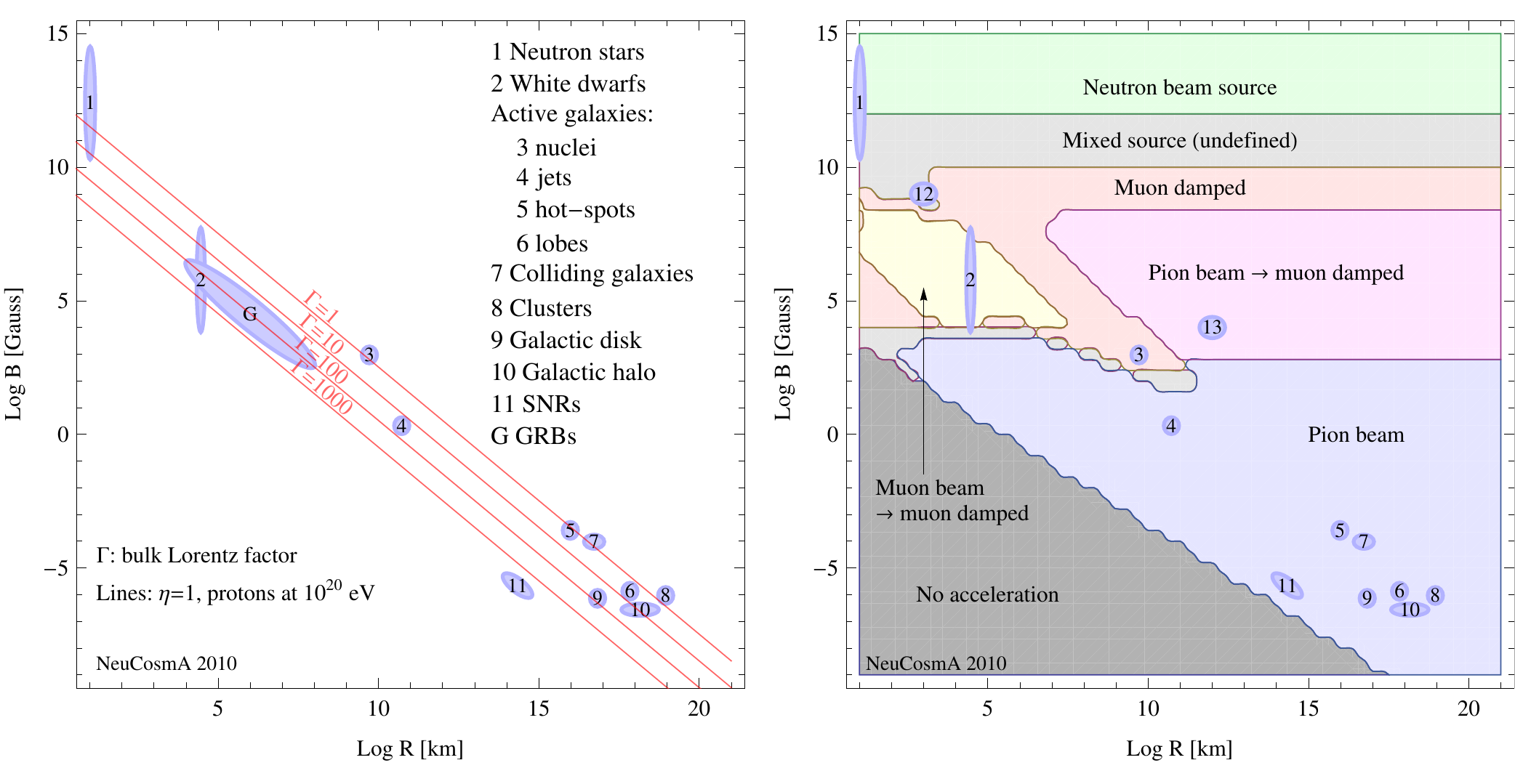}
\caption{\label{fig:hillas} Left panel: Possible acceleration sites in Hillas plot as a function of $R$ and $B$ (version adopted from M. Boratav). Right panel: Classification of sources for injection index $\alpha=2$ in this plot (see main text). Some points from left plot are shown for orientation, as well as two new points (12 and 13). Figure taken from \Ref~\cite{Hummer:2010ai}.}
\end{figure}

The main parameters in this model are $\alpha$, $R'$, and $B'$. For the sake of simplicity, we assume in the following that the source is only moderately Lorentz boosted with respect to the observer's frame, \ie, $R' \simeq R$ and $B' \simeq B$. In this case, 
a convenient description of the parameter space of interest is the Hillas plot~\cite{Hillas:1985is}. In order to confine a particle in a magnetic field at the source, the Larmor radius has to be smaller than the extension of the acceleration region $R$. This can be translated into the Hillas condition for the maximal particle energy
\begin{equation}
E_{\mathrm{max}} \, [\mathrm{GeV}] \simeq 0.03 \cdot \eta \cdot Z \cdot  R \, [\mathrm{km}] \cdot B \, [\mathrm{G}] \, .
\label{equ:hillas}
\end{equation}
Here $Z$ is the charge (number of unit charges) of the accelerated particle, $B$ is the magnetic field in Gauss, and $\eta$ can be interpreted as an efficiency factor or linked to the characteristic velocity of the scattering centers. Potential cosmic ray sources are then often shown in a plot as a function of $R$ and $B$, as it is illustrated in \figu{hillas}, left panel, by the numbered disks (see legend for possible source correspondences). Assuming that a source produces the highest energetic cosmic rays with $E \simeq 10^{20} \, \mathrm{eV}$, one can interpret \equ{hillas} as a necessary condition excluding the region below the $\Gamma=1$ line in \figu{hillas} (for protons with $\eta=1$). If one allows for relativistic boosts between source and observer, this condition is relaxed, as one can read off from the figure (in this case, $R$ and $B$ in the plot have to interpreted as $R'$ and $B'$).
However, this method does not take into account energy loss mechanisms, which  lead to a qualitatively different picture, see, \eg, \Refs~\cite{Medvedev:2003sx,Protheroe:2004rt}, and which are implied in our model. In the following, we will study the complete parameter space covered by \figu{hillas} (left) without any prejudice. Since the location of the sources in \figu{hillas} cannot be taken for granted, we will refer to the individual sources as ``test points'' (TP), and leave the actual interpretation to the reader.

Concerning the limitations of the model, it certainly does not apply exactly to all types of sources.  
For example, in supernova remnants, $pp$ (proton-proton) or $pA$ (proton-nucleus) interactions may dominate the neutrino production, which would require additional parameters to describe the target protons or nucleons. In addition, at ultra-high energies, heavier nuclei may be accelerated. 
The spirit of this model is different:  It is developed as  the simplest (minimal) possibility including non-trivial magnetic field and flavor effects. 
Another ingredient is the  target photon density, which is assumed to come from synchrotron emission of co-accelerated electrons here. In more realistic models, typically a combination of different radiation processes is at work. However,
in many examples with strong magnetic fields, the specific shape of  the photon spectrum is less  important for the neutrino spectral shape than the cooling and decay of the secondaries, which depend on particle physics only. To check this, we have tested the hypothesis that acceleration and radiation zones of the electrons are identical (\ie, the electron spectrum is not loss-steepened by synchrotron losses, which is actually a simpler version of this model). Thus, while it is unlikely that the model applies exactly to a particular source, it may be used as a good starting hypothesis. 

\subsection{Flavor composition at the source}

We discussed the flavor composition at the source already in \Sec~\ref{sec:cooling}, where we also showed several examples for the energy dependence in \figu{flratio}. Let us now approach this from a systematical point of view as a function of $R$ and $B$ (for $\alpha=2$). In this case, a qualitative classification of the sources can be found in \figu{hillas}, right panel, where it is implied that a certain flavor ratio can be clearly identified over one order of magnitude in energy close enough to the spectral peak; see \Ref~\cite{Hummer:2010ai}. From this figure, we find that for $B \gtrsim 10^{12}$~G, all charged species lose energy so rapidly that the neutron decays dominate the neutrino flux. For $B \gtrsim 10^{10}$~G, several processes compete, leading to an undefined source. For $B \lesssim 1$~kG, the sources behave as classical pion beams, which typically applies to sources on galactic scales. In the intermediate range, $1$~kG~$\lesssim B \lesssim 10^{10}$~G, the source classification somewhat depends on the spectral shape, since this affects possible muon pile-up effects, the energy range close to the spectral peak, and the competition of several effects. Depending on $R$, muon beam to muon damped sources (as a function of energy), muon damped sources, and pion beam to muon damped sources are found. There is some dependence on $\alpha$, which affects the spectral shapes. For instance, for $\alpha \sim 4$ a similar pion spectrum as for GRBs is obtained, which leads to a pion beam to muon damped sources for typical GRB parameters in consistency with \Refs~\cite{Kashti:2005qa,Lipari:2007su}. In summary, the pion beam source assumption is safe if  $B \lesssim 1$~kG, whereas for stronger magnetic fields the magnetic field effects on the secondaries have to be taken into account. However, for given parameters, this flavor composition can be predicted. For instance, $R$ may be estimated from the time variability of the source, $B$ from energy equipartition arguments, and $\alpha$ from the observed photon spectrum.

\subsection{Interplay between spectral shape and  detector response}

We discussed already earlier in \Sec~\ref{sec:detector} the interplay between spectral shape and detector response, see \Ref~\cite{Winter:2011jr}. From \figu{spectra}, it is clear that spectral shapes with a peak at the position of the differential limit minimum can be better limited than others. This is especially clear if two event types or detectors are compared for the same spectrum.  However, how should one quantify that for different spectra seen by the same detector?  Consider, for example, the fluxes \#2 and \#11 in the left panel in \figu{spectra}, both leading to the same event rate by definition.  Which of the two neutrino sources is the detector more sensitive to in terms of the physics of the source? In order to address this question, it is useful to assign a single number to each spectrum which measures how much energy in neutrinos can be tested for a specific spectrum and event type. We choose the energy flux density
\begin{equation}
F_\beta = \int E_\nu \, \phi_\beta \, d E_\nu \,  \label{equ:eflux}
\end{equation}
as this quantity, which we show in units of $\mathrm{erg \, cm^{-2} \, s^{-1}}$ for point sources in order to distinguish it from $E_\nu^2 \phi$ in units of $\mathrm{GeV \, cm^{-2} \, s^{-1}}$ ($1 \, \mathrm{erg}\simeq624 \, \mathrm{GeV}$). 

This quantity measures the total energy flux in neutrinos, and it is useful as a performance indicator measuring the efficiency of neutrino production in the source. In order to see that, consider the alternative derivation of $F_\beta$  from \equ{boost} (neglecting a possible Lorentz boost of the source)
\begin{equation}
 F_\beta = \sum\limits_{\alpha=e,\mu,\tau} \frac{L_{\nu_\alpha}}{4 \pi d_L^2}  \, , \quad \text{where} \quad L_{\nu_\alpha}=V' \int E_\nu' Q'_{\nu_\alpha} dE_\nu' \
\end{equation}
is the ``neutrino luminosity'' and $V'$ is the  volume of the interaction region. Since the neutrinos originate mostly from pion decays and take a certain fraction of the pion energy (about $1/4$ per produced neutrino for each charged pion), the neutrino luminosity is directly proportional to the (internal) luminosity of protons $L_{\mathrm{int}}$ (or the proton energy dissipated within a certain time frame $\Delta T$) and the fraction of the proton energy going into pion production, commonly denoted by $f_\pi$ (if the energy losses of the secondaries can be neglected).  Since a possibly emitted photon flux can be often linked to $L_{\mathrm{int}}$  by energy partition arguments,  one has $F_\beta \propto f_\pi \times L_{\mathrm{int}} \propto f_\pi \times L_{\gamma}$, and $F_\beta$  is a measure for $f_\pi \times L_{\mathrm{int}}$ of the source (if no photon counterpart is observed), or even $f_\pi$ itself (if a photon counterpart is observed). 

\begin{figure}[t]
\begin{center}
\includegraphics[width=0.47\textwidth]{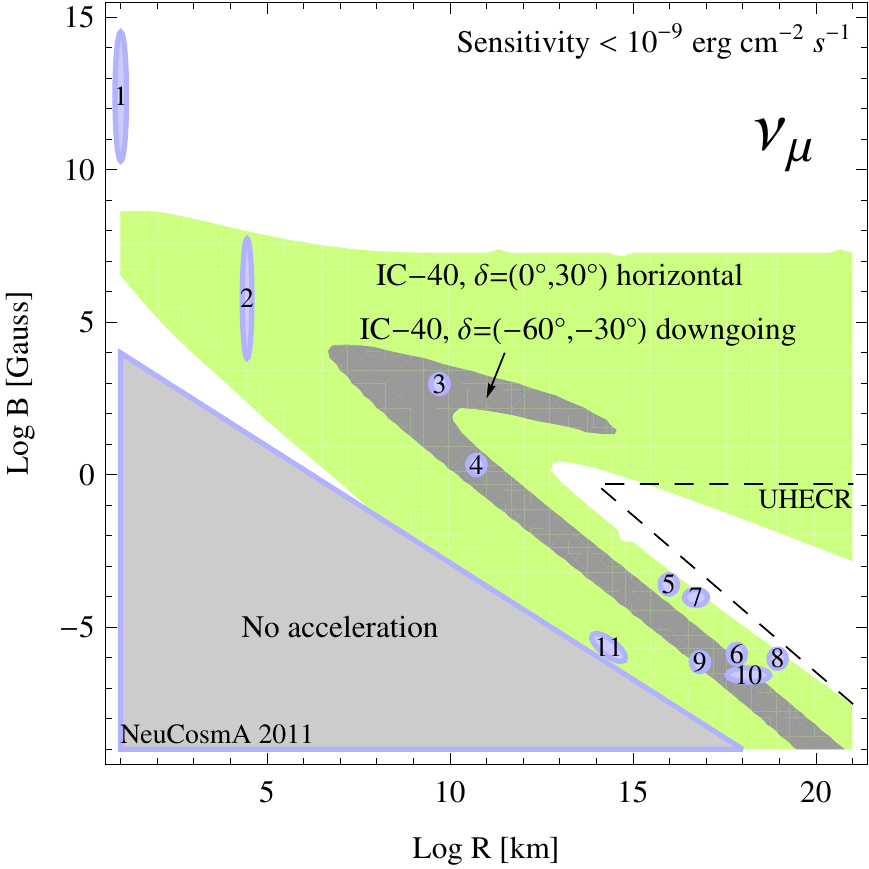} \hspace*{0.03\textwidth}
\includegraphics[width=0.47\textwidth]{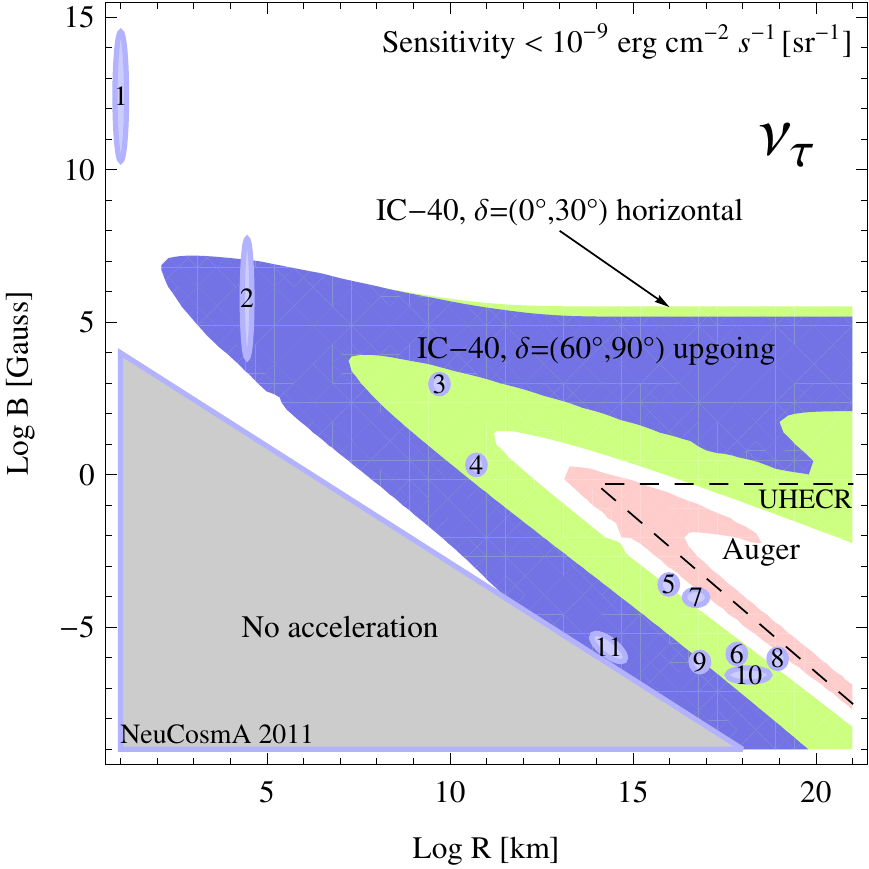}
\end{center}
\caption{\label{fig:comp} Regions where the sensitivity in $F_\beta$ exceeds $10^{-9} \, \mathrm{erg \, cm^{-2} \, s^{-1} \, [sr^{-1}]}$ for $\beta=\mu$ (left panel) and $\beta=\tau$ (right panel) for several selected data samples (90\% CL, $\alpha=2$). The dashed regions ``UHECR'' indicate where $10^{20} \, \mathrm{eV}$ cosmic ray protons are expected to be produced in the model. Figure taken from \Ref~\cite{Winter:2011jr}.}
\end{figure}

Regions for a specific sensitivity to $F_\beta$ are shown as a function of $R$ and $B$ in \figu{comp} for $\nu_\mu$ (left panel) and $\nu_\tau$ (right panel), for several source declinations in IceCube and for Earth-skimming tau neutrinos in Auger (in this case, for a diffuse flux). There are several conclusions from this figure: first of all, IceCube reponds very well to the usual suspects, such as AGNs (left panel). Even for $\nu_\tau$, which produce a muon track in only 17\% of all cases, most of this parameter space can be covered (right panel). Thus, it is clear that most sources will be also detected if the partition between $\nu_\mu$ and $\nu_\tau$ was heavily disturbed, such as by a new physics effect in the neutrino propagation.  The sensitive region, however, somewhat depends on the source declination. For very large values of $B$,  the  neutrino energies are lower and instruments such as the DeepCore array may respond better to the flux (both panels). On the other hand, the region where the UHECR are expected in this model (lower right corner) is better covered by Auger (right panel). This is not a big surprise: the neutrino spectrum follows the proton spectrum in the absence of strong magnetic field effects on the secondaries, which means that the spectra (\cf, \figu{spectra}) peak at high energies. A very interesting region may be the gap between the IceCube and Auger regions: perhaps a future instrument such as KM3NeT should optimize their geometry to be complementary in terms of energy range coverage.

\section{GRB neutrinos and the multi-messenger connection}
\label{sec:grb}

Recall again that, as it is illustrated in \figu{flowchart} (\cf, \equ{prodmaster}), the proton $N'_p$ and photon densities $N'_\gamma$ control neutrino production. In the previous section, we assumed that  $N'_\gamma$ is produced by synchrotron emission of co-accelerated electrons. Here we emphasize the multi-messenger connection, assuming that $N'_\gamma$ can be derived from the gamma-ray observation~\cite{Baerwald:2011ee,Hummer:2011ms}; see also \Refs~\cite{Guetta:2003wi,Becker:2005ej,Abbasi:2009ig,Abbasi:2011qc} for analytical approaches. The motivation is also different from the previous section: whereas we were interested in a systematic parameter space study of spectral shape and flavor ratio before, the main emphasis here is the prediction of spectral shape, flavor compositon, and absolute neutrino flux normalization for a specific set of GRBs observed in gamma-rays.

\subsection{Additional model ingredients}
\label{sec:grbmodel}

Here we describe the key ingredients of the conventional (numerical) fireball GRB model for neutrino emission, following \Ref~\cite{Baerwald:2011ee}, where we focus on the normalization. For a detailed comparison to the analytical calculations, see \Refs~\cite{Baerwald:2010fk,Hummer:2011ms}.
Because the model does not describe the neutrino production in a time-resolved way, it makes sense to relate the neutrino production to the (bolometric) gamma-ray fluence $S_{\text{bol}}$ (in units of $\mathrm{erg \, cm^{-2}}$) during the burst. The isotropic bolometric equivalent energy $E_{\text{iso},\text{bol}}$ (in $\text{erg}$) in the source (engine) frame can be then obtained as 
\begin{equation}
	E_{\text{iso},\text{bol}} = \frac{4\pi \, d_L^2}{(1+\textit{z})} \; S_{\text{bol}}  \quad .
\label{equ:eisobol}
\end{equation} 
It can easily be boosted into the SRF by $E'_{\text{iso},\text{bol}}= E_{\text{iso},\text{bol}}/\Gamma$.
Assuming energy equipartition between photons and electrons,  the photons carry a fraction $\epsilon_e$  (fraction of energy in electrons) of the total energy $E_{\text{iso},\text{tot}}$ and
\begin{equation}
	E_{\text{iso},\text{tot}} = \epsilon_e^{-1} \cdot E_{\text{iso},\text{bol}} \, .
\label{equ:eiso}
\end{equation}
In order to compute the photon and proton densities in the SRF, it turns out to be useful to define  an ``isotropic volume'' $V'_{\mathrm{iso}}$ 
\begin{equation}
	V'_{\mathrm{iso}} \simeq 4\pi \, \bigg( \underbrace{ 2 \, \Gamma^2 \, c \, \frac{t_v}{(1+\textit{z})} }_{R_C} \bigg)^2 \cdot \bigg( \underbrace{\Gamma \cdot c \, \bigg( \frac{t_v}{(1+\textit{z})} \bigg) }_{\Delta d'} \bigg) \propto \Gamma^5 \, ,
	\label{equ:visoISMexpl}
\end{equation}
where $R_C$ is the collision radius and $\Delta d'$ the shell thickness of the colliding shells. It can be estimated from the (observed) variability timescale $t_v$, $z$, and $\Gamma$, and can be regarded as the volume of the interaction region assuming isotropic emission by the source. 
Because of the intermittent nature of GRBs, the total fluence is assumed to be coming from $N \simeq T_{90}/t_v$ such interaction regions, where $T_{90}$ is the duration of the burst (time during which 90\% of the total energy is observed).
Now one can determine the normalization of the photon spectrum in \equ{prodmaster} from
\begin{equation}
 \int  \, \varepsilon' \, N'_{\gamma}(\varepsilon') \mathrm{d}\varepsilon' = \frac{E'_{\text{iso},\text{bol}}}{N \, V'_{\text{iso}}} \quad , \label{equ:photonorm}
\end{equation}
assuming that the spectral shape is determined by the observed spectrum.
Similarly, one can compute the normalization of the proton spectrum in \equ{prodmaster} by
\begin{equation}
\int \, E'_p \, N'_p(E'_p) \, \mathrm{d}E'_p =  \frac{1}{f_e} \frac{E'_{\text{iso},\text{bol}}}{N \, V'_{\text{iso}}}  \, ,
\label{equ:protonorm}
\end{equation}
where $f_e$ is the ratio between energy in electrons and protons ($f_e^{-1}$: baryonic loading).
Note that in the end, we will obtain the neutrino flux $\phi$ per time frame per interaction region from \equ{boost} with $\hat N = V'_{\mathrm{iso}}$. 
Assuming that the magnetic field carries a fraction $\epsilon_B$ of $E_{\text{iso},\text{tot}}$, one has in addition
\begin{equation}
 U'_B = \frac{\epsilon_B}{\epsilon_e} \cdot \frac{E'_{\text{iso},\text{bol}}}{N \, V'_{\text{iso}}} \quad \text{or} \quad B' = \sqrt{8\pi \, \frac{\epsilon_B}{\epsilon_e} \cdot \frac{E'_{\text{iso},\text{bol}}}{N \, V'_{\text{iso}}}}  .
\label{equ:B}
\end{equation}
Typical values used in the literature are $f_e \sim \epsilon_e \sim \epsilon_B \simeq 0.1$ (see, \eg, \Ref~\cite{Abbasi:2009ig}). 
An explicit calculation for $B'$ yields
\begin{eqnarray}
	B' & \simeq & 220 \, \left( \frac{\epsilon_B}{\epsilon_e} \right)^{\frac{1}{2}} \, \left( \frac{E_{\text{iso},\text{bol}}}{10^{53} \, \text{erg}}\right)^{\frac{1}{2}} \, \left( \frac{\Gamma}{10^{2.5}} \right)^{-3} \nonumber \\
  & \times &  \left( \frac{t_v}{0.01 \, \second} \right)^{-1} \, \left( \frac{T_{90}}{10 \, \second} \right)^{-\frac{1}{2}} \, \left( \frac{1+\textit{z}}{3} \right)^{\frac{3}{2}} \,  \kilo\text{G}  \, .
	\label{equ:BISM}
\end{eqnarray}

In summary, once the photon (from observation) and proton (typically $E_p^{-2}$) spectral shapes are determined, the proton and photon densities in the source and $B'$ can be calculated with the above formulas from the observables (gamma-ray fluence, $\Gamma$, $z$, $t_v$). 
 \equ{visoISMexpl} implies that for fixed $t_v$, the larger $\Gamma$, the larger the interaction region, and the smaller the photon density in \equ{photonorm}, which directly enters the fraction of fireball proton energy lost into pion production $f_\pi \propto \Gamma^{-4}$~\cite{Guetta:2003wi}, or, consequently,  $E_\nu^2 \phi \propto \Gamma^{-2}$. Therefore, the main contribution to the neutrino flux is often believed to  come from bursts with small Lorentz factors, see discussions in \Refs~\cite{Guetta:2000ye,Guetta:2001cd,Guetta:2003wi}. We follow this conventional fireball approach in the following. Note that this numerical approach contains all the ingredients of \Refs~\cite{Guetta:2003wi,Abbasi:2009ig} explicitely -- such as the cooling of the secondaries. The neutrino emission can be then easily computed as shown in \figu{flowchart}.
In addition, note that there are alternatives to the model. For instance, if the bursts are alike in the comoving frame, as suggested in \Ref~\cite{Ghirlanda:2011bn}, one has  $E_\nu^2 \phi \propto \Gamma^{2}$~\cite{Baerwald:2011ee}. One can read off from \equ{visoISMexpl} that the correlation $\Gamma \propto t_\nu^{-3/5}$ is expected in that case, since $V'_{\text{iso}}$ will be similar for the bursts. See \Ref~\cite{Baerwald:2011ee} for a discussion of the neutrino flux for different model hypotheses, and how the neutrino flux can in principle be used to discriminate among these.

\subsection{Systematics in the interpretation of aggregated fluxes}
\label{sec:grbsys}

Since the number of neutrinos expected from a single GRB is small, dedicated aggregation methods are needed. For instance, one may search for the diffuse flux from GRBs, which, however, has to fight the background from atmospheric neutrinos. Another possibility is to use the gamma-ray observation to infer on time window and direction of the neutrino signal, which effectively leads to significantly reduced backgrounds. In addition, a procedure such as the one in \Sec~\ref{sec:grbmodel} may be used to predict the absolute neutrino flux, its shape (see, \eg, \figu{nuinj}), or its flavor composition (see, \eg, \figu{flrobs}) on a burst-by-burst basis. Summing over many observed bursts, such an analysis is also called {\em stacking analysis}, see, \eg, \Ref~\cite{Abbasi:2011qc} for a recent example. The diffuse limit can be extrapolated from such a stacked flux, which is also called the {\em quasi-diffuse} limit; see \Ref~\cite{Becker:2007sv} for details. Here we discuss some of the implications when a stacked neutrino flux is translated into a quasi-diffuse flux.

\begin{figure}[t]
\begin{center}
\includegraphics[width=0.425\textwidth]{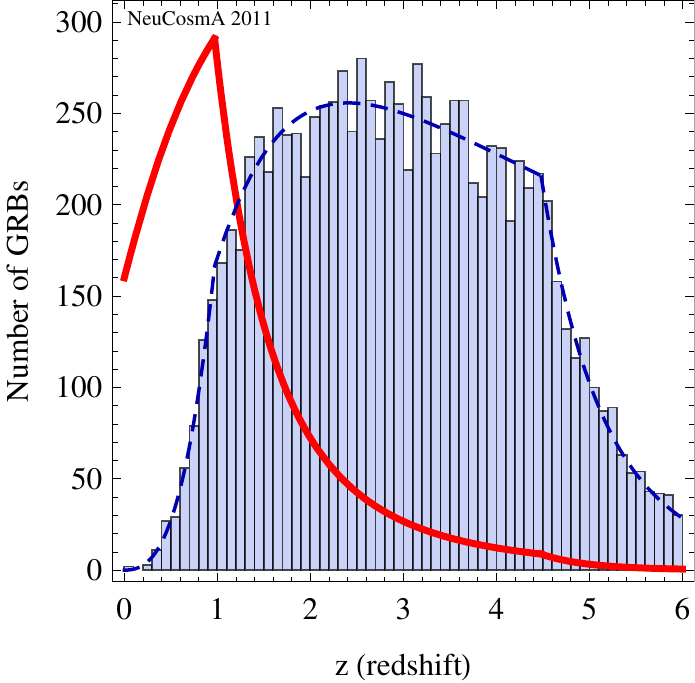} \hspace*{0.03\textwidth} 
\includegraphics[width=0.525\textwidth,viewport=0 8 210 210]{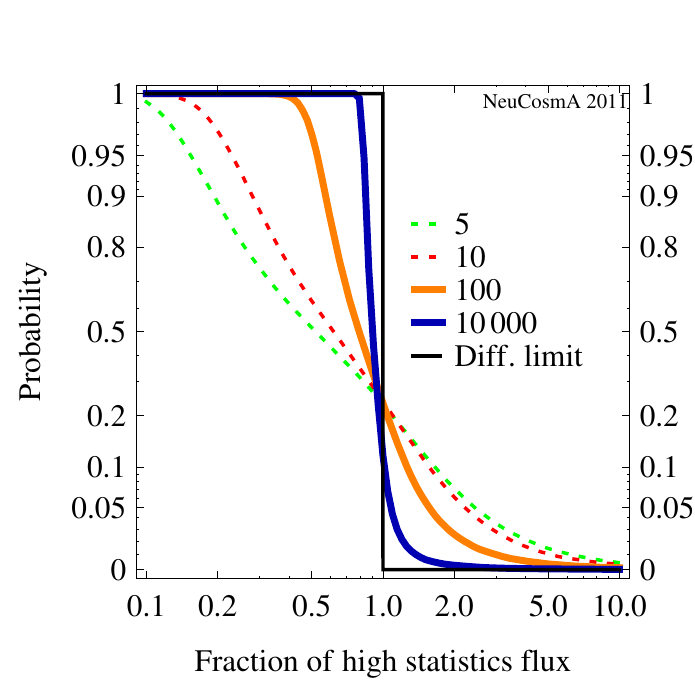}
\end{center}
\caption{\label{fig:stacking}  Left panel: Distribution of $10 \, 000$ bursts $\mathrm{d}\dot{N}/\mathrm{d}\textit{z}$ as a function of redshift  (histograms) and relative contribution of the individual GRBs $d_L^{-2} \, \mathrm{d}\dot{N}/\mathrm{d}\textit{z}$ (solid curves). The dashed curve shows the exact distribution function. Here it is assumed that the GRBs follow the star formation rate from Hopkins and Beacom~\cite{Hopkins:2006bw} with the  correction $\mathcal{E}(z)$ from Kistler et al.~\cite{Kistler:2009mv}.
Right panel:  Probability that the quasi-diffuse flux extrapolated from a low statistics sample with $n$ bursts is larger than a certain fraction of the diffuse limit (see legend for different values of $n$). This function corresponds to (one minus) the cumulative distribution function of the probability density.  Figure taken from \Ref~\cite{Baerwald:2011ee}.}
\end{figure}

In order to illustrate this problem, following \Ref~\cite{Baerwald:2011ee}, let us consider the redshift distribution of the GRBs as an example. We show in \figu{stacking}, left panel, a population of  $10 \, 000$ bursts, representative for the number of GRBs in the visible universe over about 10~years (\cf, histogram). These bursts are assumed to follow the star formation rate from Hopkins and Beacom~\cite{Hopkins:2006bw} with the  correction $\mathcal{E}(z)$ from Kistler et al.~\cite{Kistler:2009mv}. For the sake of simplicity, assume that all bursts have the same isotropic luminosity. From the discussion after \equ{boost}, we have $E_\nu^2 \phi \propto 1/d_L(z)^2$, which means that closer GRBs, which are however rarer, will lead to a larger neutrino flux. Thus, it is the product $d_L^{-2} \, \mathrm{d}\dot{N}/\mathrm{d}\textit{z}$ which determines the main contribution to the neutrino flux, shown as solid curve in \figu{stacking}. While the peak contribution in terms of the GRB distribution is at $z \sim 2-4$, this contribution function peaks at $z \simeq 1$. This observation has several implications: First of all, if the redshift $z$ is not measured, the neutrino flux may be overestimated if $z=2$ is assumed; \cf, \equ{eisobol}. Second, the number of bursts contributing in the region $z \simeq 1$ is rather small, which means that large statistical fluctuations are expected in quasi-diffuse flux estimates based on the stacking of a few bursts only. 

We quantify this systematical errors in the right panel of \figu{stacking}. In this panel, the probability that the quasi-diffuse flux extrapolated from a low statistics sample with $n$ bursts is larger than a certain fraction of the diffuse limit (see legend for different values of $n$). This function corresponds to (one minus) the cumulative distribution function of the probability density, and the step function corresponds to the diffuse limit. One can read off from this figure that for $n\simeq 100$, corresponding to the analysis in \Ref~\cite{Abbasi:2011qc}, the quasi-diffuse extrapolation will be within 50\% of the diffuse limit in the probability range corresponding to 90\% of all cases (between 0.05 and 0.95). This means that a 50\% error on the quasi-diffuse flux can be estimated from the redshift distribution only, while additional parameter variations increase this error~\cite{Baerwald:2011ee}.

\subsection{Neutrino flux predictions from gamma-ray observations}

\begin{figure}
\begin{center}
\begin{tabular}{ccc}
GRB 080916C & GRB 090902B & GRB 091024 \\
\includegraphics[width=0.25\textwidth]{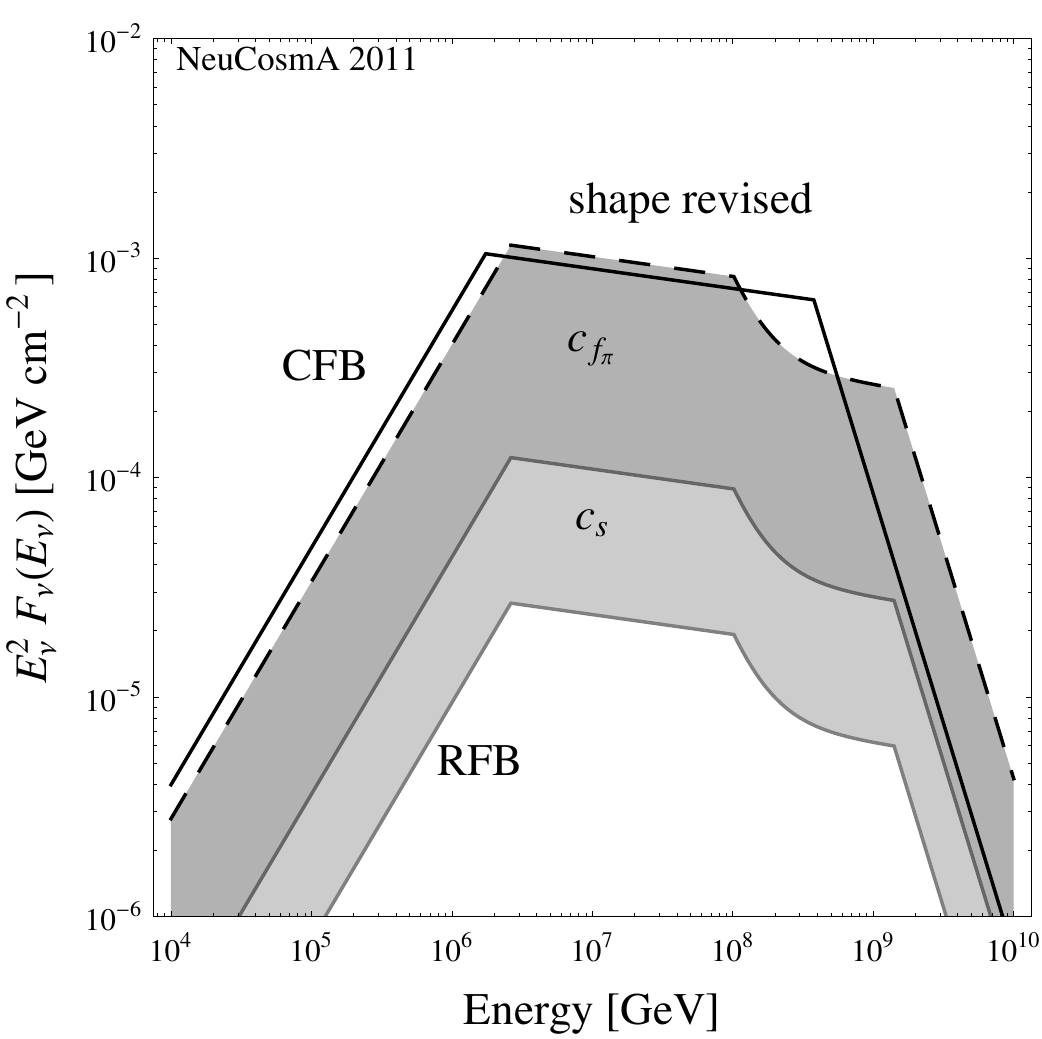} & \includegraphics[width=0.25\textwidth]{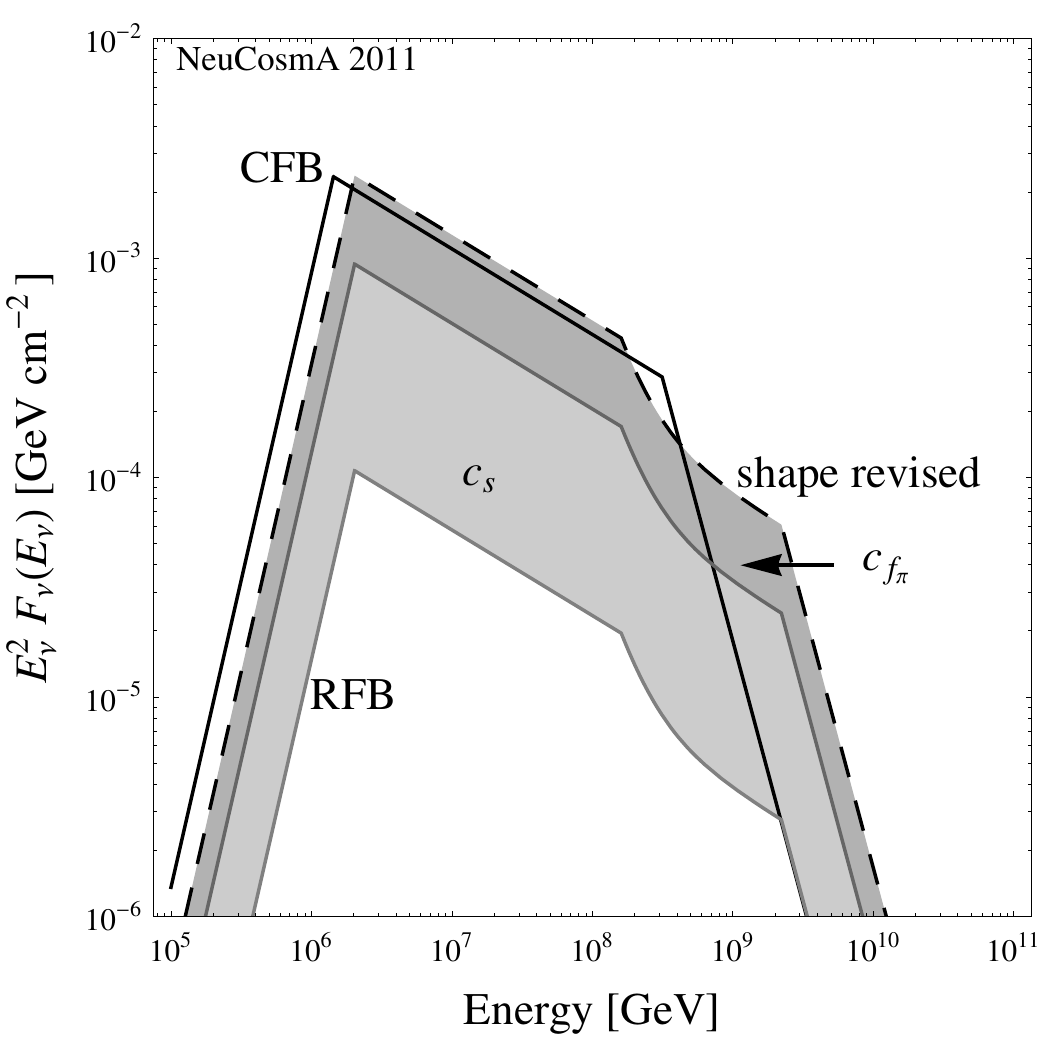} & \includegraphics[width=0.25\textwidth]{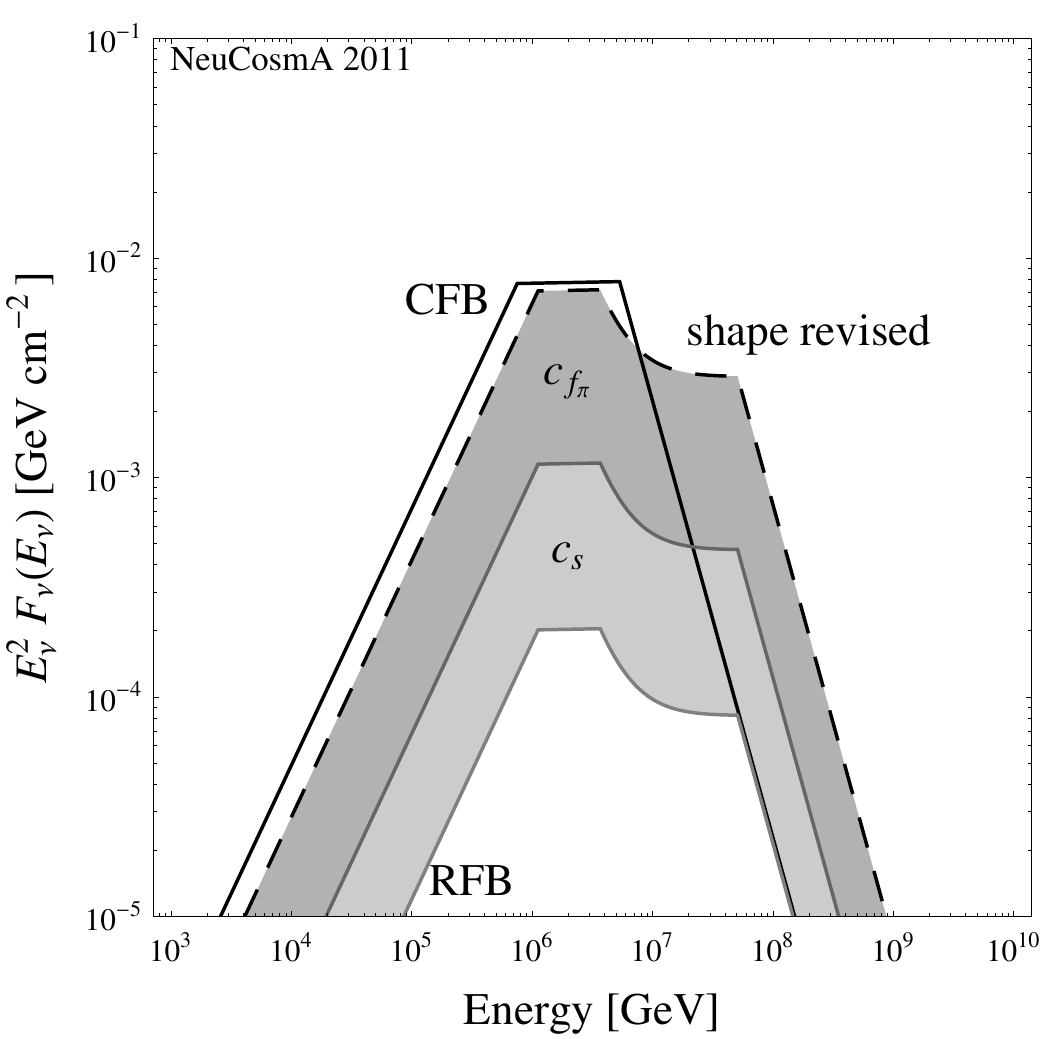} \\
\includegraphics[width=0.25\textwidth]{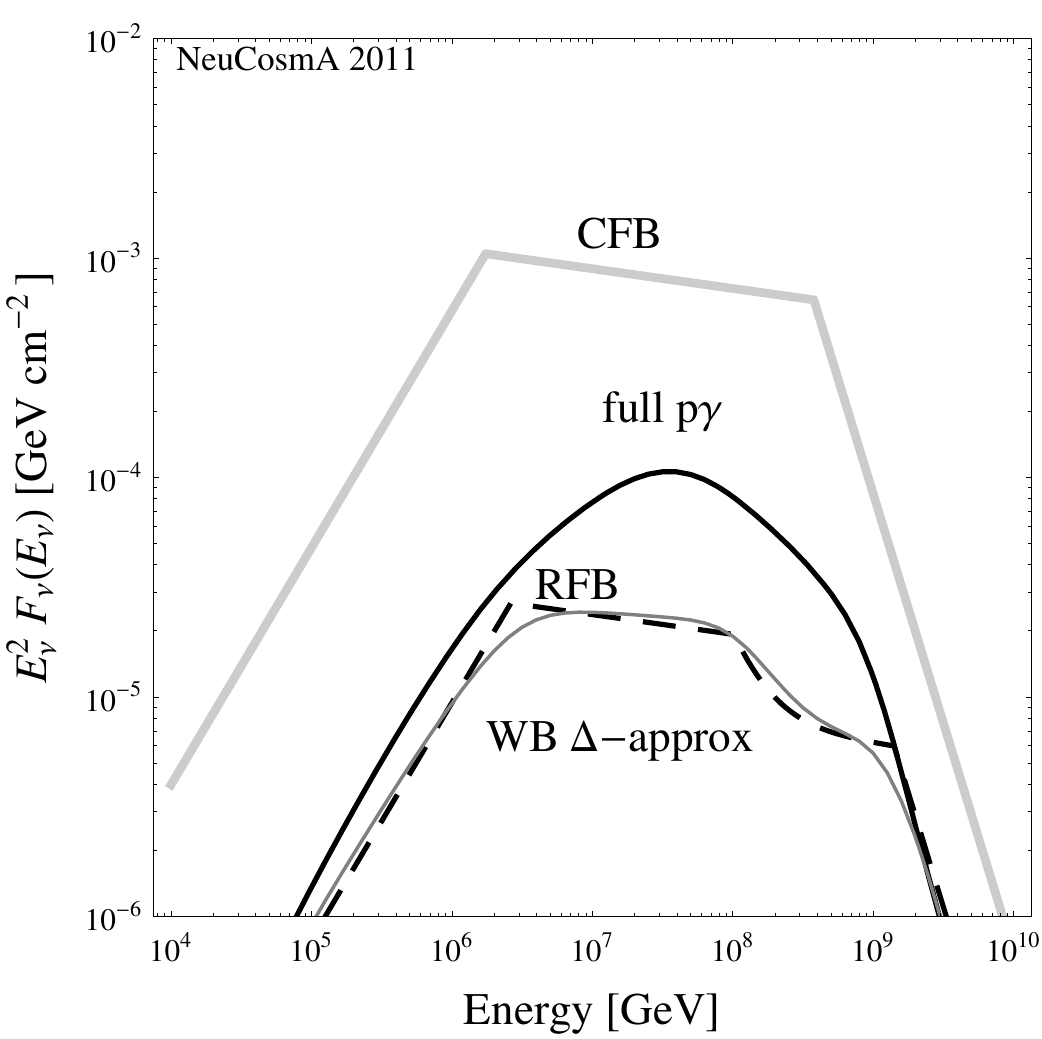} & \includegraphics[width=0.25\textwidth]{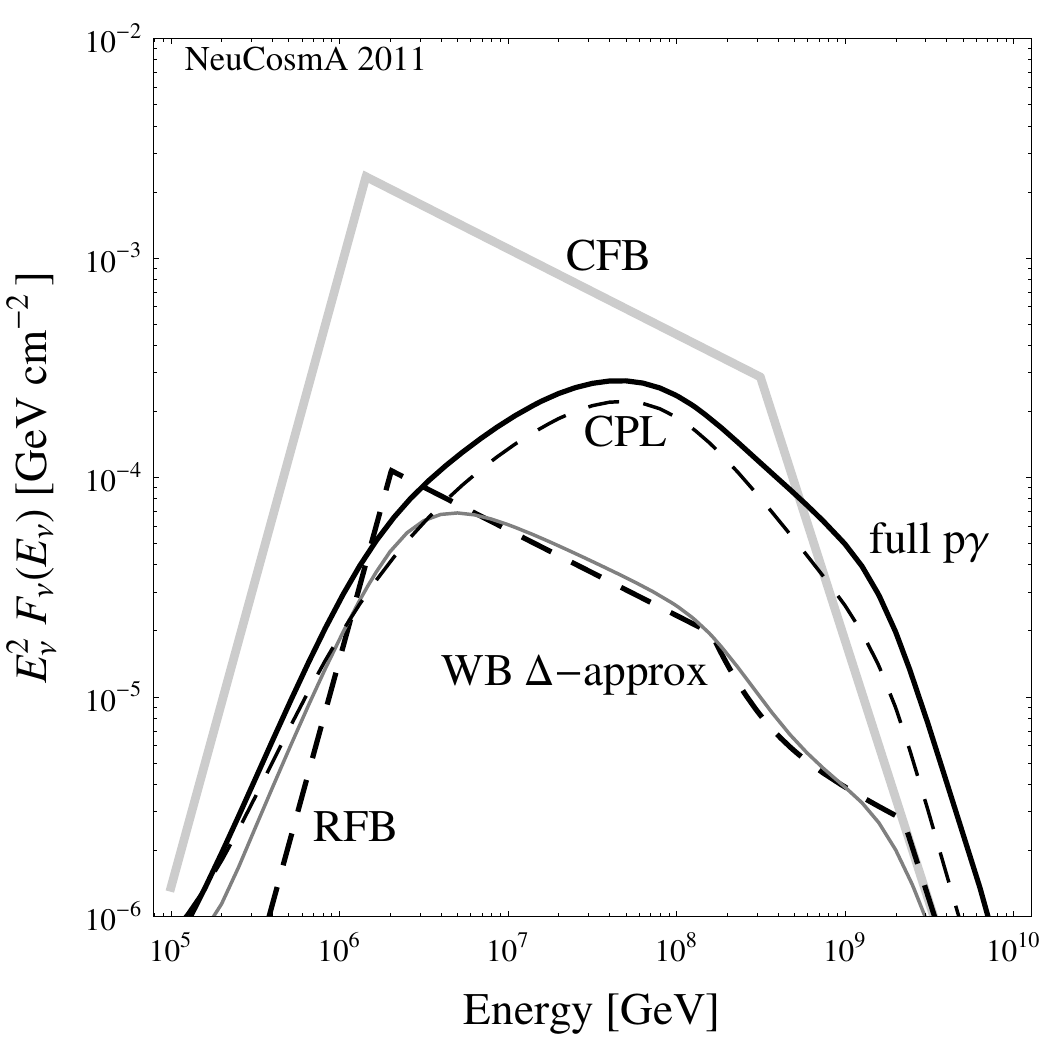} & \includegraphics[width=0.25\textwidth]{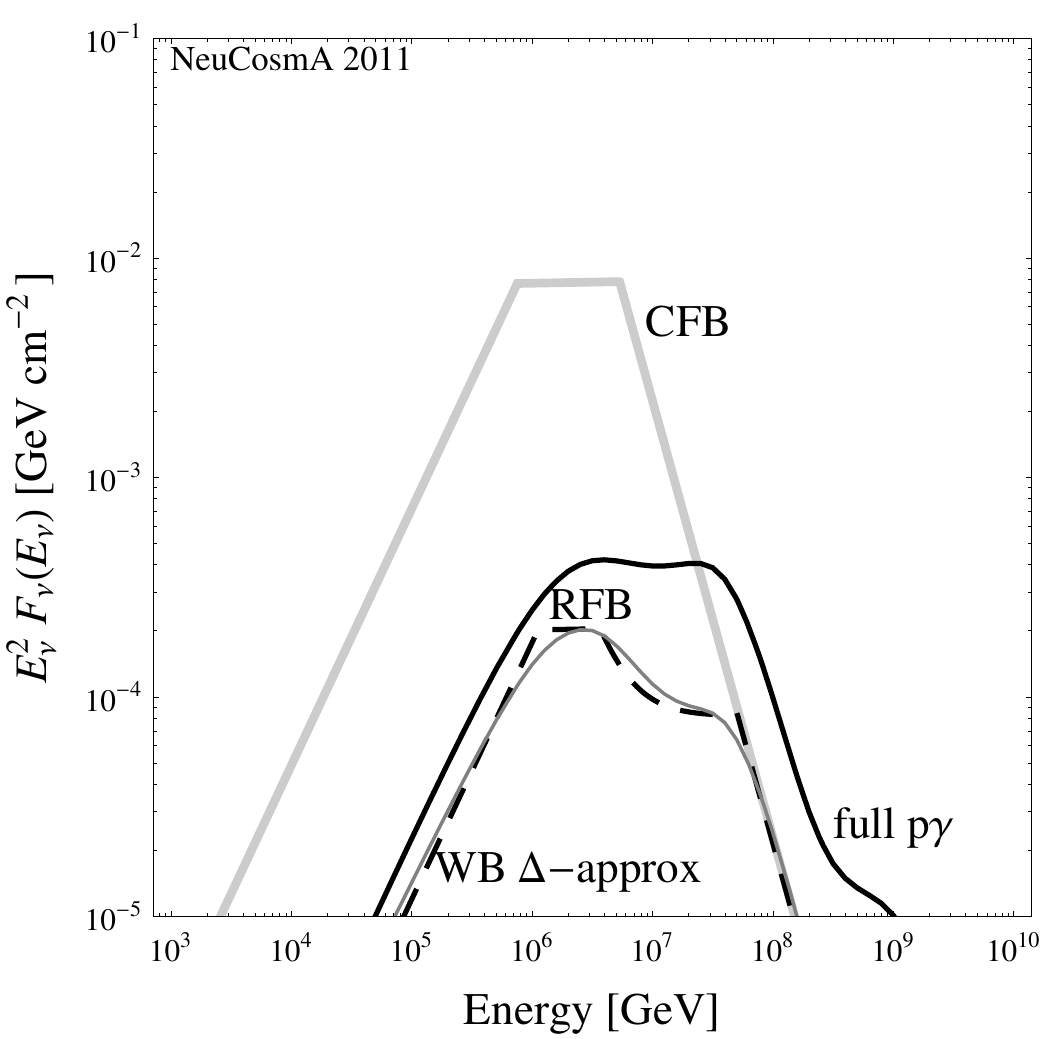} \\
\includegraphics[width=0.25\textwidth]{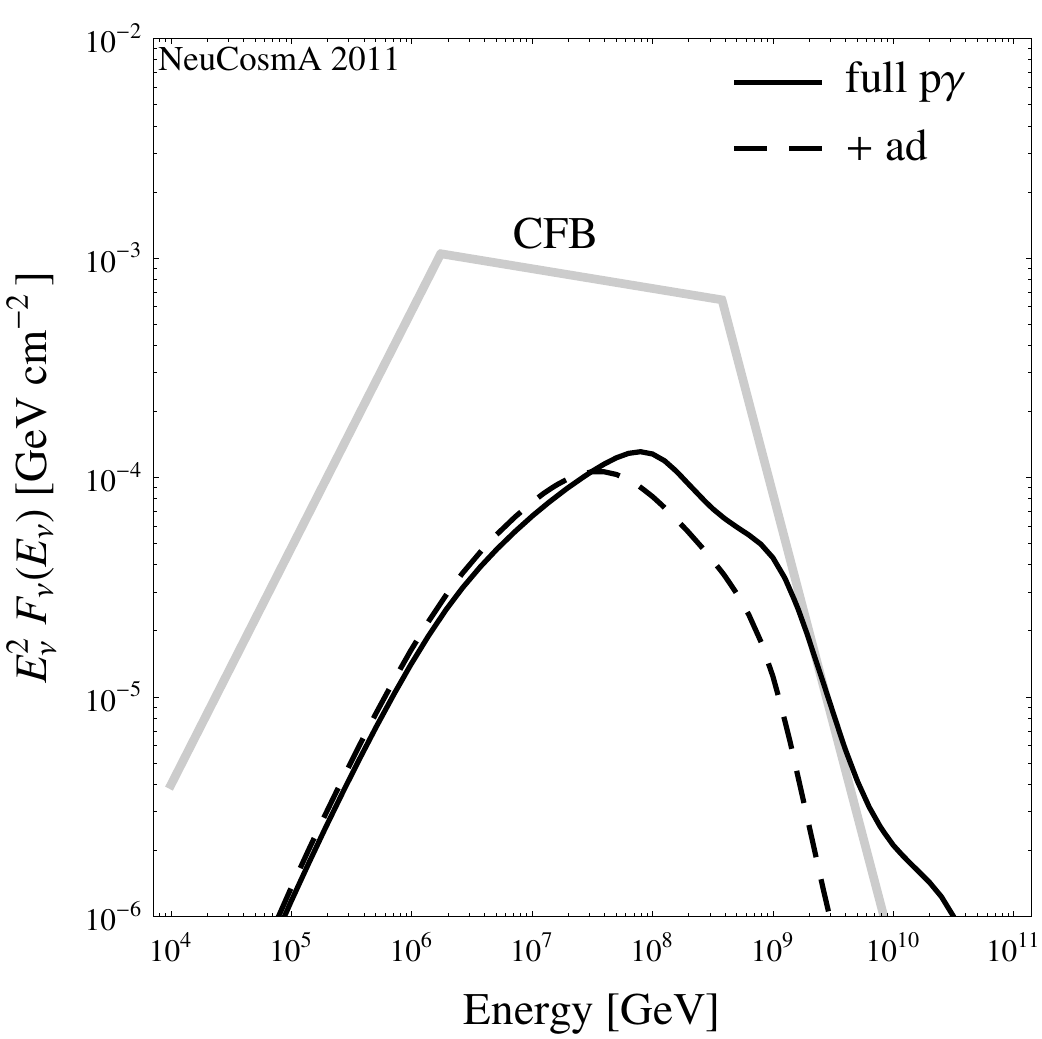} & \includegraphics[width=0.25\textwidth]{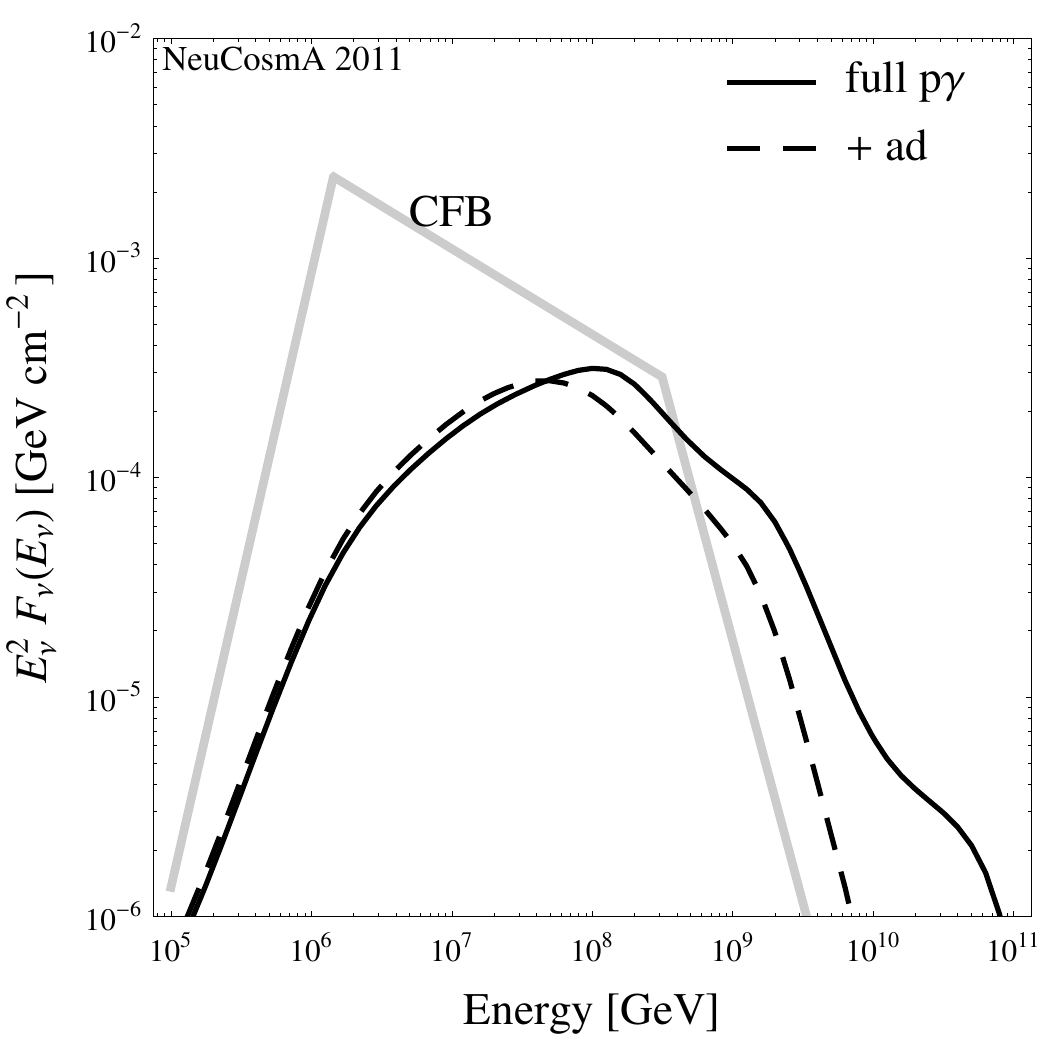} & \includegraphics[width=0.25\textwidth]{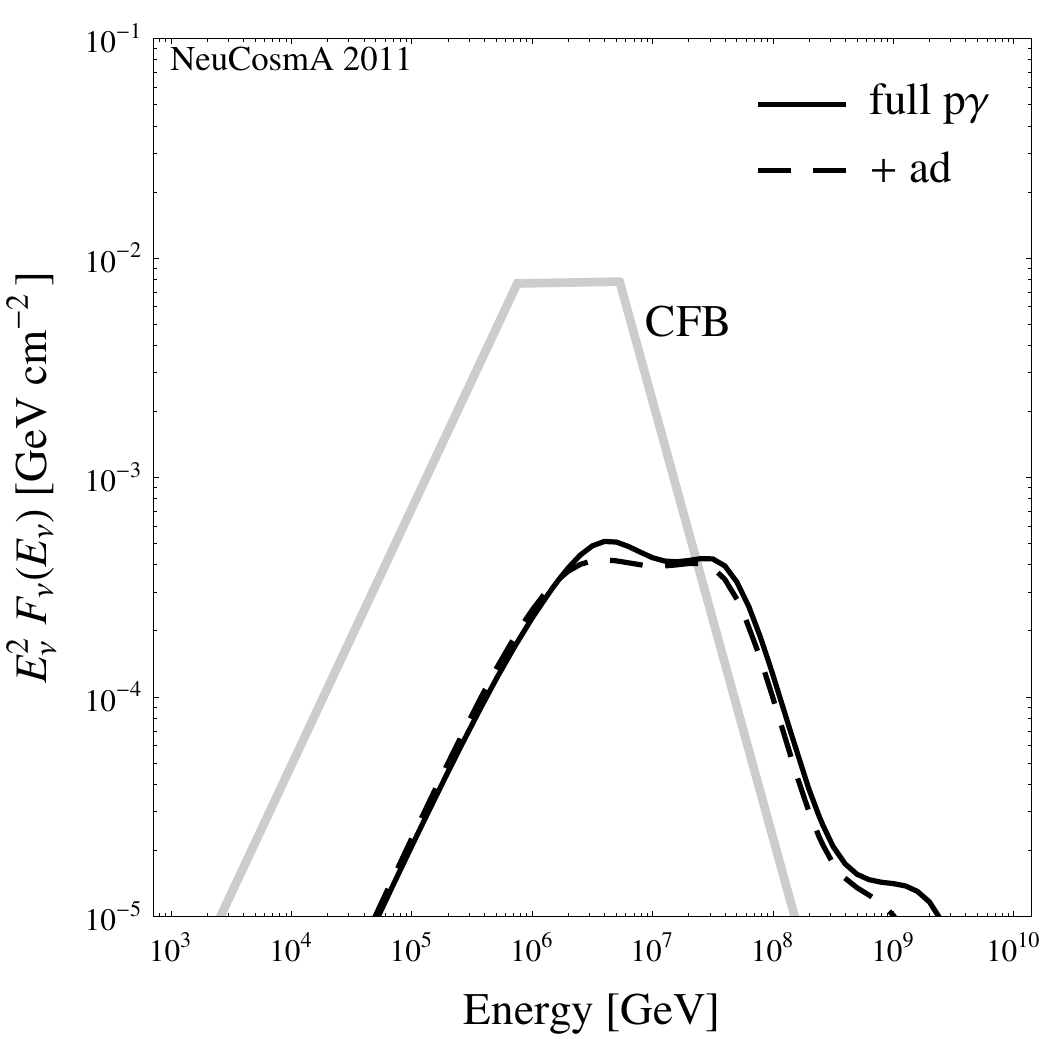}\\
\includegraphics[width=0.25\textwidth]{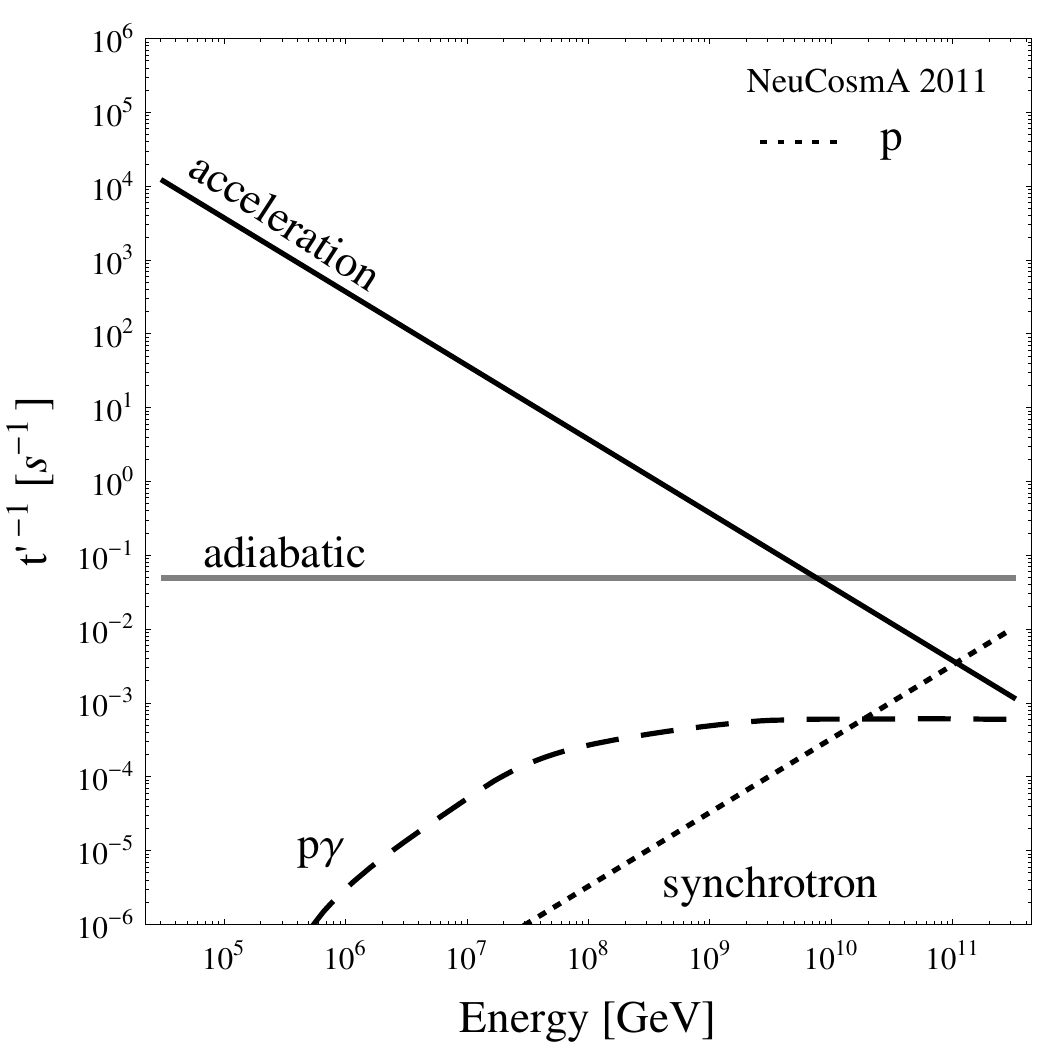} & \includegraphics[width=0.25\textwidth]{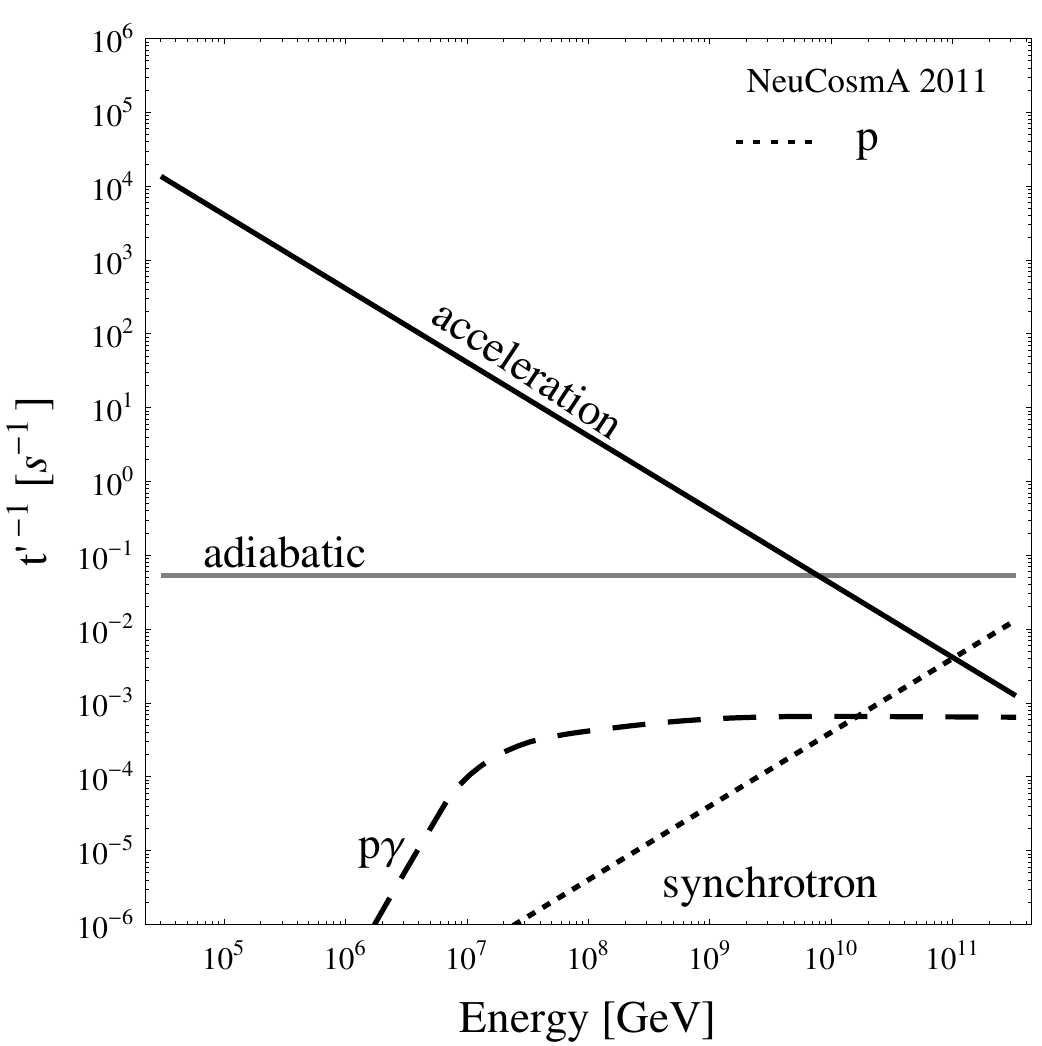} & \includegraphics[width=0.25\textwidth]{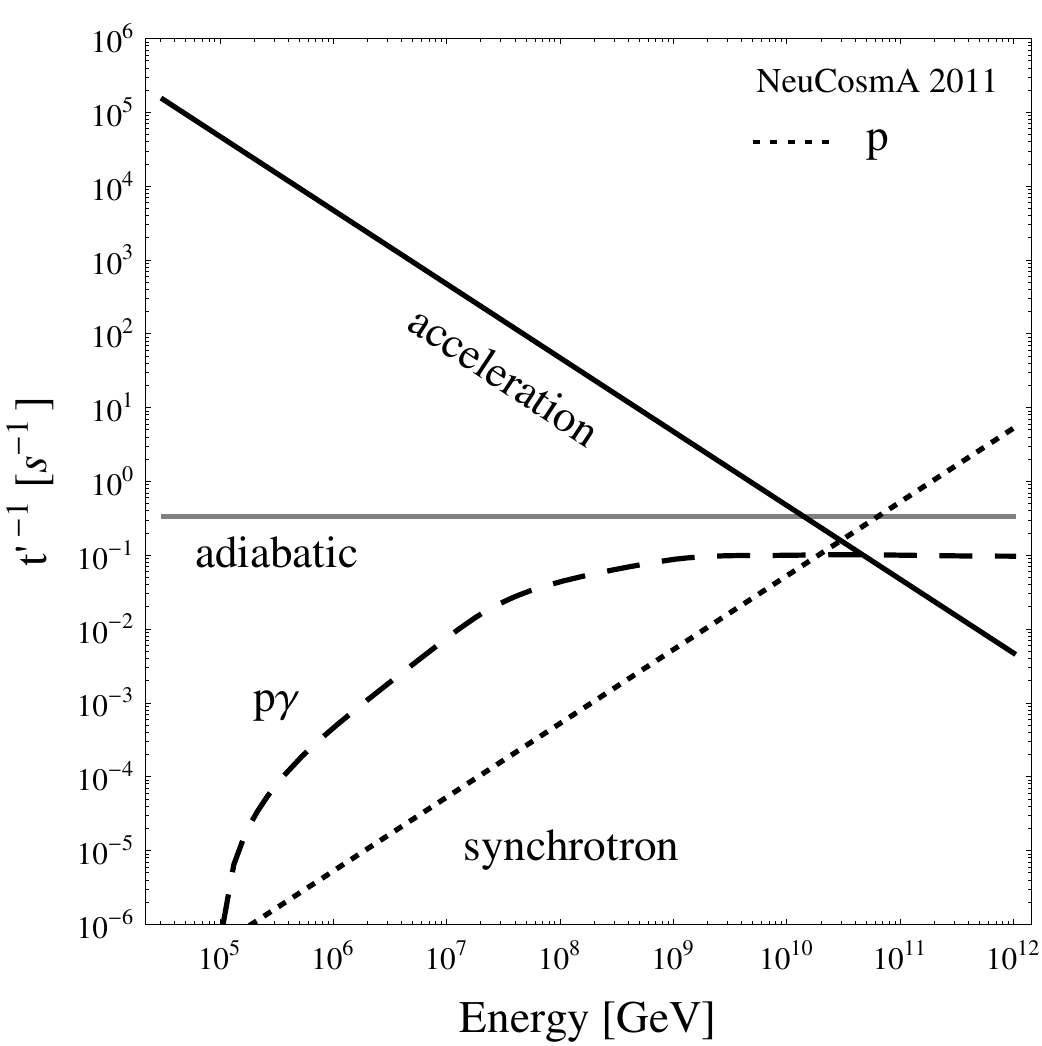}\\
\includegraphics[width=0.25\textwidth]{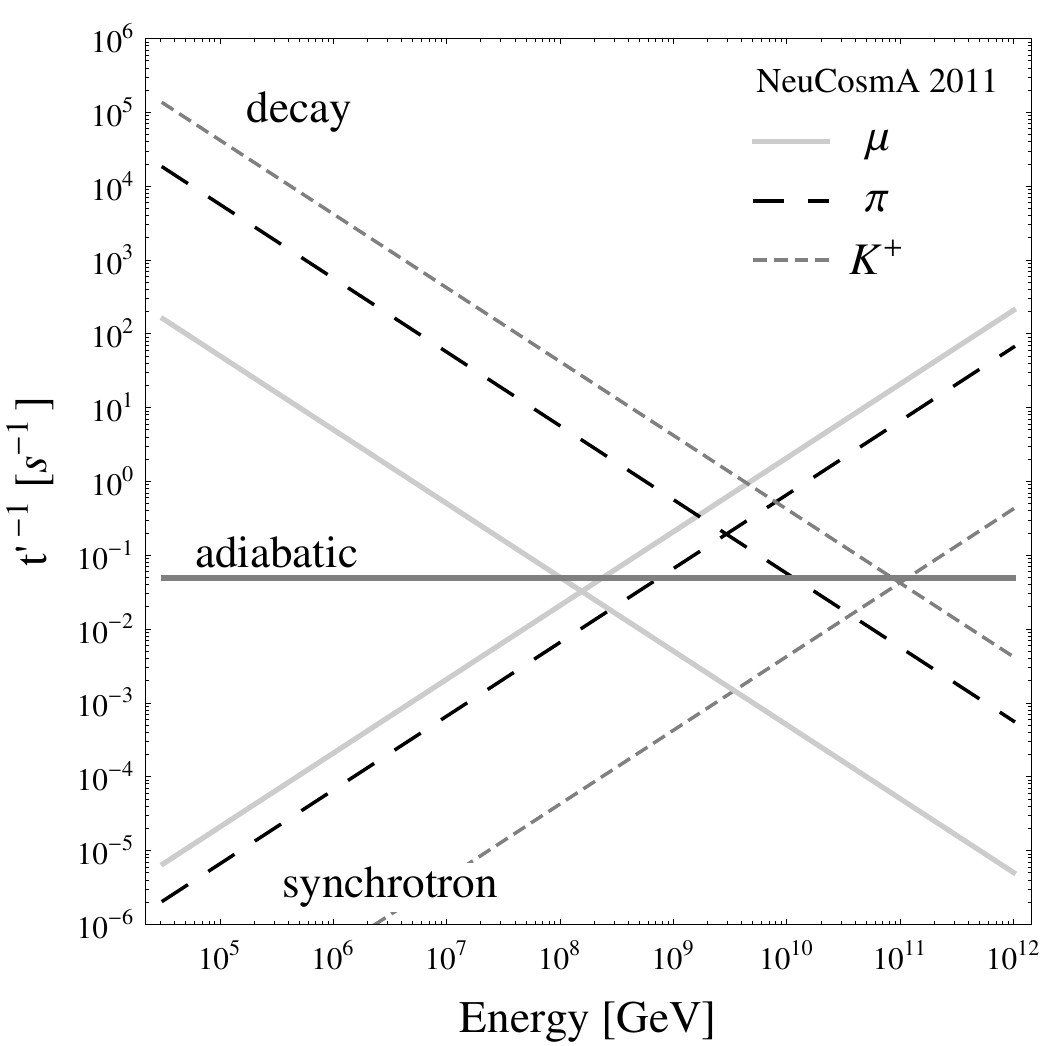}& \includegraphics[width=0.25\textwidth]{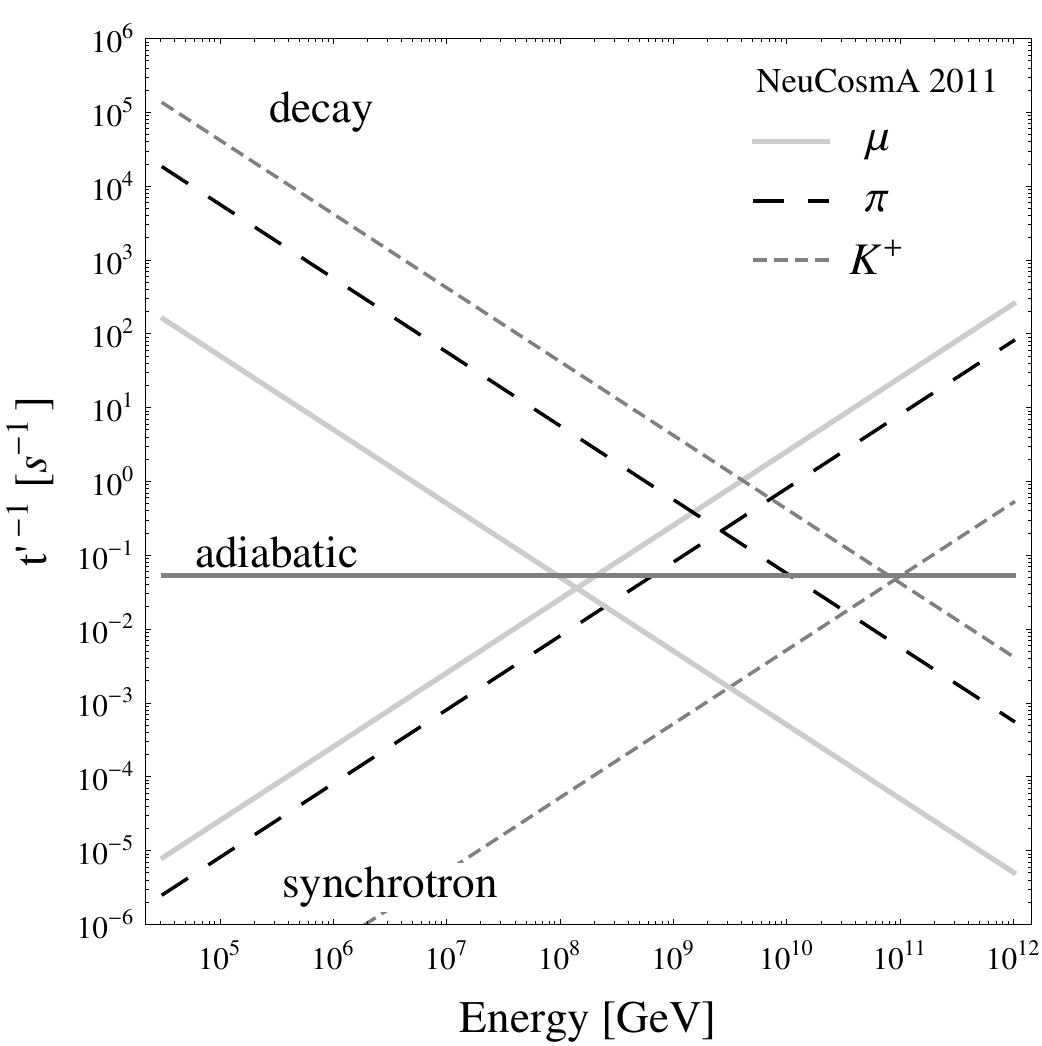}& \includegraphics[width=0.25\textwidth]{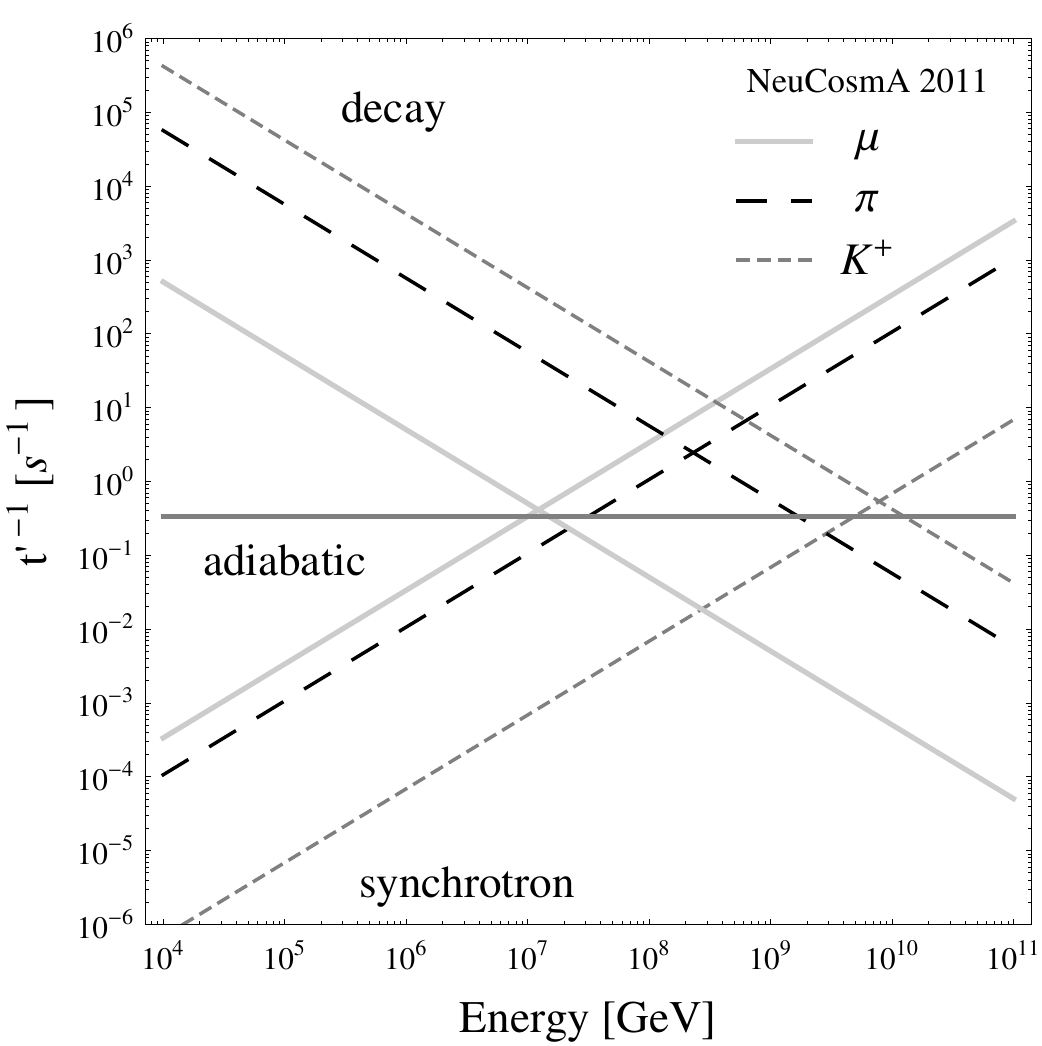}
\end{tabular}
\end{center}
 \caption{\label{fig:grbs} Expected (time-integrated) neutrino flux for three different GRBs (parameters: see \Refs~\cite{Nava:2010ig,Greiner:2009pm} for GRB 080916C, \Refs~\cite{Nava:2010ig,Abdo:2009pg} for GRB 090902B, and \Refs~\cite{Nava:2010ig,Gruber:2011gu} for GRB 091024). First row: Revision of IceCube analytical method CFB (correction of shape, normalization from pion production efficiency $c_{f_\pi}$, and normalization from neutrino versus proton spectral shape $c_S$), leading to RFB. Second row: Comparison analytical (CFB, RFB) methods with simplified numerical method ``WB $\Delta$-approximation'' \equ{photosimp} and full $p\gamma$ interactions. Third row: Impact of adiabatic cooling on protons and secondaries. Fourth and fifth row: corresponding inverse timescales (rates) for protons and secondaries, respectively. Courtesy of Svenja H{\"u}mmer~\cite{PhDHummer}.}
 \end{figure}

Since IceCube has not observed any GRB neutrino flux yet, there has been increasing tension between the model predictions~\cite{Waxman:1997ti,Guetta:2003wi,Abbasi:2009ig} and the observation~\cite{Abbasi:2011qc,IC59ProcICRC}.\footnote{Recently, the superluminal propagation of neutrinos has also been proposed as a reason why no neutrinos have been seen, see, \eg, \Ref~\cite{Autiero:2011hh}.} On the other hand, a direct comparison with the photo-meson production in \Ref~\cite{Waxman:1997ti} was performed in \Ref~\cite{Baerwald:2010fk} (see also \figu{photo}), demonstrating that the neutrino flux is actually underestimated in the analytical approaches. Therefore, the differences between the numerical approach in \Sec~\ref{sec:grbmodel} and the analytical models in \Refs~\cite{Guetta:2003wi,Abbasi:2009ig} (based on \Ref~\cite{Waxman:1997ti}), have been identified in \Ref~\cite{Hummer:2011ms} by a re-computation of the analytical models and the analytical computation of a simplified version of the numerical code. As far as the astrophysics ingredients are concerned, these approaches can be shown to be equivalent, based on the same logic; see \Sec~\ref{sec:grbmodel}. The main differences are: magnetic field and flavor-dependent effects are explicitely included in numerical approach, additional pion, neutron, and kaon production modes are computed, and the full energy dependencies are taken into account. 

We illustrate in \figu{grbs} the comparison between analytical and numerical approaches, where three different recent Fermi-measured GRBs have been chosen as examples: GRB 080916C, GRB 090902B, and GRB 091024. GRB 080916C has been selected, because it is one of the brightest bursts ever seen, although at a large redshift, and one of the best studied Fermi-LAT bursts. The gamma-ray spectrum of GRB 090902B can be fit by a Band function and a cutoff power law (CPL), which means that it can be used to illustrate the difference. GRB 091024 can be regarded as a typical example representative for many Fermi-GBM bursts~\cite{Nava:2010ig}, except from the long duration. Note that the first two bursts have an exceptionally large $\Gamma \gtrsim 1000$, whereas $\Gamma \simeq 200$ for the third burst. All three bursts have in common that that the required parameters for the neutrino flux computation can be taken from the literature, in particular, the properties of the gamma-ray spectrum (including fluence), $\Gamma$, $t_v$, $z$, and $T_{90}$, see figure caption for the references. 

In \figu{grbs}, we show in the first row the computation of the predicted neutrino flux with the IceCube analytical method~\cite{Abbasi:2009ig}, called CFB (conventional fireball calculation) here. As described in \Ref~\cite{Hummer:2011ms}, the corrections of shape, normalization from pion production efficiency $c_{f_\pi}$, and normalization from neutrino versus proton spectral shape $c_S$ (see also \Ref~\cite{Li:new}) lead to a revised analytical calculation RFB (revised fireball calculation). From the figure, one can easily read off that these revisions strongly depend on the burst parameters, especially the photon spectral shape. Comparing the predicted CFB fluxes with the bursts used for the IC40 analysis (Fig.~1 in \Ref~\cite{Abbasi:2011qc}), one can easily see that the expected fluxes of the first two bursts are about a factor 50 below that of the most luminous bursts in that analysis, and the third example about a factor of 5 below. This is expected from the scaling of the pion production efficiency $\propto \Gamma^{-4}$ in that approach. 

In the second row of  \figu{grbs}, we show the comparison of the analytical (CFB, RFB) methods with a simplified numerical method ``WB $\Delta$-approximation'', \cf, \equ{photosimp}, and the full $p\gamma$ interactions. In most cases, the simplified numerical approach matches the method RFB rather well, which proofs the validity of the derived corrections. However, for GRB 090902B (middle panel), the spectrum below the first break is different from the analytical estimate because the scalings of the weak decays limit the steepness of the spectrum there.
The final numerical calculation including all production modes is then significantly enhanced again, especially due to multi-pion production~\cite{Baerwald:2010fk}. 
 Using a cutoff power law for GRB 090902B for the gamma-ray spectral fit (middle panel), the normalization of the prediction slightly reduces in that example because the photon density above the photon break is suppressed. Comparing the original CFB method with the final numerical computation ``full p$\gamma$'', it is interesting that this can significantly deviate in both normalization and shape, but this deviation depends on the burst parameters. Very interestingly, a similar neutrino flux normalization for all three bursts is obtained, which means that its probably not warranted to say that the neutrino flux from high-$\Gamma$ bursts such as GRB 080916C is expected to be small. However, note that such extreme bursts only make up for a small fraction of the observed bursts, and the conclusions from neutrinos will be determined by the statistical properties of the burst sample in the stacking analysis. Finally, note that the numerical calculations in \figu{grbs} do not depend on any approximations,  whereas different analytical methods lead to different predictions, similar to CFB. Therefore, the numerical computations should be regarded as the benchmark which defines the corrections, not vice versa. Within the simplest fireball neutrino model, there are only small model dependencies within the numerical approach. For instance, the integral limits in \equ{protonorm} (the minimal and maximal proton energies) have to be specified,\footnote{Note, however, that for an $(E'_p)^{-2}$ injection spectrum, the energy partition only logarithmically depends on the minimal and maximal proton energies.} and a bolometric correction may have to be applied to \equ{photonorm} -- which typically has small effects.

One of the effects included in the final result of the second row of \figu{grbs} is the adiabatic cooling of primaries (protons) and secondaries (muons, pions, kaons). We illustrate the effect of this cooling component in the lower three rows of the figure, where we show the impact on the final result (third row) and the respective inverse timescales (rates) for protons (fourth row) and secondaries (fifth row). For the protons (fourth row), synchrotron losses are assumed to determine the maximal energies in the absence of adiabatic cooling, whereas the larger of the synchrotron or adiabatic cooling loss rates determines the maximal proton energy otherwise. From the fourth row, one can also easily read off that the energy losses due to $p\gamma$ interactions are typically sub-dominant. The comparison with the third row illustrates that, depending on the burst parameters, the kaon hump (the rightmost one) can be suppressed by the adiabatic cooling of the protons, whereas the normalization of the spectra is hardly affected. From the fifth row, one can also read off that adiabatic cooling may have a small effect on the muons, for which it sometimes dominates in a small energy range. This leads to a small suppression of the first hump, coming from the muon decays (see left and middle panels of third row). 

\begin{figure}[t]
\begin{center}
\includegraphics[width=0.5\textwidth]{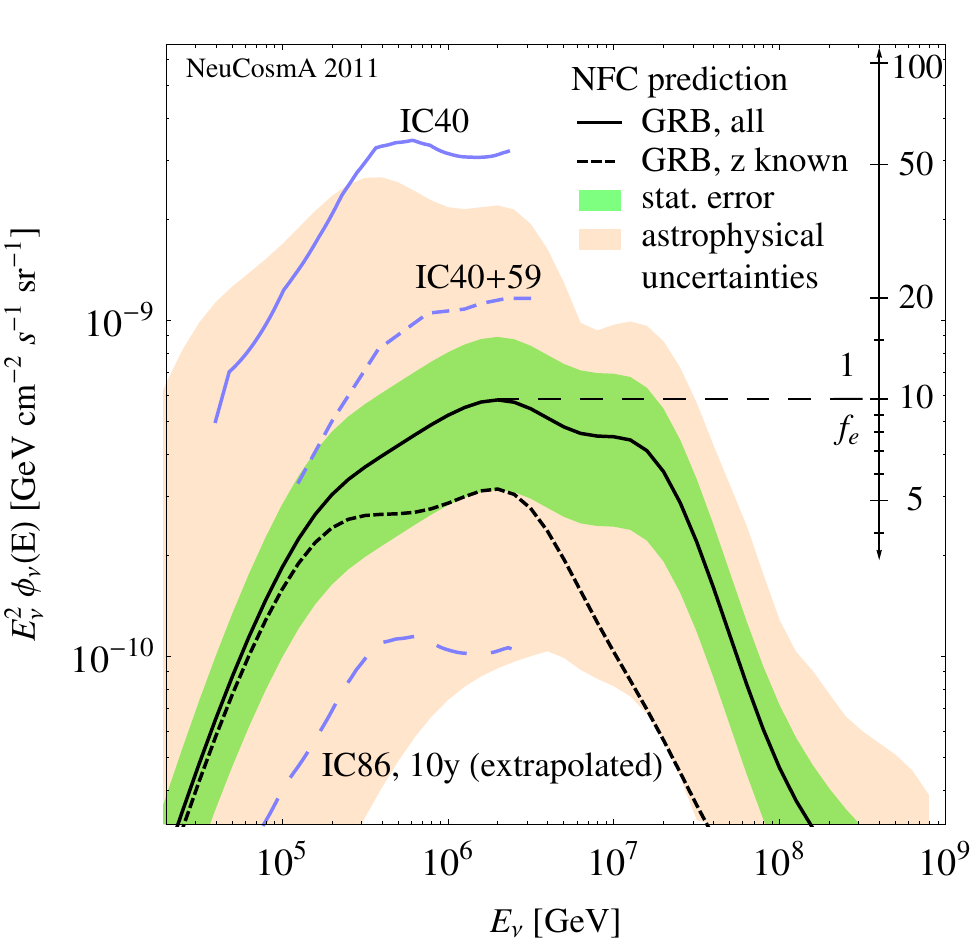}
\end{center}
\caption{\label{fig:syst} Numerical prediction (NFC) of the quasi-diffuse flux expected from the bursts used in the IC40-analysis, including the estimates for several model- or method-specific systematical uncertainties (see main text). In addition, the IC40 limit is shown, and two expectations are shown for comparison (IC59+40 from \Ref~\cite{IC59ProcICRC} and IC86 extrapolated  for $A_{\mathrm{eff}}^{\mathrm{IC86}} \simeq 3 \times A_{\mathrm{eff}}^{\mathrm{IC40}}$ from IC40; see, \eg, \Ref~\cite{Karle:2010xx}). Figure taken from \Ref~\cite{Hummer:2011ms}.}
\end{figure}

We show in \figu{syst} the predicted quasi-diffuse neutrino flux from the above numerical method to the IC40 bursts for the same bursts and parameters (solid black curve) used in that analysis, which is about one order of magnitude lower than the IC40 limit and a factor of two below the current limit. In this figure a number of systematical errors are shown as well, such as the statistical error discussed in the previous subsection, and the estimated astrophysical uncertainty (by varying the unknown parameters, such as proton injection index $\alpha_p = 1.8 \hdots 2.2$, variability timescale $t_v$ by one order of magnitude around the IceCube standard values, $t_v=0.01 \, \mathrm{s}$ for long bursts, $\Gamma$ from 200 to 1000, and the ratio $\epsilon_e/\epsilon_B$ from 0.1 to 10). In addition,
note that $z$ has only been measured for a few bursts used in the IC40 analysis, whereas $z \simeq 2$ has been assumed for the long bursts with unknown $z$. As we illustrated in the previous subsection, this is potentially problematic, which means that a solid lower limit for the prediction can be only obtained for bursts with measured $z$ (dashed black curve in \figu{syst}). 
Note that our prediction varies not as strong as one may expect from $f_\pi \propto t_v^{-1} \, \Gamma^{-4}$~\cite{Guetta:2003wi}. First of all, it is clear from \equ{boost} that $E_\nu^2 \phi_\nu(E) \propto \Gamma^{-2}$ because $E_\nu \propto \Gamma$. Second, the synchrotron losses of the secondaries damp this variation~\cite{Guetta:2001cd}: for larger $\Gamma$, the energy densities in the source will decrease because of the energy equipartition, and consequently $B'$ in \equ{B}. This reduces the energy losses of the secondaries, which means that more energy goes into the neutrinos. 

From the extrapolated IceCube limit for IC86 in \figu{syst}, it is obvious that IceCube will finally test the parameter space of the GRB fireball models, whereas the current limits already start to enter the meaningful parameter space. On the other hand, it is not clear what values of $\Gamma$ and $t_v$ most bursts actually have, and what would contribute most to the neutrino flux. For example, a theoretical study of the dominant $\Gamma$ to the diffuse neutrino flux has been performed in \Ref~\cite{Baerwald:2011ee} for different model hypotheses. The conventional fireball approach, which is presented above, leads to $\Gamma \sim 200$, which means that bursts such as GRB 091024 would dominate the neutrino flux. In this case, one would expect the prediction rather in the upper half of the shaded region of  \figu{syst}, and a near-future detection in IceCube may be rather likely. On the other hand, recent studies for $t_\nu$ seem to prefer larger average $t_\nu \simeq 0.1 \, \mathrm{s}$ than the $t_\nu \simeq 0.01 \, \mathrm{s}$ used by IceCube for long bursts~\cite{MacLachlan:new}, which points towards the lower half of the shaded region. 
Independent of these parameters, the neutrino flux prediction is proportional to $1/f_e$, which means that the final IceCube results will limit the baryonic loading as key parameter.

Finally, note that similar considerations as for the GRBs may apply to AGN models, such as \Refs~\cite{Stecker:1991vm,Mannheim:1995mm,Stecker:2005hn}, and it is yet to be seen what the impact on the cosmic ray connection is, see \Ref~\cite{Ahlers:2011jj}.

\section{Summary and conclusions}
\label{sec:summary}

We have discussed the impact of particle physics effects on the neutrino fluxes from cosmic accelerators, based on a generic numerical approach of the photohadronic charged meson production in sources optically thin to neutron escape; \cf, \figu{flowchart}. The starting point have been the proton and photon densities (spectra) within the source, which lead to the secondary meson production. We have included additional $t$-channel, higher resonance, and multi-pion production modes, neutrinos from kaon and neutron decays, the synchrotron cooling of the secondaries (pions, muons, and kaons), the helicity dependence of the muon decays, and the full spectral dependencies. While this approach may not be able to describe every source accurately, it can be regarded as the minimal approach to neutrino production by $p\gamma$ interactions using  at least the well-known particle physics ingredients, to be extended by further components if applied to particular source classes. We have applied this framework to GRBs in \Sec~\ref{sec:grbmodel}, where the target photons are inferred from the gamma-ray observation. We have demonstrated that it contains all the ingredients in frequently used analytical models, such as \Refs~\cite{Waxman:1997ti,Guetta:2003wi,Abbasi:2009ig,Abbasi:2011qc}, which means that it can be regarded as a numerical refined version of the conventional fireball neutrino model. In addition, we have discussed a generic AGN-like application where the target photons are produced by the synchrotron emission of co-accelerated electrons in \Sec~\ref{sec:gensource}. From the presentation in this review, it must be clear that this application is at a similar level as the GRB approach in terms of the ingredients. Applications to specific neutrino flux normalization predictions are, however, yet to be performed.

The main consequences of the particle physics effects discussed in this review can be summarized as follows:
\begin{itemize}
\item
 Magnetic field effects and flavor mixing change the flavor composition. The assumption of an $\nu_e:\nu_\mu:\nu_\tau$ flavor composition at the source of $1:2:0$ (pion beam) can only be justified for $B' \lesssim 1$~kG, while it will normally change as a function of energy for larger values of $B'$. 
\item
 The  neutrino spectral shape for $p\gamma$ interactions depends on proton and photon injection spectra and the magnetic field and flavor effects. Roughly speaking, the $E_\nu^{-2}$ assumption only holds for the special case  $(\varepsilon')^{-1}$ for the photon spectrum. This means that this assumption is too simple for many sources, since the detector response depends on the interplay between spectral shape and differential limit. For example, the detector may miss a neutrino signal because the spectrum does not peak at the right energy, although the fraction of energy going into pion production in the source is the same as in another detectable case.
\item
 Additional pion production processes increase the neutrino production significantly, and lead to an impact on the spectral shape. Approximations using the $\Delta$-resonance approximation are useful for analytical estimates, but they do not take into account the cross section dependence at high center-of-mass energies. The impact on the neutrino flux is a factor of a few.
\item
 Spectral effects, such as the energy dependence of the proton interaction length or the energy dependence of the photon spectrum, lead a significant reduction of the GRB neutrino flux prediction of about one order of magnitude in comparison to analytical estimates. As a consequence, IceCube has just entered the relevant part of the parameter space to test the simplest GRB fireball model. Similar effects on AGN flux models are yet to be tested.
\item
 Neutrino fluxes from kaon and neutron decays are generically expected in any source. However, the neutrinos from neutron decays show typically up below the peak, the neutrinos from kaon decays above the peak. The relative importance of these effects depends on the strength of $B'$ (which may separate the spectra from the different parents) and the maximal proton energy (which controls the neutron decay contribution).
\item
 Using the Glashow resonance for the discrimination of $pp$ and $p\gamma$ interactions in the source is challenging. Especially, it is very hard to infer any physics information from the discovery of Glashow resonant events, whereas the absence of Glashow resonant events (in presence of a neutrino signal) can be interpreted as a $p \gamma$ source optically thin to neutron escape.
\item
 Already from the current IceCube limits one can extrapolate that conclusions for astrophysical sources will mostly like be based on source classes, not individual sources.  The aggregation of fluxes from many sources, however, introduces new systematical errors. For instance, the extrapolation from the neutrino fluxes from 117 GRBs to a quasi-diffuse flux implies an error of at least 50\% (at the 90\% confidence level), from the redshift distribution only.
\item
 In the presence of a high statistics neutrino flux (close to the current bound), useful information on possible new physics effects in the neutrino propagation may be inferred from the ratio between muon tracks and cascades. In this case, the energy dependence of the flavor ratio at the source and the energy dependence of the new physics effects may provide the necessary information to identify the effect. 
\end{itemize}

In conclusion, in the presence of data, neutrino astrophysics is now at the point where rough analytical estimates for neutrino fluxes are not sufficient anymore, since in that case no reliable conclusions can be drawn for the astrophysical models and thus the origin of the cosmic rays. Especially the particle physics effects discussed in this review can be relatively easily taken into account, and they are well known. We have demonstrated with numerous examples that these effects on flux normalization, spectral shape, flavor composition, and neutrino-antineutrino composition cannot be neglected. For example, a correction of the GRB neutrino flux prediction of one order of magnitude has been identified. Therefore, a reliable treatment of the particle physics of the neutrino production should be the state-of-the-art of any neutrino data interpretation in the multi-messenger context.

{\it Acknowledgments.}

I would like to thank Philipp Baerwald, Mauricio Bustamante, Svenja H{\"u}mmer, and Guey-Lin Lin for useful comments, and Svenja H{\"u}mmer for providing some figure from her Ph.D. thesis. This publication was funded by the German Research Foundation (DFG) and the University of W{\"u}rzburg in the funding program Open Access Publishing. It has been supported by Deutsche Forschungsgemeinschaft, grants WI 2639/3-1 and WI 2639/4-1.


\begin{thebibliography}{100}

\bibitem{Ahrens:2003ix}
IceCube, J.~Ahrens {\em et~al.},
\newblock Astropart. Phys. {\bf 20}, 507 (2004), arXiv:astro-ph/0305196.

\bibitem{Aslanides:1999vq}
ANTARES, E.~Aslanides {\em et~al.},
\newblock (1999), astro-ph/9907432.

\bibitem{Learned:2000sw}
J.~G. Learned and K.~Mannheim,
\newblock Ann. Rev. Nucl. Part. Sci. {\bf 50}, 679 (2000).

\bibitem{Halzen:2002pg}
F.~Halzen and D.~Hooper,
\newblock Rept. Prog. Phys. {\bf 65}, 1025 (2002), arXiv:astro-ph/0204527.

\bibitem{Chiarusi:2009ng}
T.~Chiarusi and M.~Spurio,
\newblock Eur.Phys.J. {\bf C65}, 649 (2010), arXiv:0906.2634.

\bibitem{Katz:2011ke}
U.~F. Katz and C.~Spiering,
\newblock (2011), arXiv:1111.0507.

\bibitem{Becker:2007sv}
J.~K. Becker,
\newblock Phys. Rept. {\bf 458}, 173 (2008), arXiv:0710.1557.

\bibitem{Rachen:1998fd}
J.~P. Rachen and P.~Meszaros,
\newblock Phys. Rev. {\bf D58}, 123005 (1998), arXiv:astro-ph/9802280.

\bibitem{Waxman:1997ti}
E.~Waxman and J.~N. Bahcall,
\newblock Phys. Rev. Lett. {\bf 78}, 2292 (1997), arXiv:astro-ph/9701231.

\bibitem{Stecker:1991vm}
F.~W. Stecker, C.~Done, M.~H. Salamon, and P.~Sommers,
\newblock Phys. Rev. Lett. {\bf 66}, 2697 (1991).

\bibitem{Mannheim:1993}
K.~{Mannheim},
\newblock Astron. Astrophys. {\bf 269}, 67 (1993), arXiv:astro-ph/9302006.

\bibitem{Stecker:2005hn}
F.~W. Stecker,
\newblock Phys. Rev. {\bf D72}, 107301 (2005), arXiv:astro-ph/0510537.

\bibitem{Abbasi:2010rd}
The IceCube Collaboration, R.~Abbasi {\em et~al.},
\newblock Astrophys.J. {\bf 732}, 18 (2011), arXiv:1012.2137.

\bibitem{Abbasi:2011ara}
The IceCube Collaboration, R.~Abbasi,
\newblock (2011), arXiv:1104.0075.

\bibitem{Abbasi:2011qc}
IceCube Collaboration, R.~Abbasi {\em et~al.},
\newblock Phys.Rev.Lett. {\bf 106}, 141101 (2011), arXiv:1101.1448.

\bibitem{Abbasi:2011jx}
IceCube Collaboration, R.~Abbasi {\em et~al.},
\newblock Phys.Rev. {\bf D84}, 082001 (2011), arXiv:1104.5187.

\bibitem{Waxman:1998yy}
E.~Waxman and J.~N. Bahcall,
\newblock Phys. Rev. {\bf D59}, 023002 (1999), arXiv:hep-ph/9807282.

\bibitem{Mannheim:1998wp}
K.~Mannheim, R.~J. Protheroe, and J.~P. Rachen,
\newblock Phys. Rev. {\bf D63}, 023003 (2001), arXiv:astro-ph/9812398.

\bibitem{IC59ProcICRC}
IceCube, R.~Abbasi {\em et~al.},
\newblock arXiv:1111.2741,
\newblock Proceedings of ICRC2011.

\bibitem{Guetta:2003wi}
D.~Guetta, D.~Hooper, J.~Alvarez-Muniz, F.~Halzen, and E.~Reuveni,
\newblock Astropart. Phys. {\bf 20}, 429 (2004), arXiv:astro-ph/0302524.

\bibitem{Abbasi:2009ig}
IceCube Collaboration, R.~Abbasi {\em et~al.},
\newblock Astrophys. J. {\bf 710}, 346 (2010), arXiv:0907.2227.

\bibitem{Ahlers:2011jj}
M.~Ahlers, M.~Gonzalez-Garcia, and F.~Halzen,
\newblock Astropart.Phys. {\bf 35}, 87 (2011), arXiv:1103.3421.

\bibitem{Arguelles:2010yj}
C.~Arguelles, M.~Bustamante, and A.~Gago,
\newblock JCAP {\bf 1012}, 005 (2010), arXiv:1008.1396.

\bibitem{Mucke:1998mk}
A.~M{\"u}cke, J.~P. Rachen, R.~Engel, R.~J. Protheroe, and T.~Stanev,
\newblock Publ. Astron. Soc. Austral. {\bf 16}, 160 (1999),
  arXiv:astro-ph/9808279.

\bibitem{Mucke:1999yb}
A.~M{\"u}cke, R.~Engel, J.~Rachen, R.~Protheroe, and T.~Stanev,
\newblock Comput.Phys.Commun. {\bf 124}, 290 (2000), arXiv:astro-ph/9903478.

\bibitem{Mucke:2000rn}
A.~M{\"u}cke and R.~J. Protheroe,
\newblock Astropart. Phys. {\bf 15}, 121 (2001), arXiv:astro-ph/0004052.

\bibitem{Murase:2005hy}
K.~Murase and S.~Nagataki,
\newblock Phys. Rev. {\bf D73}, 063002 (2006), arXiv:astro-ph/0512275.

\bibitem{Hummer:2010vx}
S.~H{\"u}mmer, M.~R{\"u}ger, F.~Spanier, and W.~Winter,
\newblock Astrophys. J. {\bf 721}, 630 (2010), arXiv:1002.1310.

\bibitem{Baerwald:2010fk}
P.~Baerwald, S.~H{\"u}mmer, and W.~Winter,
\newblock Phys. Rev. {\bf D83}, 067303 (2011), arXiv:1009.4010.

\bibitem{Baerwald:2011ee}
P.~Baerwald, S.~H{\"u}mmer, and W.~Winter,
\newblock Astropart. Phys.  (to appear), arXiv:1107.5583.

\bibitem{Asano:2006zzb}
K.~Asano and S.~Nagataki,
\newblock Astrophys. J. {\bf 640}, L9 (2006), arXiv:astro-ph/0603107.

\bibitem{Kachelriess:2006fi}
M.~Kachelriess and R.~Tomas,
\newblock Phys. Rev. {\bf D74}, 063009 (2006), arXiv:astro-ph/0606406.

\bibitem{Kachelriess:2007tr}
M.~Kachelriess, S.~Ostapchenko, and R.~Tomas,
\newblock Phys. Rev. {\bf D77}, 023007 (2008), arXiv:0708.3047.

\bibitem{Hummer:2010ai}
S.~H{\"u}mmer, M.~Maltoni, W.~Winter, and C.~Yaguna,
\newblock Astropart. Phys. {\bf 34}, 205 (2010), arXiv:1007.0006.

\bibitem{Moharana:2010su}
R.~Moharana and N.~Gupta,
\newblock Phys.Rev. {\bf D82}, 023003 (2010), arXiv:1005.0250.

\bibitem{Moharana:2011hh}
R.~Moharana and N.~Gupta,
\newblock (2011), arXiv:1107.4483.

\bibitem{Kashti:2005qa}
T.~Kashti and E.~Waxman,
\newblock Phys. Rev. Lett. {\bf 95}, 181101 (2005), arXiv:astro-ph/0507599.

\bibitem{Lipari:2007su}
P.~Lipari, M.~Lusignoli, and D.~Meloni,
\newblock Phys. Rev. {\bf D75}, 123005 (2007), arXiv:0704.0718.

\bibitem{Reynoso:2008gs}
M.~M. Reynoso and G.~E. Romero,
\newblock Astron. Astrophys. {\bf 493}, 1 (2009), arXiv:0811.1383.

\bibitem{Pakvasa:2008nx}
S.~Pakvasa,
\newblock Mod. Phys. Lett. {\bf A23}, 1313 (2008), arXiv:0803.1701.

\bibitem{Barr:1988rb}
S.~M. Barr, T.~Gaisser, P.~Lipari, and S.~Tilav,
\newblock Phys.Lett. {\bf B214}, 147 (1988).

\bibitem{Barr:1989ru}
G.~Barr, T.~Gaisser, and T.~Stanev,
\newblock Phys.Rev. {\bf D39}, 3532 (1989).

\bibitem{Lipari:1993hd}
P.~Lipari,
\newblock Astropart.Phys. {\bf 1}, 195 (1993).

\bibitem{Hummer:GRBFireball}
S.~H{\"u}mmer, P.~Baerwald, and W.~Winter,
\newblock (2011), arXiv:1112.1076.

\bibitem{Li:new}
Z.~Li,
\newblock Phys.Rev. {\bf D} (to appear), arXiv:1112.2240.

\bibitem{Winter:2011jr}
W.~Winter,
\newblock Phys.Rev. {\bf D} (to appear), arXiv:1103.4266.

\bibitem{Kelner:2006tc}
S.~Kelner, F.~A. Aharonian, and V.~Bugayov,
\newblock Phys.Rev. {\bf D74}, 034018 (2006), arXiv:astro-ph/0606058.

\bibitem{Vissani:2011ea}
F.~Vissani and F.~Aharonian,
\newblock (2011), arXiv:1112.3911.

\bibitem{Abraham:2010yv}
Pierre Auger Observatory Collaboration, J.~Abraham {\em et~al.},
\newblock Phys.Rev.Lett. {\bf 104}, 091101 (2010), arXiv:1002.0699.

\bibitem{Greisen:1966jv}
K.~Greisen,
\newblock Phys.Rev.Lett. {\bf 16}, 748 (1966).

\bibitem{Zatsepin:1966jv}
G.~Zatsepin and V.~Kuzmin,
\newblock JETP Lett. {\bf 4}, 78 (1966).

\bibitem{Kelner:2008ke}
S.~R. Kelner and F.~A. Aharonian,
\newblock Phys. Rev. {\bf D78}, 034013 (2008), arXiv:0803.0688.

\bibitem{Pakvasa:2010jj}
S.~Pakvasa,
\newblock (2010), arXiv:1004.5413.

\bibitem{Koers:2007je}
H.~B. Koers and R.~A. Wijers,
\newblock (2007), arXiv:0711.4791.

\bibitem{Farzan:2008eg}
Y.~Farzan and A.~Y. Smirnov,
\newblock Nucl.Phys. {\bf B805}, 356 (2008), arXiv:0803.0495.

\bibitem{Schwetz:2011zk}
T.~Schwetz, M.~Tortola, and J.~Valle,
\newblock New J.Phys. {\bf 13}, 109401 (2011), arXiv:1108.1376.

\bibitem{Learned:1994wg}
J.~G. Learned and S.~Pakvasa,
\newblock Astropart. Phys. {\bf 3}, 267 (1995), arXiv:hep-ph/9405296.

\bibitem{Rodejohann:2006qq}
W.~Rodejohann,
\newblock JCAP {\bf 0701}, 029 (2007), hep-ph/0612047.

\bibitem{Serpico:2005sz}
P.~D. Serpico and M.~Kachelriess,
\newblock Phys. Rev. Lett. {\bf 94}, 211102 (2005), hep-ph/0502088.

\bibitem{Serpico:2005bs}
P.~D. Serpico,
\newblock Phys. Rev. {\bf D73}, 047301 (2006), hep-ph/0511313.

\bibitem{Winter:2006ce}
W.~Winter,
\newblock Phys. Rev. {\bf D74}, 033015 (2006), hep-ph/0604191.

\bibitem{Xing:2006xd}
Z.-z. Xing,
\newblock Phys. Rev. {\bf D74}, 013009 (2006), hep-ph/0605219.

\bibitem{Majumdar:2006px}
D.~Majumdar and A.~Ghosal,
\newblock Phys.Rev. {\bf D75}, 113004 (2007), arXiv:hep-ph/0608334.

\bibitem{Blum:2007ie}
K.~Blum, Y.~Nir, and E.~Waxman,
\newblock (2007), arXiv:0706.2070 [hep-ph].

\bibitem{Awasthi:2007az}
R.~L. Awasthi and S.~Choubey,
\newblock Phys.Rev. {\bf D76}, 113002 (2007), arXiv:0706.0399.

\bibitem{Hwang:2007na}
G.-R. Hwang and K.~Siyeon,
\newblock (2007), arXiv:0711.3122 [hep-ph].

\bibitem{Pakvasa:2007dc}
S.~Pakvasa, W.~Rodejohann, and T.~J. Weiler,
\newblock JHEP {\bf 02}, 005 (2008), arXiv:0711.4517.

\bibitem{Donini:2008xn}
A.~Donini and O.~Yasuda,
\newblock (2008), arXiv:0806.3029.

\bibitem{Quigg:2008ab}
C.~Quigg,
\newblock (2008), arXiv:0802.0013 [hep-ph].

\bibitem{Choubey:2008di}
S.~Choubey, V.~Niro, and W.~Rodejohann,
\newblock Phys.Rev. {\bf D77}, 113006 (2008), arXiv:0803.0423.

\bibitem{Xing:2008fg}
Z.-z. Xing and S.~Zhou,
\newblock Phys.Lett. {\bf B666}, 166 (2008), arXiv:0804.3512.

\bibitem{Beacom:2003nh}
J.~F. Beacom, N.~F. Bell, D.~Hooper, S.~Pakvasa, and T.~J. Weiler,
\newblock Phys. Rev. {\bf D68}, 093005 (2003), hep-ph/0307025,
\newblock Erratum-ibid.D72, 019901 (2005).

\bibitem{IcCascade:2011ui}
IceCube, R.~Abbasi {\em et~al.},
\newblock (2011), arXiv:1101.1692.

\bibitem{Xing:2006uk}
Z.-Z. Xing and S.~Zhou,
\newblock Phys.Rev. {\bf D74}, 013010 (2006), arXiv:astro-ph/0603781.

\bibitem{Lai:2009ke}
K.-C. Lai, G.-L. Lin, and T.~Liu,
\newblock Phys.Rev. {\bf D80}, 103005 (2009), arXiv:0905.4003.

\bibitem{Choubey:2009jq}
S.~Choubey and W.~Rodejohann,
\newblock Phys. Rev. {\bf D80}, 113006 (2009), arXiv:0909.1219.

\bibitem{Esmaili:2009dz}
A.~Esmaili and Y.~Farzan,
\newblock Nucl. Phys. {\bf B821}, 197 (2009), arXiv:0905.0259.

\bibitem{Lai:2010tj}
K.-C. Lai, G.-L. Lin, and T.~Liu,
\newblock Phys.Rev. {\bf D82}, 103003 (2010), arXiv:1004.1583.

\bibitem{Bustamante:2010nq}
M.~Bustamante, A.~Gago, and C.~Pena-Garay,
\newblock JHEP {\bf 1004}, 066 (2010), arXiv:1001.4878.

\bibitem{Schwetz:2008er}
T.~Schwetz, M.~A. Tortola, and J.~W.~F. Valle,
\newblock New J. Phys. {\bf 10}, 113011 (2008), arXiv:0808.2016.

\bibitem{Huber:2009cw}
P.~Huber, M.~Lindner, T.~Schwetz, and W.~Winter,
\newblock JHEP {\bf 11}, 044 (2009), arXiv:0907.1896.

\bibitem{Huber:2003ak}
P.~Huber and W.~Winter,
\newblock Phys. Rev. {\bf D68}, 037301 (2003), hep-ph/0301257.

\bibitem{Tang:2009na}
J.~Tang and W.~Winter,
\newblock Phys. Rev. {\bf D80}, 053001 (2009), arXiv:arXiv:0903.3039.

\bibitem{Feldman:1997qc}
G.~J. Feldman and R.~D. Cousins,
\newblock Phys.Rev. {\bf D57}, 3873 (1998), arXiv:physics/9711021.

\bibitem{Abraham:2009uy}
Pierre Auger Collaboration, J.~Abraham {\em et~al.},
\newblock Phys.Rev. {\bf D79}, 102001 (2009), arXiv:0903.3385.

\bibitem{Karle:2010xx}
A.~Karle and f.~t.~I. Collaboration,
\newblock (2010), arXiv:1003.5715.

\bibitem{Anchordoqui:2004eb}
L.~A. Anchordoqui, H.~Goldberg, F.~Halzen, and T.~J. Weiler,
\newblock Phys. Lett. {\bf B621}, 18 (2005), hep-ph/0410003.

\bibitem{Bhattacharjee:2005nh}
P.~Bhattacharjee and N.~Gupta,
\newblock (2005), hep-ph/0501191.

\bibitem{Maltoni:2008jr}
M.~Maltoni and W.~Winter,
\newblock JHEP {\bf 07}, 064 (2008), arXiv:0803.2050.

\bibitem{Xing:2011zm}
Z.-z. Xing and S.~Zhou,
\newblock Phys.Rev. {\bf D84}, 033006 (2011), arXiv:1105.4114.

\bibitem{Bhattacharya:2011qu}
A.~Bhattacharya, R.~Gandhi, W.~Rodejohann, and A.~Watanabe,
\newblock JCAP {\bf 1110}, 017 (2011), arXiv:1108.3163.

\bibitem{Farzan:2002ct}
Y.~Farzan and A.~Y. Smirnov,
\newblock Phys. Rev. {\bf D65}, 113001 (2002), hep-ph/0201105.

\bibitem{Beacom:2002vi}
J.~F. Beacom, N.~F. Bell, D.~Hooper, S.~Pakvasa, and T.~J. Weiler,
\newblock Phys. Rev. Lett. {\bf 90}, 181301 (2003), hep-ph/0211305.

\bibitem{Beacom:2003zg}
J.~F. Beacom, N.~F. Bell, D.~Hooper, S.~Pakvasa, and T.~J. Weiler,
\newblock Phys. Rev. {\bf D69}, 017303 (2004), hep-ph/0309267.

\bibitem{Meloni:2006gv}
D.~Meloni and T.~Ohlsson,
\newblock Phys. Rev. {\bf D75}, 125017 (2007), hep-ph/0612279.

\bibitem{Majumdar:2007mp}
D.~Majumdar,
\newblock (2007), arXiv:0708.3485 [hep-ph].

\bibitem{Bhattacharya:2009tx}
A.~Bhattacharya, S.~Choubey, R.~Gandhi, and A.~Watanabe,
\newblock Phys. Lett. {\bf B690}, 42 (2010), arXiv:0910.4396.

\bibitem{Bhattacharya:2010xj}
A.~Bhattacharya, S.~Choubey, R.~Gandhi, and A.~Watanabe,
\newblock JCAP {\bf 1009}, 009 (2010), arXiv:1006.3082.

\bibitem{Mehta:2011qb}
P.~Mehta and W.~Winter,
\newblock JCAP {\bf 1103}, 041 (2011), arXiv:1101.2673.

\bibitem{Lindner:2001fx}
M.~Lindner, T.~Ohlsson, and W.~Winter,
\newblock Nucl.Phys. {\bf B607}, 326 (2001), arXiv:hep-ph/0103170.

\bibitem{Lindner:2001th}
M.~Lindner, T.~Ohlsson, and W.~Winter,
\newblock Nucl. Phys. {\bf B622}, 429 (2002), arXiv:astro-ph/0105309.

\bibitem{Blennow:2005yk}
M.~Blennow, T.~Ohlsson, and W.~Winter,
\newblock JHEP {\bf 06}, 049 (2005), hep-ph/0502147.

\bibitem{Muecke:2002bi}
A.~M{\"u}cke, R.~J. Protheroe, R.~Engel, J.~P. Rachen, and T.~Stanev,
\newblock Astropart. Phys. {\bf 18}, 593 (2003), arXiv:astro-ph/0206164.

\bibitem{Razzaque:2004yv}
S.~Razzaque, P.~Meszaros, and E.~Waxman,
\newblock Phys. Rev. Lett. {\bf 93}, 181101 (2004), arXiv:astro-ph/0407064.

\bibitem{Ando:2005xi}
S.~Ando and J.~F. Beacom,
\newblock Phys. Rev. Lett. {\bf 95}, 061103 (2005), arXiv:astro-ph/0502521.

\bibitem{Razzaque:2005bh}
S.~Razzaque, P.~Meszaros, and E.~Waxman,
\newblock Mod. Phys. Lett. {\bf A20}, 2351 (2005), arXiv:astro-ph/0509729.

\bibitem{Razzaque:2009kq}
S.~Razzaque and A.~Y. Smirnov,
\newblock JHEP {\bf 03}, 031 (2010), arXiv:0912.4028.

\bibitem{Melrose:1980gb}
D.~B. {Melrose},
\newblock {\em {Plasma astrophysics: Nonthermal processes in diffuse magnetized
  plasmas. Volume 1 - The emission, absorption and transfer of waves in
  plasmas}} (Gordon and Breach Science Publishers, 1980).

\bibitem{Hillas:1985is}
A.~M. Hillas,
\newblock Ann. Rev. Astron. Astrophys. {\bf 22}, 425 (1984).

\bibitem{Medvedev:2003sx}
M.~V. Medvedev,
\newblock Phys. Rev. {\bf E67}, 045401 (2003), arXiv:astro-ph/0303271.

\bibitem{Protheroe:2004rt}
R.~J. Protheroe,
\newblock Astropart.Phys. {\bf 21}, 415 (2004), arXiv:astro-ph/0401523.

\bibitem{Hummer:2011ms}
S.~H{\"u}mmer, P.~Baerwald, and W.~Winter,
\newblock (2011), arXiv:1112.1076.

\bibitem{Becker:2005ej}
J.~K. Becker, M.~Stamatikos, F.~Halzen, and W.~Rhode,
\newblock Astropart. Phys. {\bf 25}, 118 (2006), arXiv:astro-ph/0511785.

\bibitem{Guetta:2000ye}
D.~{Guetta}, M.~{Spada}, and E.~{Waxman},
\newblock Astrophys. J. {\bf 557}, 399 (2001), arXiv:astro-ph/0011170.

\bibitem{Guetta:2001cd}
D.~Guetta, M.~Spada, and E.~Waxman,
\newblock Astrophys. J. {\bf 559}, 101 (2001), arXiv:astro-ph/0102487.

\bibitem{Ghirlanda:2011bn}
G.~Ghirlanda {\em et~al.},
\newblock (2011), arXiv:1107.4096.

\bibitem{Hopkins:2006bw}
A.~M. Hopkins and J.~F. Beacom,
\newblock Astrophys. J. {\bf 651}, 142 (2006), arXiv:astro-ph/0601463.

\bibitem{Kistler:2009mv}
M.~D. Kistler, H.~Y{\"u}ksel, J.~F. Beacom, A.~M. Hopkins, and J.~S.~B. Wyithe,
\newblock Astrophys. J. {\bf 705}, L104 (2009), arXiv:0906.0590.

\bibitem{Nava:2010ig}
L.~Nava, G.~Ghirlanda, G.~Ghisellini, and A.~Celotti,
\newblock A\&A {\bf 530}, A21+ (2011), arXiv:1012.2863.

\bibitem{Greiner:2009pm}
J.~Greiner {\em et~al.},
\newblock A\&A {\bf 498}, 89 (2009), arXiv:0902.0761.

\bibitem{Abdo:2009pg}
The Fermi/GBM, A.~A. Abdo {\em et~al.},
\newblock Astrophys. J. {\bf 706}, L138 (2009), arXiv:0909.2470.

\bibitem{Gruber:2011gu}
D.~Gruber {\em et~al.},
\newblock A\&A {\bf 528}, A15 (2011), arXiv:1101.1099.

\bibitem{PhDHummer}
S.~H{\"u}mmer,
\newblock {\em in preparation},
\newblock PhD thesis, W{\"u}rzburg university, 2012.

\bibitem{Autiero:2011hh}
D.~Autiero, P.~Migliozzi, and A.~Russo,
\newblock JCAP {\bf 1111}, 026 (2011), arXiv:1109.5378.

\bibitem{MacLachlan:new}
G.~A. MacLachlan {\em et~al.},
\newblock (2012), arXiv:1201.4431.

\bibitem{Mannheim:1995mm}
K.~Mannheim,
\newblock Astropart.Phys. {\bf 3}, 295 (1995).

\end{thebibliography}

\end{document}